\documentclass[useAMS,usenatbib,times,letter,amssymb]{mnras}
\usepackage{epsfig,times,amssymb,amsmath,verbatim,xspace}
\usepackage[usenames,dvipsnames,svgnames,table]{xcolor}
\usepackage{tablefootnote}
\usepackage{todonotes}
\usepackage{tabularx}
\usepackage{hyperref}
\usepackage{float}
\usepackage{hhline}
\usepackage{lineno}
\usepackage{changepage}
\usepackage{natbib,ifthen,soul} 

\usepackage{eso-pic}% http://ctan.org/pkg/eso-pic

\AddToShipoutPictureBG*{%
  \AtPageUpperLeft{%
    \hspace{0.72\paperwidth}%
    \raisebox{-4.5\baselineskip}{%
      \makebox[0pt][l]{\textnormal{DES-2021-0657}}
}}}%

\AddToShipoutPictureBG*{%
  \AtPageUpperLeft{%
    \hspace{0.72\paperwidth}%
    \raisebox{-5.5\baselineskip}{%
      \makebox[0pt][l]{\textnormal{FERMILAB-PUB-22-914-PPD}}
}}}%

\newcommand{\omegam}{\ensuremath{\Omega_\mathrm{m}}}
\newcommand{\omegab}{\ensuremath{\Omega_\mathrm{b}}}

\newcommand{\ns}{\ensuremath{n_\mathrm{s}}}
\newcommand{\lcdm}{$\Lambda$CDM}

\newcommand{\mpc}{\ensuremath{h^{-1} \mathrm{Mpc}}}
\newcommand{\rp}{\ensuremath{r_{\rm p}}}

\DeclareRobustCommand{\PRF}[3]{#2} 

\newcommand{\blockfont}[1]{{\textsc{#1}}\xspace}

\newcommand{\redmagic}{redMaGiC}
\newcommand{\metacal}{\blockfont{Metacalibration}}

\title[IAs with DES \& eBOSS]{The Dark Energy Survey Year 3 and eBOSS: constraining galaxy intrinsic alignments across luminosity and colour space}
\author[S.~Samuroff et al]{
\parbox{\textwidth}{
\Large
S.~Samuroff,$^{1,2}$\thanks{s.samuroff@northeastern.edu}
R.~Mandelbaum,$^{2,3}$
J.~Blazek,$^{1}$
A.~Campos,$^{2,3}$
N.~MacCrann,$^{4}$
G.~Zacharegkas,$^{5}$
A.~Amon,$^{6,7}$
J.~Prat,$^{5,8}$
S.~Singh,$^{2,3}$
J.~Elvin-Poole,$^{9}$
% ...
\begin{center} (DES Collaboration) \end{center}
}
\vspace{0.4cm}
\\
Author affiliations are listed at the end of the paper.
}

\pubyear{2022}

\begin{document}
\label{firstpage}
\pagerange{\pageref{firstpage}--\pageref{lastpage}}
\maketitle

\begin{abstract}
\vspace{-3mm}
\begin{adjustwidth}{15pt}{15pt}
We present direct constraints on galaxy intrinsic alignments using the Dark Energy Survey Year 3 (DES Y3), the Extended Baryon Oscillation Spectroscopic Survey (eBOSS) and its precursor, the Baryon Oscillation Spectroscopic Survey (BOSS). Our measurements incorporate photometric red sequence (redMaGiC) galaxies from DES with median redshift $z\sim0.2-1.0$, luminous red galaxies (LRGs) from eBOSS at $z\sim0.8$, and also a SDSS-III BOSS CMASS sample at $z\sim0.5$. We measure two point intrinsic alignment correlations, which we fit using a model that includes lensing, magnification and photometric redshift error. Fitting on scales $6<r_{\rm p} < 70$ Mpc$/h$, we make a detection of intrinsic alignments in each sample, at $5\sigma-22\sigma$ (assuming a simple one parameter model for IAs). Using these red samples, we measure the IA-luminosity relation. Our results are statistically consistent with previous results, but offer a significant improvement in constraining power, particularly at low luminosity. With this improved precision, we see detectable dependence on colour between broadly defined red samples. It is likely that a more sophisticated approach than a binary red/blue split, which jointly considers colour and luminosity dependence in the IA signal, will be needed in future. We also compare the various signal components at the best fitting point in parameter space for each sample, and find that magnification and lensing contribute $\sim2-18\%$ of the total signal. As precision continues to improve, it will certainly be necessary to account for these effects in future direct IA measurements. Finally, we make equivalent measurements on a sample of Emission Line Galaxies (ELGs) from eBOSS at $z\sim 0.8$. We report a null detection, constraining the IA amplitude (assuming the nonlinear alignment model) to be $A_1=0.07^{+0.32}_{-0.42}$ ($|A_1|<0.78$ at $95\%$ CL).
\end{adjustwidth}
\end{abstract}
\begin{keywords}
\vspace{-8mm}
cosmological parameters - cosmology: observations - gravitational lensing: weak - galaxies: statistics
\end{keywords}

% ------------- 1. Introduction ------------------------------------

\section{Introduction}\label{sec:intro}

The study of cosmic shear as a probe of the large scale structure of the Universe has developed rapidly over the past decade. Although its potential was recognised some time ago (see e.g. \citealt{jain97, hu99}), only more recently have high-precision cosmological constraints been possible. In the past ten years, data sets have grown to the point where weak lensing measurements alone have roughly comparable power to constrain certain cosmological parameters as the Cosmic Microwave Background (CMB) temperature fluctuations. Galaxy weak lensing and the CMB are both sensitive to the amplitude of the matter power spectrum in the low-redshift Universe, $S_8$. Whereas lensing allows one to probe the late-time matter field directly, the primary temperature anisotropies of the CMB provide a somewhat more complicated route, relying on an extrapolation from the surface of last scattering to the present day. Ever since the results of \citet{heymans13} lensing measurements have given a typically lower $S_8$ than the CMB; interestingly, this finding holds across multiple lensing surveys, whose members have implemented their own independent, well tested, blind analyses (\citealt{kilbinger13,svcosmology,jee16,y1cosmicshear,hildebrandt17,hildebrandt20,hikage19, hamana20, asgari20}; \citealt*{y3-cosmicshear2}; \citealt{y3-cosmicshear1}). The current level of (dis)agreement in the full parameter space, as assessed using various different metrics \citep{y3-tensions}, is at the level of up to $\sim 2.5\sigma$ (although it differs significantly between surveys and probe combinations).

Future lensing surveys will have much smaller statistical uncertainties compared with the current generation, which will greatly increase the precision of weak lensing measurements. This will, in turn, improve our constraining power and help us make sense of the apparent tensions in the literature. It will also, however, require a much tighter control of modelling errors in order to avoid our analyses becoming dominated by systematic uncertainties. Although much progress has been made in recent years, and the methods for mitigating systematics are highly sophisticated, we still have some way to go, as a field. One outstanding gap in our understanding is the treatment of intrinsic alignments (IAs; \citealt{joachimi15,troxel15,kirk15,kiessling15}). 

IAs are shape correlations induced not by cosmological lensing, but by local interactions, which can mimic cosmic shear. Most obviously, galaxies that are physically close by to each other experience the same background tidal field, which couples their intrinsic shapes, inducing what are known as II correlations. Additionally, GI (shear-intrinsic) correlations are generated by the fact that the same foreground matter that interacts with foreground galaxies also lenses background sources. A significant amount of literature over the past few years has focused on developing analytic models for IAs, which allow them to be forward modelled and marginalised in cosmological analyses \citep{catelan01,mackey02,hirata04,bridle07,blazek15,blazek17,vlah20,fortuna20}. 

Perhaps the most well established approach is an analytic formalism that assumes the intrinsic shapes of galaxies are linear in the background tidal field, and frozen at the point of galaxy formation \citep{catelan01,hirata04}. What became known as the Linear Alignment  (LA) model predicts both the GI and II power spectra and is, by convention, normalised such that the free amplitude $A_1$ is very roughly one for a typical lensing source sample (i.e. a mixed colour sample, dominated by blue galaxies at $z\lesssim1$). A few years later, \citet{hirata07} and \citet{bridle07} introduced a modification, whereby the full nonlinear matter power spectrum is used in place of the linear version in the LA model equations. This has been shown to improve the performance of the model on scales $\sim$ a few \mpc~ \citep{blazek11}. More recently, \citet{blazek15} and \citet{blazek17} take further steps along this route. The perturbative model developed in those papers (known as the Tidal Alignment and Tidal Torque model, TATT) extends the LA model to include higher order terms. Although in principle there are specific physical mechanisms for how correlations that are, for example, quadratic in the tidal field arise, in practice the model is agnostic to the underlying physics. An alternative approach, which is more closely connected with the physics on sub-halo scales, is to use a version of the halo model. The basic concept was introduced a decade ago \citep{schneider10}, and more recently \citet{fortuna20} took significant steps towards developing a practical implementation. 

Although a useful tool for learning about IAs, pure theory cannot provide a complete picture. Real data is very much necessary for properly understanding their behaviour in the real universe. Broadly, measurements can be classified as \emph{direct} (i.e. using a statistic that is dominated by IAs, with little or no contribution from lensing), or \emph{simultaneous} (i.e. where IAs contribute only a small part of the total signal, and are inferred alongside cosmological and other parameters). There have also been studies that have sought to do something in between, using particular combinations of lensing data to try to isolate an IA signal (e.g. \citealt{zhang10,blazek12}). By this definition, almost all cosmic shear studies to date are simultaneous IA measurements. Although comparison is complicated by non-trivial differences in the samples and measurement methods, as well as the high-dimensional model space, such constraints have typically found IA amplitudes for mixed lensing samples in the range $A_1\sim 0.1-1$ (\citealt{hildebrandt20,asgari20,y1cosmicshear}; \citealt*{y3-cosmicshear2}). A smaller number of works have attempted to understand how IAs enter simultaneous measurements in more detail. For example \citet{heymans13} split the CFHTLenS source catalogue into early and late types and perform independent cosmic shear analyses; they report $A_1\sim0$ in the bluer population, and $A_1\sim5$ in early types (albeit with large error bars). Several years later, \citet{y1coloursplit} implemented a similar colour-split methodology to explore IAs in DES Y1, this time analysing red and blue galaxies along with their cross-correlations simultaneously. Assuming the NLA model, that work found qualitatively similar results, with blue galaxies consistent with zero alignments and $A_1\sim3$ in the red population. Using the TATT model, it found the quadratic alignment amplitude $A_2$ to be $<0$ at the level of $\sim 2\sigma$ in both colour samples.

Direct measurements are typically restricted by the need for precise estimates for the redshifts of individual galaxies, and for this reason have tended to focus on bright red samples. A number of such studies have been carried out over the years \citep{hirata07,joachimi11,singh15,johnston19,fortuna21}, and the alignment strength as a function of luminosity is relatively well measured in brighter populations. Since these samples tend to have compact redshift distributions, any given study only weakly (if at all) constrains the redshift evolution of IAs. In the case of bluer galaxies, a handful of direct measurements have been attempted \citep{mandelbaum10,tonegawa17,johnston19}, but the samples here are typically small; though they make null detections and place upper limits on the IA amplitude, the error bars are wide enough to allow for significantly non-zero values. Although analogous IA measurements can (and have) been made on hydrodynamic simulations, these are limited by finite box size, difficulty in constructing realistic galaxy samples and the accuracy of the simulations themselves \citep{codis15a,hilbert16,samuroff21}. 

This paper falls into the category of direct measurements, and represents the first such exercise using DES. We use a combination of DES redMaGiC (or the ``red sequence Matched filter Galaxy Catalog"; photometric, but with precise per-galaxy redshift estimates) and the overlapping BOSS and eBOSS surveys (spectroscopic) to measure IA correlations in physical space, which we then fit using a range of IA models. This work follows implicit IA constraints from DES cosmic shear (\citealt*{y3-cosmicshear2}; \citealt{y3-cosmicshear1}) and galaxy-galaxy lensing (\citealt{y3-gglensing}; \citealt*{y3-shearratio}). We should note that, while they make use of the same DES catalogues, the samples in these earlier works are significantly different from ours, and so we do not expect the IA signal to be the same.

The paper is structured as follows. In Section \ref{sec:data} we describe the various data sets used in this work. Section \ref{sec:measurements} then outlines measurements on these data, including redshifts, calibrated galaxy shapes, and two point correlations. In Section \ref{sec:theory} we set out the model used fit those measurements, and discuss our analysis choices such as priors and scale cuts; a range of validation tests of that theory pipeline, using real and simulated data, are outlined in Section \ref{sec:validation}. Our main results are discussed in Section \ref{sec:results}. We conclude in Section \ref{sec:conclusion}.

% ------------- 2. Data ------------------------------------

\section{Data}\label{sec:data}

In this section we briefly describe the data sets used in this work, and how the various galaxy samples are defined.

\begin{figure*}
\includegraphics[width=2.\columnwidth]{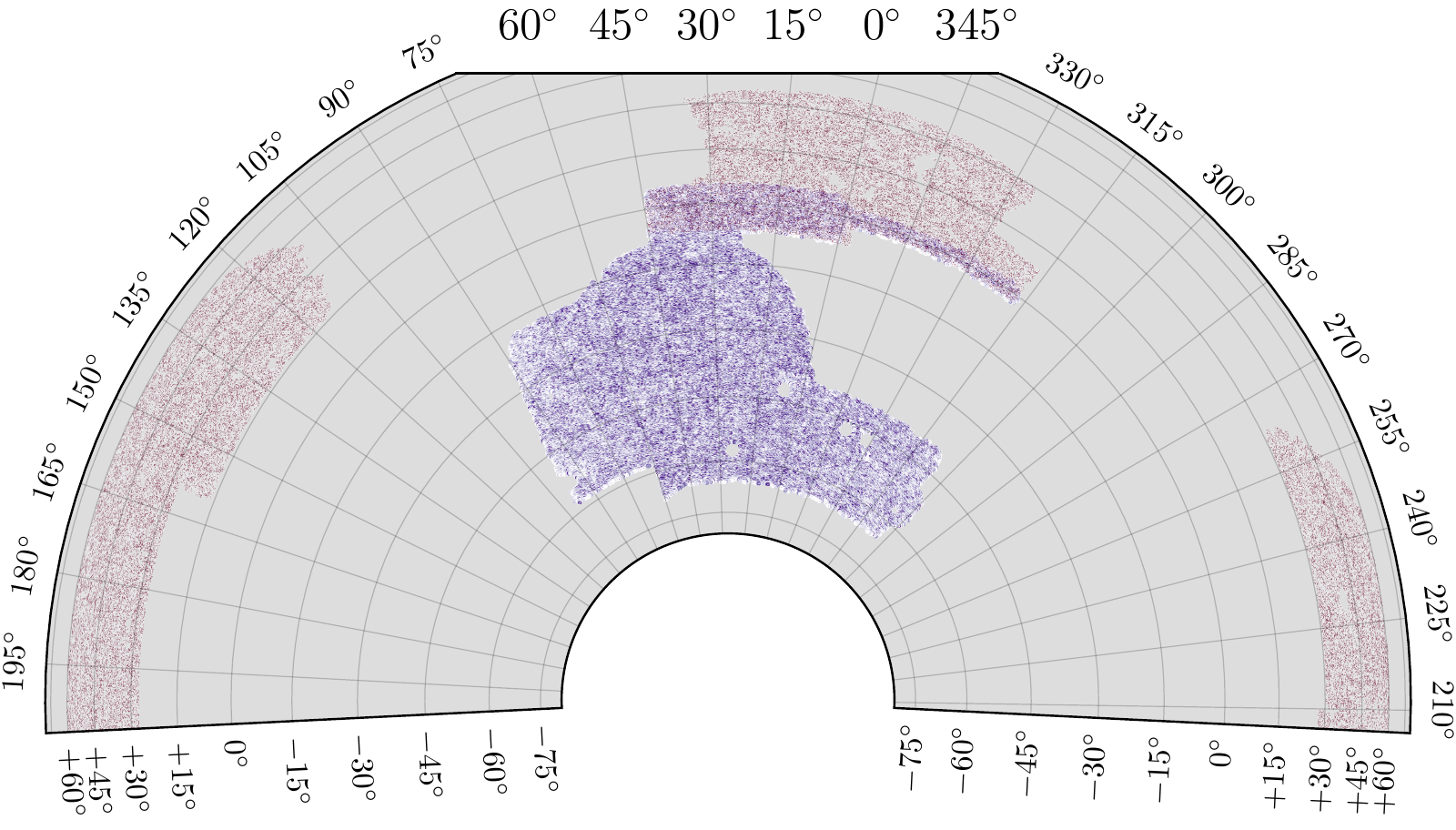}
\caption{The approximate footprint of the Dark Energy Survey (purple) and SDSS eBOSS (pink). The overlap in the Stripe 82 region (across the beak and head of the hummingbird shape) is roughly 600 square degrees.
}
\label{fig:data:footprint}
\end{figure*}

\begin{table*}
\begin{center}
\begin{tabular}{c|ccccccccc}
\hline
                   & No. of Galaxies & $f_{D}$ & Area [sq. deg.] & $\bar{n}_{\rm c}$ $[10^{4} h^{3}$Mpc$^{-3}]$ & $\sigma_e$ & Mean redshift $\langle z \rangle$ & $R_S  \times 10^4$ & $\langle M_r-M_z \rangle$ & $\langle M_r \rangle$  \\
\hhline{==========}
\redmagic~high-$z$ & 754,574 & 1.13 & 4203            & $3.22$                                 & 0.28       & 0.78      & 4.8                & 0.43       & -21.6   \\
\redmagic~low-$z$  & 1.61 M & 1.14  & 4203            & $7.64$                                 & 0.26       & 0.46      & 23.1               & 0.43       & -20.9  \\
eBOSS ELGs         & 92,954 & 1.10  & 604             & $1.95$                                 & 0.24       & 0.84      & -4.2               & -       & -  \\
eBOSS LRGs         & 22,244 & 3.03   & 604             & $0.55$                                 & 0.26       & 0.75      & -4.7               & 0.37       & -21.9  \\
CMASS              & 49,820 & 3.35  & 604             & $4.11$                                 & 0.23       & 0.52      & -4.0               & 0.36           & -21.8 \\
\hline
\end{tabular}
\caption{A summary of the properties of the shape samples used in this work. From left, we show the total number of galaxies after cuts; the area of the footprint in square degrees and the mean comoving number density (averaged over redshift). The value $f_{D}$ indicates the number of galaxies in the density tracer sample relative to the size of the shape sample. Note that this reflects both the impact of \blockfont{metacalibration} cuts and, in the SDSS samples, the geometric selection. The ellipticity dispersion $\sigma_e$ is defined according to \citealt{heymans12} (see also \citealt*{y3-shapecatalog} Eq. 13). Also shown are the ensemble mean redshift $\langle z \rangle$; the selection response for each sample (the shear response due to the galaxy shape catalogue cuts; see \citealt{sheldon17}) and the mean rest frame colour and $r-$band magnitude. The means incorporate the weights described in Section \ref{sec:measurements:weights}.
}\label{tab:data:summary1}
\end{center}
\end{table*}

\subsection{The Dark Energy Survey Year 3}

The Dark Energy Survey is a six-year programme encompassing $\sim5000$ square degrees of the Southern sky using the V\'{i}ctor Blanco telescope in Chile. The approximate footprint is shown (purple) in Figure \ref{fig:data:footprint}. Images were taken in five photometric bands ($g$, $r$, $i$, $z$, and $Y$) with a nominal depth of magnitude $r\sim24.1$ (at full Y6 depth); although all five are used for redshift estimation, galaxy shape measurements are limited to $riz$ due to difficulties in accurately estimating the point spread function in the $g$-band \citep{jarvis21}, and shallower imaging in the $Y$-band (45s exposures as opposed to 90s in $griz$). The work described in this paper is based on data collected during the first three years of operation (the Y3 data). These data cover the full area at slightly less than the full depth, with an average of about 5 exposures per galaxy. A description of the image processing and reduction pipeline, including background subtraction and object detection can be found in \citet{morganson18} and \citet*{y3-gold}. The photometric data set, before any further cuts, is known as the \blockfont{gold} catalogue \citep*{y3-gold}. In Y3, this has a limiting magnitude $i\sim23$. Per-galaxy photometry  measurements are obtained using multi-object fitting \citep[MOF;][]{y1gold}, and shapes are measured using the \metacal~algorithm (\citealt*{y3-shapecatalog}; see Section \ref{sec:measurements:shapes}). 

\subsection{BOSS \& eBOSS}

We also make use of galaxies from both the Baryon Oscillation Spectroscopic Survey (BOSS) and the Extended Baryon Oscillation Spectroscopic Survey (eBOSS) in this paper, and so we discuss both briefly here. The former is a spectroscopic sample collected as part of SDSS-III \citep{eisenstein11}. Imaging in five photometric bands ($ugriz$) and spectroscopy for BOSS were performed using the 2.5 metre Sloan Telescope at Apache Point Observatory in New Mexico \citep{gunn06,smee13}. The observing program took place between autumn 2009 and spring 2014, and covered more than 1.5M galaxies across 10,000 square degrees of high latitude sky. One can find a description of the BOSS spectrographs and other details in \citet{dawson13}. 

The eBOSS data are slightly more recent, being taken over the period between summer 2014 and spring 2019, as part of SDSS-IV (\citealt{blanton17}; see also \citealt{dawson16} for a discussion of the differences between BOSS and eBOSS). Again, spectroscopy relied on the BOSS spectrographs on the Sloan Telescope. The method for target selection differs slightly depending on the nature of the galaxy sample. Emission line galaxies were targeted from the DECam Legacy Survey (DECaLS), which is deeper than SDSS, detections and photometry. DECaLS was carried out using the DECam on the Blanco telescope, and covers an area of 6700 square degrees in the region $-20 < \delta < +30$ degrees to a $5 \sigma$ limiting magnitude of 24.7, 23.9, and 23.0 in the $g$, $r$ and $z$ bands respectively (compared with 22.8, 22.3 and 20.4 for SDSS; \citealt{delubac17}). Luminous red galaxy targets, on the other hand, were selected using SDSS $riz$ imaging and infrared sky maps from the Wide-field Infrared Survey Explorer (WISE; \citealt{wright10}). 

The BOSS and eBOSS footprints are divided in two approximately equal area regions; the emission line galaxy (ELG) and luminous red galaxy (LRG) samples used in this work come from the Southern Galactic Cap (SGC), which contains roughly $600$ square degrees of overlap with the DES footprint.

\subsection{Primary Galaxy Samples}

In this paper we consider five distinct samples (for a summary, see Table \ref{tab:data:summary1}).
These are:
\begin{itemize}
	\item \textbf{eBOSS Emission Line Galaxies}: Our eBOSS ELG sample contains $\sim 100,000$ galaxies in the SGC region. The target selection process in described in more detail in \citet{raichoor17}, and further discussion can be found in \citet{tamone20,raichoor21}. The cuts that define the sample are relatively complex, and include a $g-$band magnitude limit at $g=22.8$ mag, as well as selection in colour space designed to limit the redshift range. In total, we have 92,954 galaxies, with a mean redshift of about 0.8. Although this sample covers a similar range in redshift to the LRGs described below, they are significantly bluer than any of our other samples (both in apparent and rest frame magnitudes), and are thus not expected to exhibit strong intrinsic alignments.
	\item \textbf{eBOSS Luminous Red Galaxies}: A sample of luminous red galaxies (LRGs) from eBOSS DR16, drawn from the SGC region. Selection is performed using the criteria outlined in \citet{prakash16} (see also \citealt{ross20,bautista21,rossi21} for details on the LRG sample); the cuts are primarily on extinction corrected magnitudes and colours ($z<19.95$ mag). The redshift coverage is similar to that of the ELG sample, with a mean of $z\sim0.8$, and the total number of galaxies is 22,244.
	\item \textbf{BOSS CMASS}: The CMASS selection algorithm is described in detail by \citet{reid16} (see their Sec 3.3 and references therein). We additionally impose redshift cuts at $z<0.6$ to ensure there is no overlap with the eBOSS LRGs, and at $z>0.4$ to remove outliers. This leaves us with 49,820 galaxies. 
	\item \textbf{\redmagic~high-$z$ (RMH)}: A sample of red sequence galaxies from the DES Y3 \redmagic~catalogue. These objects are selected using the algorithm outlined in \citet{rozo16}. In brief, all detected galaxies are fitted using a red sequence template, yielding a best fitting redshift, $z_{\rm redmagic}$, and a derived luminosity $L$, as well as a corresponding $\chi^2$. If the galaxy falls above a minimum $L$ and below a maximum $\chi^2$, it is included in the catalogue. The $\chi^2$ threshold is a function of redshift, such that the comoving density is constant (\citealt{rozo16} Sec. 3.3). This process gives a set of bright red galaxies with both well constrained per-object photometric redshifts ($\sigma_z/(1+z)<0.02$) and well understood redshift error. Our high-$z$ sample consists of the upper two redshift bins of the lens sample used in \citet{y3-3x2ptkp}, cut at $z>0.6$. The luminosity threshold is $L_{\rm min}=L_*$, where $L_*$ is a characteristic luminosity, as defined in \citet{rozo16} Sec. 3.1. The sample before shape cuts comprises $\sim 0.8$ M galaxies over 4203 square degrees. The redshift distribution is relatively compact and peaks at a similar value to our eBOSS samples at $z\sim0.8$.
	\item \textbf{\redmagic~low-$z$ (RML)}: Our low-$z$ \redmagic~sample is defined in a similar way to \redmagic~high-$z$, with key differences. Primarily, the luminosity threshold is lower at $L_{\rm min}=0.5 L_{*}$ \citep{y3-3x2ptkp}. A cut on $z_{\rm redmagic}$ is imposed at $z<0.6$,  equivalent to the three lower lens bins from \citet{y3-3x2ptkp}. Without shape cuts, the catalogue contains $1.84$ M objects, with a median redshift of $z\sim0.5$.
\end{itemize}

In each case we define density and shape tracer selections. The former make use of all galaxies passing the baseline cuts described above, and also are not required to be within the DES-SDSS overlap. In each sample, we obtain galaxy shape estimates by matching galaxies to the DES Y3 \metacal~catalogue \citep*{y3-shapecatalog}. We construct a KDTree of \metacal~galaxy angular positions, which is used to locate the nearest DES neighbour for each eBOSS or \redmagic~object. A match tolerance of 1 arcsecond is imposed to exclude spurious matches, and objects outside the overlap region between the two surveys. To obtain a subset of galaxies with reliable shapes, we then impose the cuts recommended by \citet*{y3-shapecatalog} (their Section 4.2), which includes selections based on size and signal-to-noise ratio, as well as a cut designed to remove binary star contamination. We show the estimated redshift distributions $n(z)$ for each sample in the top panel of Figure \ref{fig:data:nofz} (see Section \ref{sec:measurements:redshifts} for more detail about how these are estimated), and the comoving number density $n_c(z)$ in the bottom.      

\subsection{Comparison Sample: LOWZ}

In addition to the five catalogues discussed above, we also make use of BOSS LOWZ \citep{dawson13} as a reference sample. The point of including these data is to test our measurement and inference pipelines by comparing against the baseline analysis of \citet{singh15}. LOWZ is a convenient choice for this, since there are relatively detailed published results using a very similar methodology to our own. 

LOWZ is a sample of LRGs from BOSS DR11. The sample covers a footprint of 9243 square degrees and is approximately volume-limited over the redshift range $0.16<z<0.36$; a sharp cut off is imposed at these bounds. Unlike the other samples, we do not match to Y3 \metacal~to obtain shape estimates, but simply use the pre-existing catalogues \citep{reyes12}. For all other catalogue-level quantities (redshifts, $k+e$-corrected magnitudes etc), we likewise use the pre-computed columns (see \citealt{singh15} for details). After cuts, the LOWZ shape and density tracer samples contain 159,620 and 173,854  galaxies respectively.

% ------------- 3. Measurements ------------------------------------

\section{Measurements}\label{sec:measurements}

\subsection{Shapes}\label{sec:measurements:shapes}

The galaxy shapes for all samples except LOWZ are obtained by matching to the DES Y3 \metacal~catalogue. Discussion of the shape measurement algorithm, and catalogue level tests, can be found in \citet*{y3-shapecatalog}. The basic measurement is a maximum likelihood fit of an elliptical Gaussian to each galaxy. This process uses an MCMC, and is performed over multiple exposures and in bands $riz$ simultaneously. In order to calibrate biases due to image noise, model bias and other effects, the fit is repeated four times using artificially PSF-deconvolved and resheared images, a technique known as \metacal. For details of how the \metacal~corrections are applied in this particular context see Section \ref{sec:measurements:2pt}; for the general case and validation on simulations see \citet{y1shearcat,huff17,sheldon17}.  

\subsection{Galaxy Weights}\label{sec:measurements:weights}

For galaxy clustering and galaxy-shape measurements, we use the recommended weights for each sample respectively. Descriptions of these can be found in \citet{raichoor17} and \citet{ross20} (for eBOSS), \citet{reid16} (for CMASS) and \citet{y3-galaxyclustering} (redMaGiC). These are designed to correct for correlations between the observed galaxy number density and various survey properties, which can be induced by systematics. For the SDSS samples, there are additional weights designed to account for fibre collisions and redshift failures, which are combined as per the references above. 

It is worth noting briefly that previous works (see e.g. \citealt{ross20}) identified a possible systematic due to variations in the redshift distributions of the eBOSS samples within the SGC and NGC regions, which is not explicitly corrected by the weights. Although relatively mild for eBOSS LRGs, it was found to be significant enough to need correcting for in an RSD analysis using the ELG sample (\citealt{tamone20,bautista21, deMattia21}). We do not, however, believe this to be a significant concern for our analysis, given the fact that our IA constraint (from $w_{g+}+w_{++}$) is constrained to the DES-eBOSS overlap region, which is a relatively small part of the overall eBOSS SGC footprint. Although we do use the full area for $w_{gg}$, given that the result from ELGs is essentially a null detection (see Section \ref{sec:results:elgs}), we do not expect a small systematic affecting the galaxy bias to be a significant factor.

The shape catalogues for the different samples are all ultimately subsets of Y3 \metacal, and so we adopt the inverse variance weights discussed in \citet*{y3-shapecatalog}.

\subsection{Magnification Coefficients}\label{sec:measurements:magnification}

In addition to imprinting a coherent pattern in their shapes, lensing by large scale structure also modulates the observed brightness and size of galaxies, an effect known as magnification. In order to model the impact on our galaxy number counts, we require an estimate for the slope of the faint end of the galaxy luminosity function for each of our density tracer samples (see e.g. \citealt{mandelbaum05}; \citealt*{y3-2x2ptmagnification} and \citealt{joachimi10}'s Appendix A). Our fiducial estimates are derived via what we refer to as the ``flux-only'' method \citep*{y3-2x2ptmagnification}. In the cases of eBOSS, CMASS and LOWZ the process is straightforward. For a particular catalogue containing $N_0$ galaxies, with a given pre-existing selection function, we apply a small achromatic shift $\delta m$ to the observed magnitudes. We reapply the magnitude cuts using this perturbed catalogue, and count how many galaxies are lost to the bright-end cut $\delta N_-$. The sign of $\delta m$ is then flipped, and the process repeated to estimate the number shifted up over the faint-end threshold $\delta N_+$. The total change in number counts is then simply:
\begin{equation}
    \delta N (\delta \kappa) = \delta N_+ - \delta N_-
\end{equation}
\noindent
with $\delta \kappa = 0.5(10^{-\delta m / 2.5} - 1)$.
For small perturbations we can measure the slope of $\delta N (\delta \kappa)/N_0$ with $\delta \kappa$ numerically. This gives us a quantity \citet*{y3-2x2ptmagnification} refer to as $C_{\rm sample}$, which describes the linear response of the observed galaxy number density to a small change in $\kappa$. We define a quantity referred to as the magnification coefficient as $\alpha=C_{\rm sample}/2$ (see Table \ref{tab:theory:cls} and Section \ref{sec:theory} for how this enters the theory predictions).

For \redmagic, the sample selection is more complex. For this reason, we start with the Y3 \blockfont{GOLD} catalogue \citep*{y3-gold}, perturb the magnitudes, and re-run the \redmagic~algorithm for each $\delta m$. We then estimate $\alpha$ in the same way as before. We find $\alpha^{\rm RML} = 1.101$ for \redmagic~low-$z$ and $\alpha^{\rm RMH} = 1.719$ for \redmagic~high-$z$. For our LRG, ELG and CMASS samples we find $\alpha^{\rm LRG}=2.020$, $\alpha^{\rm ELG}=1.177$ and $\alpha^{\rm CMASS}=0.529$ respectively.

In addition to the flux-only estimates, we have alternative values, derived using an algorithm called \blockfont{balrog} \citep{suchyta16,y3-balrog}: $\alpha^{\rm RMH}=2.11\pm0.32$ and $\alpha^{\rm RML}=0.20\pm0.29$ for \redmagic~high-$z$ and low-$z$. \blockfont{balrog} works by inserting additional synthetic galaxies into real photometric images. By running the detection and measurement processes on the altered \blockfont{balrog} images, one can sample the selection function of the survey and explore effects such as magnification and blending. Although these, in principle, capture size selection effects that the flux-only numbers cannot (see  \citealt*{y3-2x2ptmagnification} for discussion), they are also relatively noisy. We also have \blockfont{balrog} estimates for the \redmagic~samples only, and not CMASS/eBOSS. We thus use the flux-only estimates as our fiducial choice; we do, however, confirm that in the two redMaGiC samples our basic conclusions are unaffected by this choice (see Section \ref{sec:results:lensmag} and Figure \ref{fig:results:magnification} specifically).  

\subsection{Redshift Distributions}\label{sec:measurements:redshifts}

\begin{figure}
\includegraphics[width=\columnwidth]{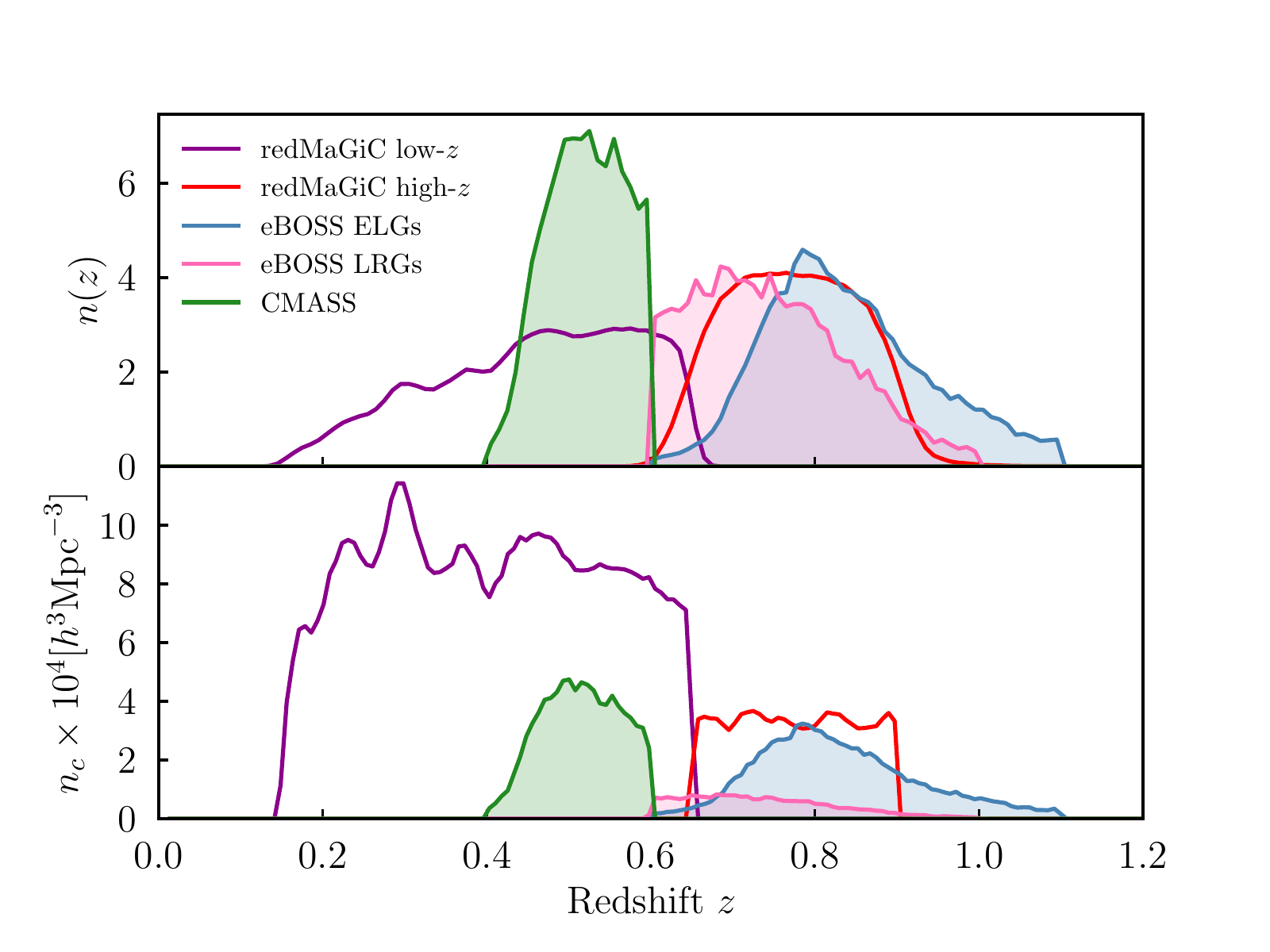}
\caption{\textbf{Top:} The estimated redshift distributions of the various shape samples used in this work. Each $n(z)$ is normalised to integrate to 1 over the redshift range shown. The $n(z)$s for the spectroscopic samples are shown as shaded curves, and are estimated as the histogram of single-galaxy $z$ estimates. The unshaded $n(z)$s are estimated by stacking random samples from the \redmagic~redshift PDFs. 
\textbf{Bottom:} The same, but showing comoving number density as a function of redshift. Note that $n_c$ is weakly cosmology-dependent, and so we assume the fiducial cosmology specified in Section \ref{sec:theory}. Note that both the $n(z)$ and $n_c(z)$ are qualitatively the same for the density tracer samples. The shape cuts remove galaxies, but do not change the shape or mean redshift of these distributions significantly.
}
\label{fig:data:nofz}
\end{figure}

\begin{figure}
\includegraphics[width=0.9\columnwidth]{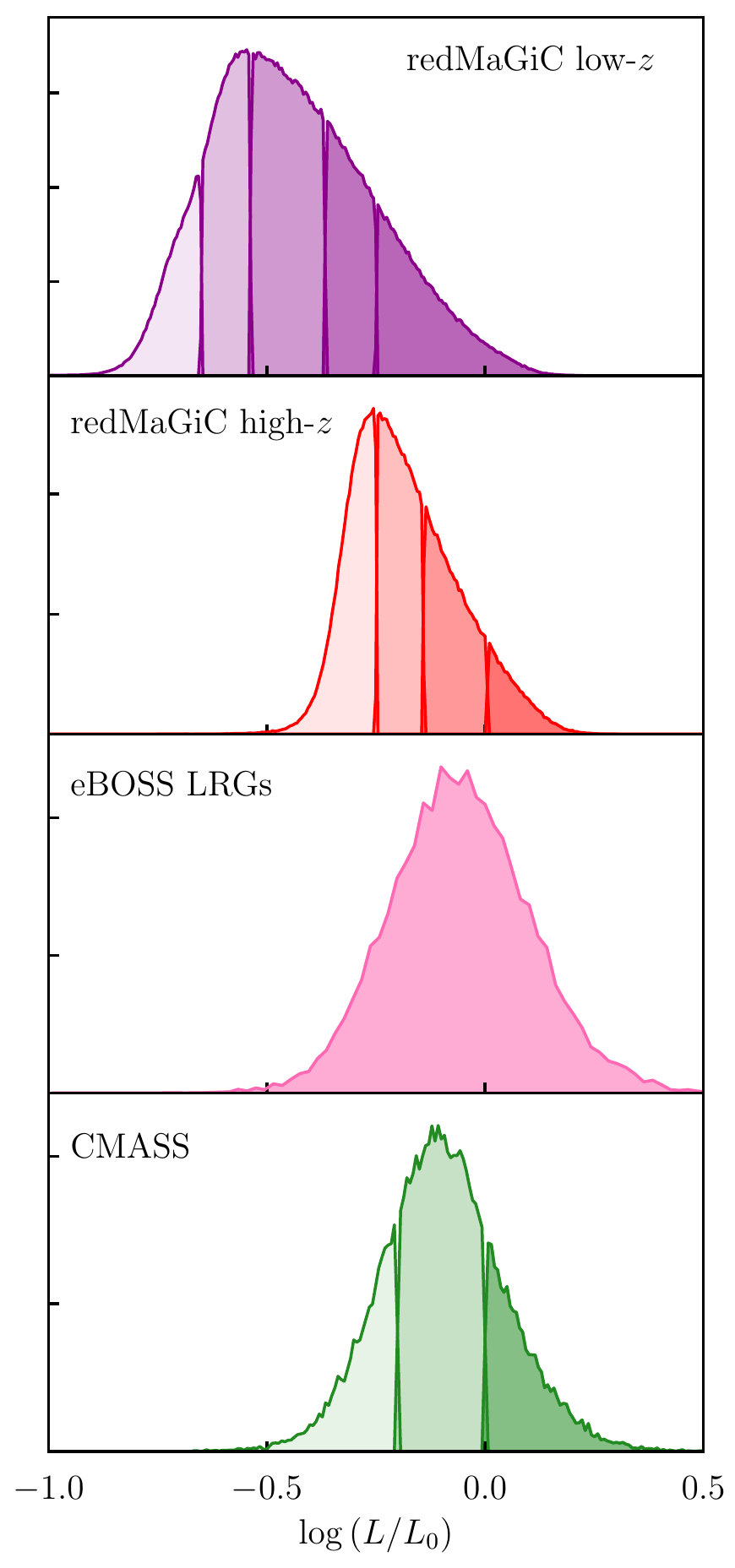}
\caption{The $r$-band luminosity distributions of the red samples used in this work. We divide each sample, with the exception of eBOSS LRGs, into roughly equal number bins in luminosity, as indicated by the shading. See Section \ref{sec:measurements:absmag} for details.
}
\label{fig:data:nofL}
\end{figure}

\subsubsection{Spectroscopic Redshifts for BOSS \& eBOSS}

For details on the BOSS and eBOSS spectroscopic redshift pipelines see \citet{comparat16}, \citet{hutchinson16} and \citet{bolton12}. In brief, galaxy spectra are collected using the BOSS spectrographs on the Sloan Telescope \citep{smee13}; the instrument has two identical spectrographs,  each of which has a red and a blue camera, collectively covering the wavelength range $360-1040$ nm, and 1000 optical fibres, 3 arcsec in diameter, and with a collision scale of 62 arcsec (corresponding to a physical scale of $\sim$ 0.6 \mpc~at $z=0.8$). Each object is observed in multiple exposures, which are 15 minutes in duration and can be distributed across several nights. All good data for a particular galaxy are co-added together during the spectroscopic data reduction process. Fits are made to each observed spectrum using a number of templates and combinations of templates evaluated for all allowed redshifts. A point estimate redshift is then obtained by maximising the likelihood. The estimated redshift distributions used in our theory modelling of the CMASS and eBOSS samples (the shaded curves in the top panel of Figure \ref{fig:data:nofz}) are, then, histograms of these point estimates. Note that in making these histograms, we apply the \metacal~weights described in Section \ref{sec:measurements:weights}.

\subsubsection{Photometric Redshifts}

Unlike with the SDSS samples, we do not have spectroscopic redshifts for our DES \redmagic~samples. Rather, for each galaxy, we have a redshift PDF, which is obtained using DES photometry. The \redmagic~algorithm \citep{rozo16} relies on the fact that red sequence galaxies have a relatively tight magnitude-colour-redshift relation, which can be calibrated using overlapping spectroscopic data \citep{y3-wpzlens}. Each candidate galaxy above some fixed luminosity threshold $L_{\rm min}$ is fit to obtain a likelihood $\mathcal{L}(z)$ and best $\chi^2$. The latter acts as a selection criterion, with $\chi^2_{\rm max}$ adjusted as a function of redshift to ensure approximately constant comoving density. Where it is necessary to have point redshift estimates (e.g. for the binning in Section \ref{sec:measurements:2pt}), we use the value that maximises the likelihood, $z_{\rm redmagic}$. We follow \citet{y3-3x2ptkp} and estimate the ensemble $n(z)$s by stacking samples from the full non-Gaussian redshift PDFs (see also \citealt{y3-2x2maglimforecast} for discussion). These are shown in Figure \ref{fig:data:nofz} (upper panel). 

In addition to the $n(z)$ for each sample and point estimates themselves, our modelling also requires an estimate for the per-galaxy redshift \emph{uncertainty} as a function of redshift. In the cases of eBOSS and CMASS, the spectral resolution allows very precise redshift estimates, and so we can assume this to be negligible. In the case of \redmagic~we obtain error estimates using a representative subsample of the Y3 \redmagic~catalogues with spectra (see \citealt{y3-2x2ptbiasmodelling}). Specifically, we divide the sample into bins of $z_{\rm spec}$, and within each bin we evaluate the histogram $p([z_{\rm samp} - z_{\rm spec}] / [1+z_{\rm spec}] )$, where $z_{\rm samp}$ are the PDF draws used in estimating the $n(z)$ above. Since we have four PDF samples per galaxy, we compute four histograms, and average them, giving us a noisy estimate for the redshift error in the bin centred on $z_{\rm spec}$. We find that the histograms are well approximated by a Gaussian distribution, and so we fit each histogram to obtain a width $\sigma_z$. This process leaves us with $\sigma_z(z)$, an estimate for the redshift scatter as a function of redshift, which we interpolate and incorporate into the modelling described in Section \ref{sec:modelling:2pt_theory:photoz}. Although there is some slight variation with redshift, a constant $\sigma_z/(1+z)\sim0.01$ is a reasonable approximation, with $\sigma_z/(1+z)<0.02$ over the range $z=0.2-1.1$ (see \citealt{y3-2x2maglimforecast}, and in particular their Fig. 1).

\subsection{Luminosities, Colours \& Absolute Magnitudes}\label{sec:measurements:absmag}

To obtain rest frame absolute magnitudes for our galaxy catalogues, we first convert the best-fit $r$-band fluxes from \metacal~to apparent magnitudes, $r = 30 - 2.5\mathrm{log}f_r$. The corresponding absolute magnitude is then given by:
\begin{equation}
    M_r^i = r^i - 5 \left ( \mathrm{log}D_l(z^i)  - 1 \right ) - K(z^i),
\end{equation}
where the index $i$ denotes a galaxy, $z^i$ is the best point estimate redshift for that galaxy, and $D_l$ is the corresponding luminosity distance. Note that $D_l$ \emph{is in units of pc$/h$}. We calculate a $k+e-$correction $K$ for each galaxy based on the redshift using the stellar synthesis models of \citet{bruzual03}. In brief, we employ two models: one assuming a passively evolving Spectral Energy Distribution (SED), and the other passive but with a single instantaneous burst of star formation at $z=9.84$. These models give us predicted colours and a $k+e-$correction as a function of $z$. For each galaxy $i$, we then compare the observed $r-i$ colour with the model predictions; if the observed colour is redder than the predicted one from the passive model, we use that model. If it is bluer than the one from the passive plus star formation burst model, then we use that one. Otherwise, we calculate a weighted average of the two $k+e-$corrections. In all cases, we correct the magnitudes to $z=0$. Note that these star formation models are designed to describe elliptical galaxies, and we do not apply them to our ELG sample.

The above procedure is based on the assumption that the overall stellar population in a given galaxy sample is a mixture of two sub-populations, such that the observed colours are a linear combination of the colours of those components; these observed colours are therefore subject to a linear combination of the associated $k+e-$corrections. Note that in practice the templates do not differ enormously over the redshift range of our samples. Indeed, we recompute the $k+e-$corrections using the two models separately, and find no significant change in the distributions shown in Figure \ref{fig:data:nofL}.

\begin{figure}
\includegraphics[width=\columnwidth]{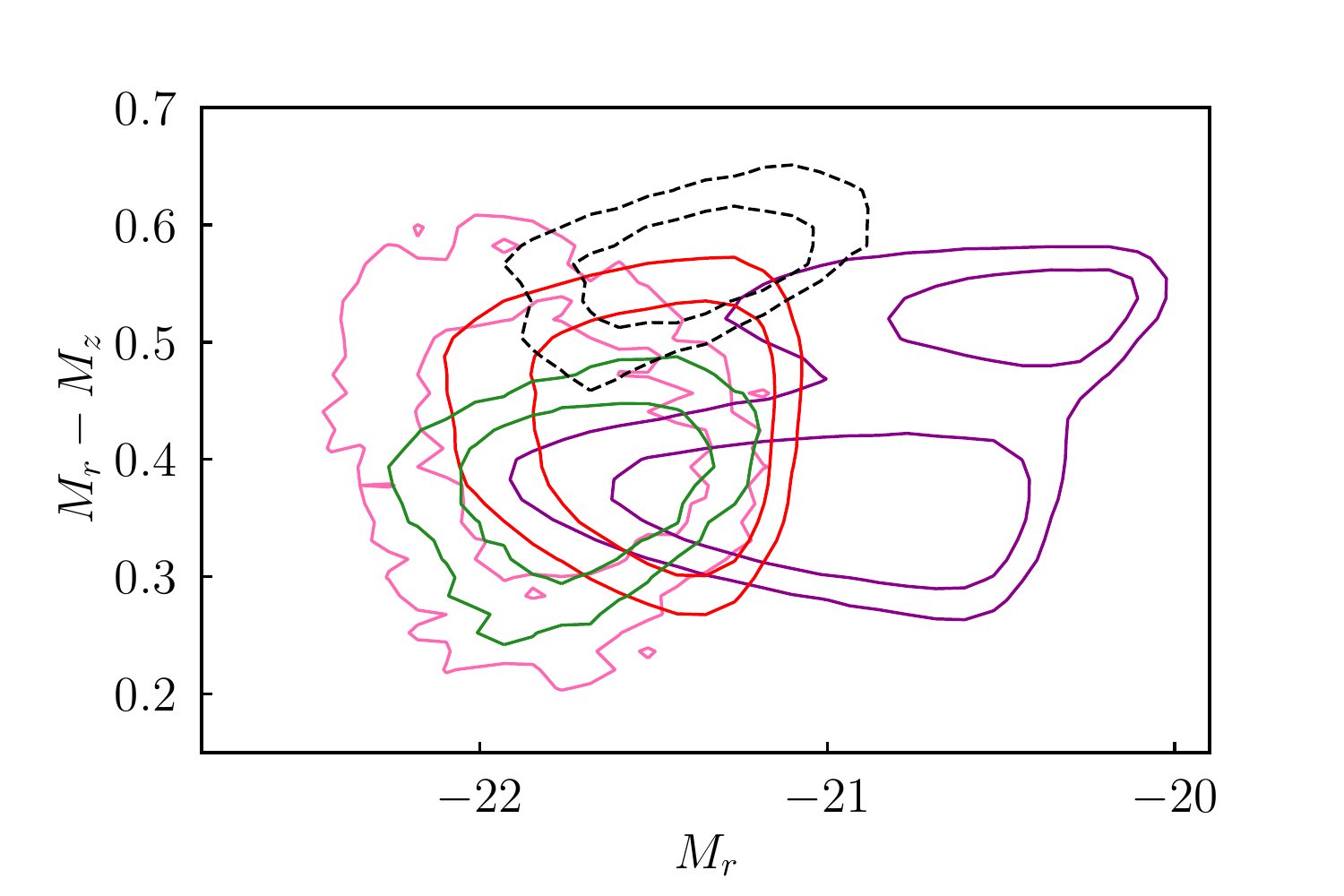}
\includegraphics[width=\columnwidth]{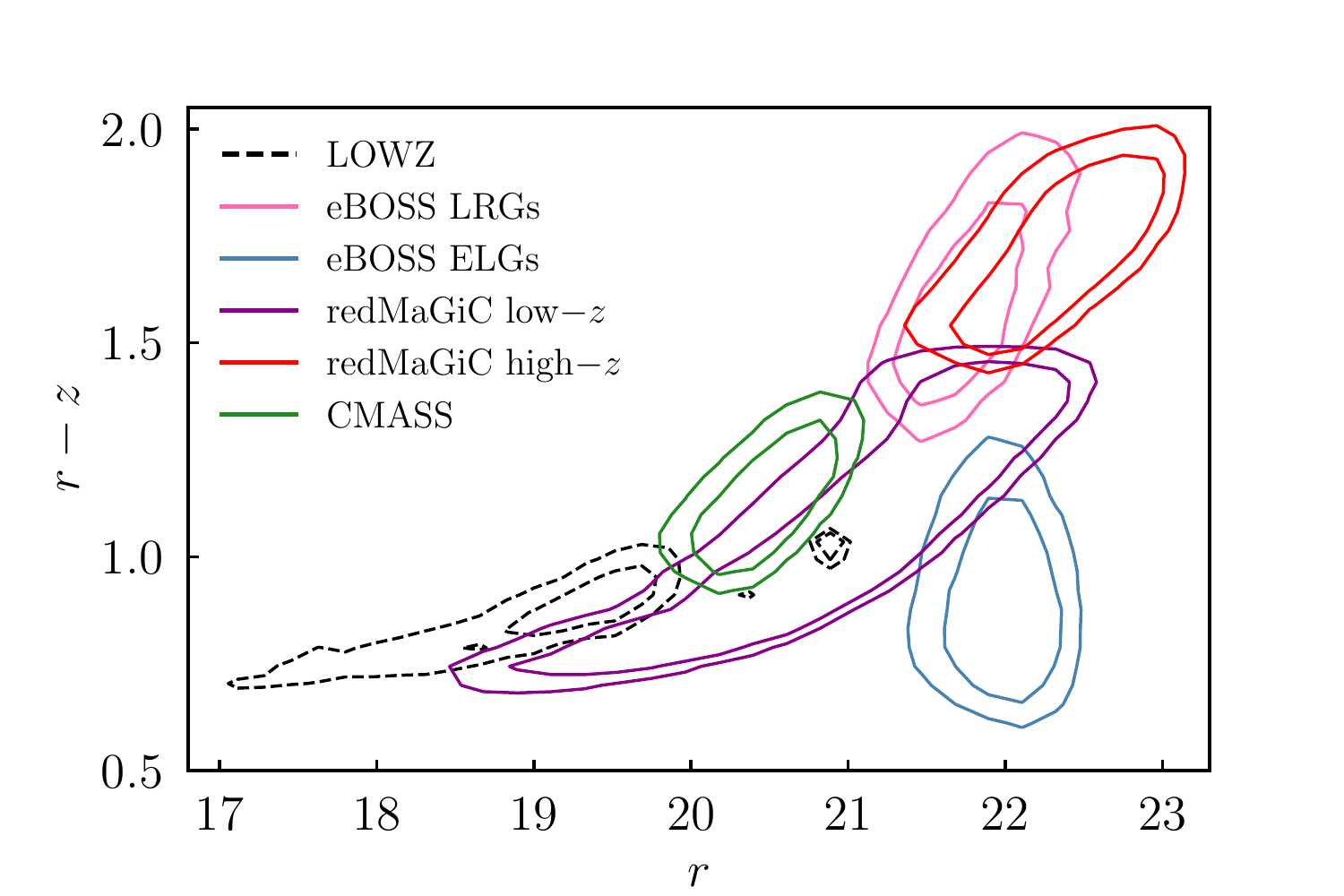}
\caption{\textbf{Top:} The rest frame colour magnitude diagram for the samples used in this work. The quantities here are $k+e-$corrected magnitudes in the DES filters. In the main galaxy samples (solid lines), these are estimated using \metacal~fluxes. For LOWZ, which is kept as a validation sample, we use the pre-computed absolute magnitudes described in \citet{singh15}. The contours are defined relative to the peak, at $0.5\times$ and $0.25\times$ the maximum density. \textbf{Bottom:} The equivalent colour-magnitude space, but using apparent magnitudes. Note that we do not have $k+e-$corrections for eBOSS ELGs, so they appear in the lower panel only.
}
\label{fig:data:colour_mag}
\end{figure}

The luminosity relative to a pivot $L_0$ is then given by $\log{(L_r/L_0)} = -(M_r-M_0)/2.5$, where $M_0$ is a fixed reference magnitude; we adopt a value $M_0=-22$ for the sake of comparability with previous results. For the purposes of constraining trends in alignment properties, we subdivide our red galaxy samples into luminosity bins. These are shown in Figure \ref{fig:data:nofL}, and are defined such that they contain roughly equal numbers of galaxies (with the exception of the bright end of \redmagic~high$-z$, where the signal-to-noise was sufficient to allow us to further split the highest $L$ bin in two.). Between them, our four samples cover a range of roughly $\log{(L_r/L_0)} =[-0.9,0.4]$, with \redmagic~low-$z$ in particular providing excellent coverage of the fainter end. We also show the rest frame colour magnitude diagram for these red samples (as well as LOWZ) in the top panel of Figure \ref{fig:data:colour_mag}. As can be seen here, although we group these samples together as ``red", there is some significant variation in colour at fixed luminosity. We will return to this in the context of our main results in Section \ref{fig:results:luminosity_dependence}. The lower panel shows the same colour-magnitude space, but using apparent magnitudes. Here the distributions are relatively elongated, primarily due to the colour-redshift degeneracy; that is, a galaxy of given rest-frame magnitude and colour observed at high redshift will appear both fainter and redder than the same object observed at low redshift.

\subsection{Two-Point Correlations}\label{sec:measurements:2pt}

Following a number of previous intrinsic alignment studies, our primary measurements are constructed using a modified Landy-Szalay estimator \citep{landy93}. For two-point galaxy clustering, this has the form:
\begin{equation}
\xi_{gg}(r_{\rm p}, \Pi) = 
\frac{(D - R_D)(D - R_D)} {R_D R_D}. \\
\end{equation}
The measurement is made on a grid of line-of-sight and perpendicular (comoving) separation, \rp~and $\Pi$. For a particular sample of galaxies, we have a density tracer catalogue and a second catalogue of random points tracing the same footprint and redshift distribution. $DD$, $R_DD$ and $R_D R_D$ are the weighted counts of galaxy-galaxy, galaxy-random and random-random pairs in a given bin of \rp~and $\Pi$. To reduce shot noise, the randoms $R_D$ are oversampled relative to the actual data by a factor of $>10$.

Similarly one can estimate the shape-density cross correlation:

\begin{equation}
\xi_{g+}(\rp, \Pi) = \frac{S_{+}(D - R_D)}{R_D R_S}, \\
\end{equation}

\noindent
where again $D$ represents the density sample, and $R_S$ and $R_D$ are randoms matching the shape and density samples respectively. The shape-shape correlation is constructed in a similar way: 

\begin{equation}
\xi_{++}(r_{\rm p}, \Pi) = \frac{S_{+}S_{+}}{R_S R_S}.
\end{equation}

\noindent
We also define:

\begin{equation}
S_+S_+ \equiv \sum_{\alpha,\beta; \beta\neq \alpha} e_{+}(\beta|\alpha)e_{+}(\alpha|\beta),
\end{equation}
\begin{equation}
S_+D \equiv \sum_{\alpha,\beta; \beta\neq \alpha} e_{+}(\beta|\alpha).
\end{equation}

\noindent
Here the sum runs over galaxies (or random points) at a given separation drawn from the two catalogues; $e_{+}(\beta|\alpha)$ is the tangential ellipticity component of galaxy $i$, defined by the separation vector with galaxy $\beta$. One can write down a set of analogous equations for $\xi_{g\times}$, $\xi_{+\times}$ and $\xi_{\times\times}$, which are identical to the above, but with galaxy ellipticities rotated by 45 degrees. Any astrophysical contribution to these, however, is expected to be negligible (due to parity arguments) and for this reason they are commonly used for null testing. 

The $\xi(\rp, \Pi)$ measurements are then projected along the line of sight as
\begin{equation}
w_{ab}(r_{\rm p}) = \int_{-\Pi_{\rm max}}^{\Pi_{\rm max}} \xi_{ab}(r_{\rm p}, \Pi) \mathrm{d}\Pi.
\end{equation} 

\begin{figure*}
\includegraphics[width=2.2\columnwidth]{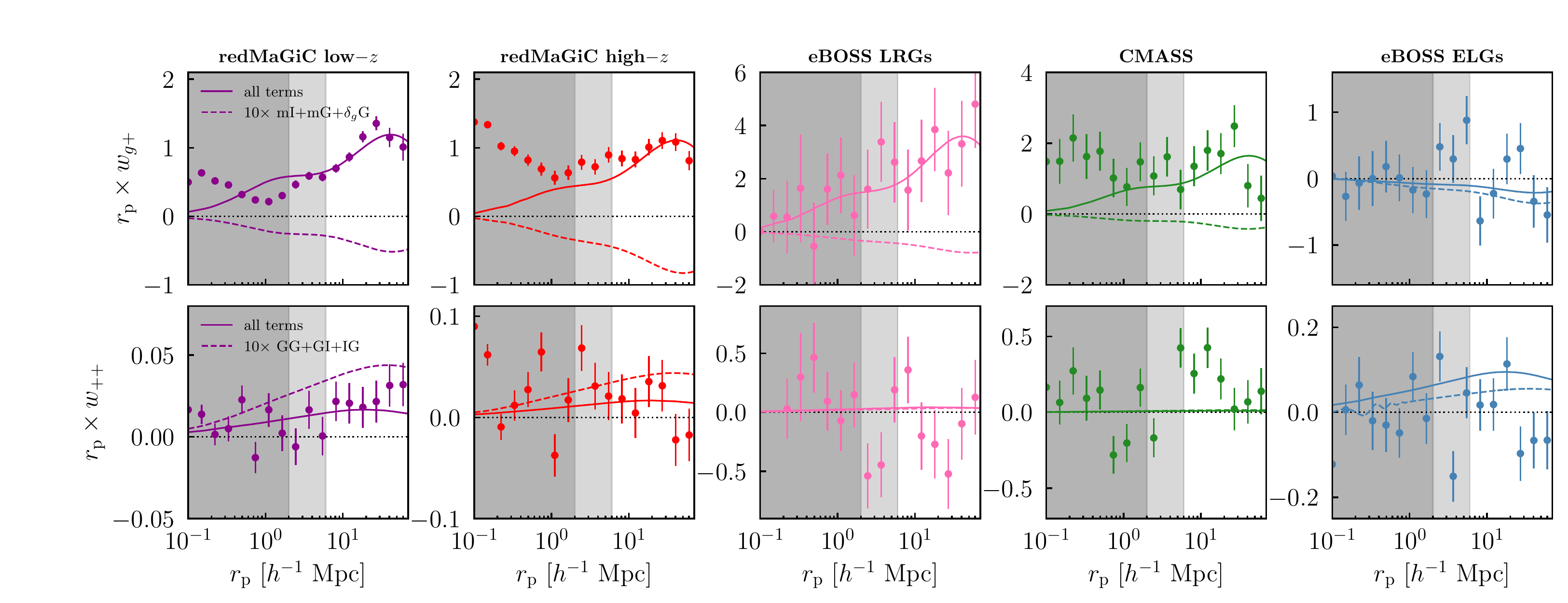}
\caption{Intrinsic alignment correlations measured from DES Y3 and eBOSS. The rows show (top/bottom) position-shape and shape-shape correlations, as defined in the text. The shaded bands indicate scales excluded from our NLA (light grey) and TATT (darker grey) fits. The solid lines are the best fitting NLA predictions for each data set at $\rp>6\mpc$. For reference, we also show the various terms arising from IAs, lensing and magnification separately. For clarity, the total lensing + magnification contribution, shown as a dashed line, is scaled up by a factor of 10. Note that the vertical axes vary between panels.}
\label{fig:data:2pt}
\end{figure*}

\begin{figure*}
\includegraphics[width=2\columnwidth]{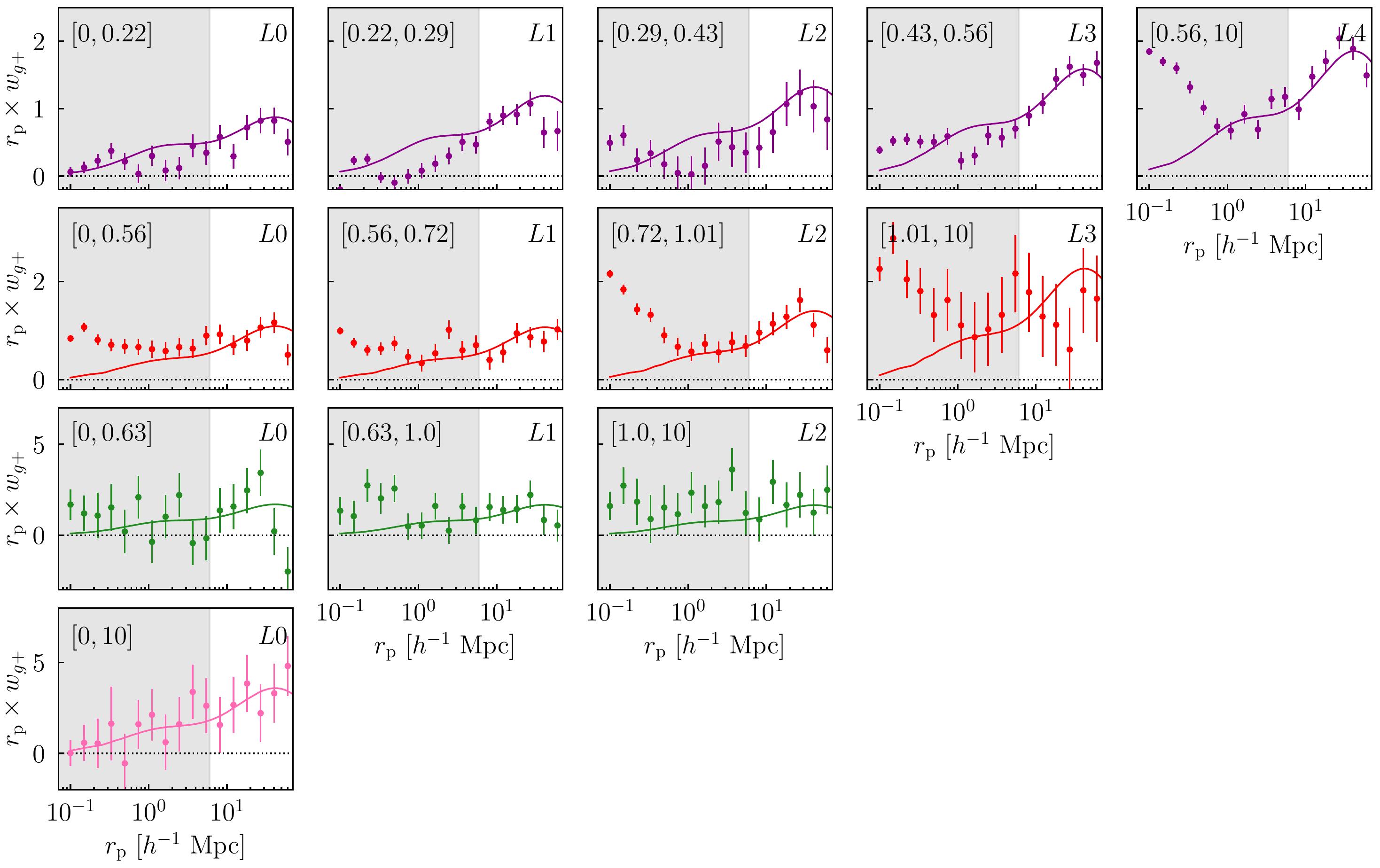}
\caption{The measured galaxy-shape correlations from \redmagic~and eBOSS LRGs. The columns (left to right) show bins of $k+e-$corrected $r$-band luminosity (as defined in Table \ref{tab:results:nla_luminosity}). The rows show (top to bottom, colours the same as in Figure \ref{fig:data:2pt}) \redmagic~low-$z$ (purple), \redmagic~high-$z$ (red), CMASS (green) and LRGs (pink). Note that the luminosity bins are \textbf{not} the same in the three cases; we use the $LX$ notation for convenience, but the bin edges and widths are defined for a particular sample (shown in the upper left of each panel; see also Section  \ref{sec:measurements:absmag} for discussion of how the luminosity bins are defined).
We also show the best fitting NLA model prediction for each measurement (solid line). As above, points within the shaded grey regions are excluded from the fits. Note that we do not fit the TATT model to our luminosity-binned measurements and so, unlike in Figure \ref{fig:data:2pt}, there is only one set of grey bands shown here.
}
\label{fig:data:2pt_lbins}
\end{figure*}

\noindent
We use \blockfont{TreeCorr}\footnote{\url{http://rmjarvis.github.io/TreeCorr} (version 4.1.1)
} \citep{jarvis04} for 
all two point measurements with $\mathrm{bin\_slop}=0.0$. We use 20 logarithmically spaced bins in $r_{\rm p}$, over the range $0.1-200$ \mpc. For the line of sight binning we set $\Pi_{\rm max}=100$ \mpc, with 20 linearly spaced bins between $\pm \Pi_{\rm max}$. The resulting data vectors are shown in Figure \ref{fig:data:2pt}. For parts of our analysis, we also make use of data vectors in bins of luminosity. These are shown in Figure \ref{fig:data:2pt_lbins}. The shaded regions here indicate physical scales excluded from our fiducial analysis. Further discussion of the fits to these data can be found in Section \ref{sec:results}. The choice of $\Pi_{\rm max}$ is driven by signal-to-noise considerations in the photometric samples, and follows \citet{singh15}. We also note that since we are including a galaxy-galaxy lensing (Section \ref{sec:theory}) term in our model, our choice here is not limited by the need to suppress such contributions.

The DES shape catalogues make use of a technique called \metacal~for accurately inferring an underlying shear signal from galaxy shape estimates. We apply response corrections in exactly the same way as in the DES Y3 cosmology analyses \citep*{y3-shapecatalog}. That is, we have a mean scale-independent factor $\langle R \rangle$, which is applied at the level of the two-point measurements as:

\begin{equation}
\xi_{g+} \rightarrow \frac{1}{\langle R \rangle} \xi_{g+},
\;\;\;\;
\xi_{++} \rightarrow \frac{1}{\langle R \rangle^2} \xi_{++},
\end{equation}

\noindent
where the angular brackets indicate an average over galaxies, and $\langle R \rangle = \langle R_\gamma + R_{\rm sel} \rangle$, or the sum of a shear response and a selection term for the shape sample in question. 

Note that $R_{\rm sel}$ corrects only for shape catalogue selection cuts. Since we do not have either \redmagic~re-runs or the eBOSS selection on sheared images, any bias induced by the basic sample selection is not included in this correction. That said, the estimated selection response for an early/late split of the DES Y1 catalogues was found to be of the order of $10^{-4}$, which is easily subdominant to our uncertainties (see \citealt{y1coloursplit} Section 4.1). We thus judge it safe to ignore this missing correction for the purposes of our analysis.  

% ---------- Theory stuff ------------

\section{Modelling \& Analysis Choices}\label{sec:theory}

The following section sets out our theoretical modelling choices. Our aim here is to connect an observed joint data vector, $w_{gg} + w_{g+} + w_{++}$, with underlying physical quantities, which can be calculated from theory. Each of these data vector components is a combination of two observable fields $\hat{\delta_{g}}$ and $\hat{\gamma}$, or the observed galaxy overdensity in counts and shapes. If we assume the former is the sum of contributions from gravitational clustering and magnification, $\hat{\delta}_{g} = \delta_g+\delta_\mu$, and the latter is the combination of intrinsic shape alignments and lensing, $\hat{\gamma} = \gamma^I+\gamma^{\rm G}$, we have a total of four correlations contributing to each observable. The sections below will set out how we evaluate these model ingredients

In reality, the observed shear is weighted by the overdensity of shape galaxies, $\hat{\gamma}\rightarrow (1+\delta_{g,\mathrm{S}})\hat{\gamma}$ (see e.g. \citealt{hirata04} eq. (6)). This contributes an additional intrinsic alignment term, which is explicitly included in TATT (although not NLA; \citealt{blazek17}). For conciseness, we absorb this factor into the definition of $\gamma^{I}$ when discussing the TATT model. Note that since the overdensity weighting applies to the total observed shear, not just the intrinsic component, it also gives rise to terms that scale as $\gamma^{\rm G} \times \delta_{g,\mathrm{S}}$, an effect known as source clustering. An analogous effect called source magnification enters in a similar way. These extra terms, however, are expected to be small at the level of projected observables, and so we neglect them here (see e.g. \citealt{y3-generalmethods} Sec 5B and \citealt{schmidt09} for discussion).

We start in Sections \ref{sec:theory:IA_power_spectra}- \ref{sec:theory:magnification} by describing how we calculate the 3D power spectra that enter each of our models. Section \ref{sec:modelling:2pt_theory} then sets out how these are combined and projected to give predictions for the observable correlations. We discuss how the covariance matrix of the data is estimated in Section \ref{sec:measurements:covmat}. Finally, Section \ref{sec:theory:scale_cuts} discusses how we choose a set of scale cuts, which restrict our analysis to the regime where our model is thought to be sufficient.

When it is necessary to assume a background cosmology, we use a flat \lcdm~model $\mathbf{p}_{\rm cos} = (\omegam, \omegab, \sigma_8, h, \ns, \Omega_\nu h^{2}) = (0.3, 0.048, 0.82, 0.69, 0.97, 0.00083$\footnote{This corresponds to a total sum of the neutrino masses, $\sum m_\nu = 0.077$ eV.}). Although our results are not strongly cosmology dependent, we do quantify the impact of this choice in Section \ref{sec:results}. The linear matter power spectrum is computed using the Boltzmann code \blockfont{CAMB}\footnote{\url{http://camb.info/}} \citep{lewis00}, with nonlinear corrections using \blockfont{halofit} \citep{takahashi12}. Parameter inference is performed within \blockfont{CosmoSIS}\footnote{\url{https://bitbucket.org/joezuntz/cosmosis/wiki/Home}; v1.6, ``des-y3" branch of cosmosis-standard-library, ``develop" branch of cosmosis} \citep{zuntz14} using the \blockfont{MultiNest} nested sampling algorithm\footnote{v3.6; $\mathrm{efficiency}=0.3$, $\mathrm{live\_points}=500$} \citep{feroz13}. 

\begin{table}
\begin{center}
\begin{tabular}{c|ccc}
\hline
Parameter & Description & TATT Prior & NLA Prior \\ 
\hhline{====}
$A_1$           & Lin. IA amplitude & $\mathrm{U}[-8,8]$ & $\mathrm{U}[-8,8]$ \\ 
$A_2$           & Quadratic IA amplitude & $\mathrm{U}[-8,8]$ & $\delta[0]$ \\ 
$b_{\rm TA}$     & Density wt. coefficient  &  $\mathrm{U}[-6,6]$ & $\delta[0]$ \\ 
$b_1$           & Lin. galaxy bias & $\mathrm{U}[0,3]$ & $\mathrm{U}[0,3]$ \\ 
$b_2$           & 2nd order galaxy bias & $\mathrm{U}[-3,3]$ & $\mathrm{U}[-3,3]$ \\ 
\hline
\end{tabular}
\caption{The free parameters and priors for the models discussed in this paper. The upper three rows are IA model parameters, while the lower two describe galaxy bias. We include two sets of priors here: one for our TATT model analyses (which extends to $\rp>2\mpc$), and one for NLA ($\rp>6\mpc$).}\label{tab:theory:models}
\end{center}
\end{table}

\subsection{Intrinsic Alignment Power Spectra}\label{sec:theory:IA_power_spectra}

To model the power spectra of the intrinsic alignment GI and II signals (respectively the correlation between $\gamma^I$ and $\gamma^{\rm G}$, and $\gamma^I$ with itself), we use the Tidal Alignment and Tidal Torque model (TATT; \citealt{blazek17}). The basic idea is that the intrinsic shape field $\gamma^I$ can be expressed as an expansion in powers of the background tidal field $s$ and matter overdensity $\delta$:
\begin{align}\label{eq:theory:gammaI_terms}
\gamma^{I}_{ij} = C_1 s_{ij} +C_{1\delta} \delta s_{ij} + C_2 \sum_{k}s_{ik}s_{kj} + \cdots
\end{align}
\noindent
Note that while $\delta$ is a scalar at any given position $\mathbf{x}$, $\gamma^I$ and $s$ are $3\times3$ tensors. The above expansion can be propagated to the two-point level to give expressions for $P_{\rm GI}$ and $P_{\rm II}$ (see \citealt{blazek17}). Our implementation of the TATT model is identical to that of \citet*{y3-cosmicshear2} and \citet{y3-generalmethods}. We refer the reader to those papers for specifics, and in particular Sec. D2 and eq. (21) and (22) of \citet*{y3-cosmicshear2} for the full expressions. 

The TATT model has three free parameters, which we refer to as $A_1$, $A_2$ and $b_{\rm TA}$. One can also parameterise the redshift dependence of all the contributions, if desired, as in previous cosmological analyses. Since our individual samples do not have a particularly wide redshift range, however, this is not an especially useful thing to do in our case. On the other hand, one can look at the evolution \emph{between} samples. Considering galaxies with similar colour and luminosity properties in Appendix \ref{app:redshift_dependence}, we find no evidence for $z$ evolution over a significantly wider range than the coverage of any one of our samples alone. The two amplitudes modulate the strength of IA contributions that are linear and quadratic in $s$:
\begin{equation}
    C_1 = - A_1 \frac{\rho_{\rm crit} \omegam \bar{C}_1}{D(z)} 
    \;\;\;\;\;\;\;
    C_2 = A_2 \frac{5 \rho_{\rm crit} \omegam \bar{C}_1}{D^2(z)},
\end{equation}
\noindent
where $D(z)$ is the linear growth factor, $\rho_{\rm crit}$ is the critical density and $\bar{C}_1$ is a constant, which by convention is fixed to a value of $5 \times 10^{-14} M_\odot \mathrm{Mpc}^2/h^2$. The other parameter, $b_{\rm TA}$, is known as the density weighting coefficient, and controls the size of the $C_{1\delta}$ term above as $C_{1\delta} = b_{\rm TA} C_1$.

We also consider a nested subspace of the full TATT model. The simplest subspace, known as the nonlinear alignment model (NLA; \citealt{hirata07,bridle07}), has only one free parameter, $A_1$. The $\gamma^I$ field is assumed to be purely linear in the tidal field (effectively setting $C_2=0$, $C_{1\delta}=0$ in Equation \eqref{eq:theory:gammaI_terms}), and so the IA power spectra have the same shape as the nonlinear matter power spectra, but with a scaling factor applied. 

In all samples considered, we vary the IA parameters with wide flat priors, as given in Table \ref{tab:theory:models}.\footnote{Note that these differ slightly from those used in \citealt{y3-3x2ptkp}. Although the DES Y3 priors were chosen to be uninformative for that particular sample, we are considering significantly different (often much redder) populations of galaxies. We thus opt to allow for more extreme IA values.}

\subsection{Galaxy Power Spectrum}\label{sec:theory:galaxy_power_spectra}

Galaxy bias, or the mapping between the matter and galaxy overdensity fields, is an important source of uncertainty in any analysis that relies on galaxy-shape correlations. Similarly to $\gamma^I$ in Section \ref{sec:theory:IA_power_spectra}, one can expand the galaxy overdensity in terms of $\delta$ \citep{mcdonald06,baldauf10,saito14}:
\begin{align}\label{eq:theory:deltag_terms}
\delta_{g} = b_1 \delta + 
\frac{1}{2} b_2 \left( \delta^2 - \langle \delta^2 \rangle  \right )  +
\frac{1}{2} b_{s^2} \left( s^2 - \langle s^2 \rangle  \right )
+ b_{3\mathrm{nl}} \psi
+ \cdots
\end{align}
\noindent
Here $\psi$ is the sum of several different third-order terms with the same scaling (see \citealt{saito14}). On large enough scales, it is often sufficient to assume a simple linear relation $\delta_g = b_1 \delta$; in this case the galaxy power spectrum is simply:
\begin{equation}
P_{\delta_g}(k,z) = b_1^2 P_\delta(k,z),
\end{equation}
\noindent
where the galaxy bias $b_1$ depends on the galaxy sample, but is independent of wave number. $P_\delta$ is the nonlinear matter power spectrum. Unfortunately, we see evidence of the need for a more sophisticated approach in some of our samples. This is discussed further in Section \ref{sec:theory:scale_cuts}, where we see that \redmagic~high$-z$ and eBOSS LRGs favour a more complicated bias model, even on relatively large scales.

Using Equation \eqref{eq:theory:deltag_terms} one can write down a slightly more complete expression for $P_{\delta_g}$ (e.g. in \citealt{y3-generalmethods} Eq. 38). Our fiducial model for the galaxy power spectrum includes all terms in the expansion above, for which we use the implementation in \blockfont{FastPT} \citep{mcewen16}. Assuming co-evolution, however, we can reduce the number of free parameters to two, with $b_{s^2}=-4/7(1-b_1)$ and $b_{3\mathrm{nl}} = b_1-1$ (see \citealt{saito14}, and also \citealt{pandey20} and \citealt{y3-generalmethods} for further discussion). For all samples, we marginalise these galaxy bias parameters with wide flat priors $b_1 = [0.05,3]$, $b_2 = [-3,3]$. 

For the power spectra entering $w_{g+}$ we assume linear bias (despite using Equation \eqref{eq:theory:deltag_terms} for $P_{\delta_g}$):  
\begin{equation}
P_{\delta_g\mathrm{I}}(k,z) = b_1 P_{\rm GI}(k,z),
\end{equation}
\noindent
and similarly:
\begin{equation}\label{eq:p_gG}
P_{\delta_g\delta}(k,z) = b_1 P_{\delta}(k,z),
\end{equation}
\noindent
where $b_1$ is the same as in Equation \eqref{eq:theory:deltag_terms} above.
In principle, nonlinear galaxy bias, and also various cross terms between TATT parameters and higher order bias are expected to contribute to $w_{g+}$. In all cases considered here, however, $w_{g+}$ has significantly lower signal-to-noise than the equivalent galaxy-galaxy correlations, and thus the latter dominate the fits for galaxy bias. To within the level of uncertainty the TATT model is able to sufficiently describe the potential impact of correlations between nonlinear galaxy bias and IA through the free $b_{\rm TA}$ parameter (see the similarity of these nonlinear terms in \citealt{blazek15}). For fits using the NLA model, we exclude scales where nonlinear bias correlations are significant (see Section \ref{sec:theory:scale_cuts} for discussion of how the scale cuts are chosen). Although $b_{\rm TA}$ cannot absorb the galaxy-galaxy lensing signal $\delta_g \delta$ so easily, this term is subdominant on all scales and for all samples ($\sim 5-10\%$ of the total $w_{g+}$ signal; see Section \ref{sec:results:lensmag}). We test the impact by substituting the fully nonlinear $P_{\delta_g \delta}$ in place of the approximation in Equation \eqref{eq:p_gG} above. Using the best fit bias parameters for each of our samples, we find a roughly $10\%$ change in $w_{\delta_g \mathrm{G}}$ on scales $<6\mpc$; compared to the full signal, however, the impact is at the sub-percentage level. Implementing a fully consistent nonlinear model is a work in progress, but we do not expect this to have a significant impact given the statistical uncertainties in current data sets.

\subsection{Magnification \& Lensing Power Spectra}\label{sec:theory:magnification}

As well as contributions due to galaxy clustering and intrinsic shape correlations, magnification can have an effect on direct IA measurements. Its impact is to alter the observed galaxy number density in a patch of sky as $\hat{\delta}_{g} = \delta_g + \delta_\mu$. Similarly, the observed shear in a set of galaxies has both an IA contribution, and one from cosmological lensing: $\hat{\gamma} = \gamma^I + \gamma^{\rm G}$. At the two-point level, one has two additional terms in the $gg$ correlation (galaxy-magnification and magnification-magnification; $\delta_g \delta_\mu$ and $\delta_\mu \delta_\mu$), and two in the galaxy-shape correlation (magnification-intrinsic and magnification-lensing; $\delta_\mu \gamma^{I}$ and $\delta_\mu \gamma^{\rm G}$). Similarly, $w_{++}$ has contributions from the standard II and GI power spectra, but also a pure cosmic shear term $\gamma^{\rm G} \gamma^{\rm G}$. On large scales, the additional magnification power spectra are all related to galaxy and IA power spectra via magnification coefficients $\alpha$ (see Table \ref{tab:theory:cls}, and \citealt{joachimi10,vonWietersheim21,joachimi21}; \citealt*{y3-2x2ptmagnification} for discussion). 

A number of different methods for constraining $\alpha$ have been discussed in the literature. We describe how we estimate $\alpha$ for each density sample in Section \ref{sec:measurements:magnification}. In short, our fiducial estimates are obtained by artificially perturbing the observed galaxy magnitudes (i.e. a flux-only estimate). For the two \redmagic~samples, we have estimates from \blockfont{balrog}, which we use for validation (see Section \ref{sec:results:robustness}). 

\subsection{Modelling Projected Correlation Functions}\label{sec:modelling:2pt_theory}
\subsubsection{Modelling Spectroscopic Data}\label{sec:modelling:2pt_theory:specz}

Given power spectra from any model, one can convert into projected correlation functions of the sort discussed in Section \ref{sec:measurements:2pt} via Hankel transforms. In the case of perfect knowledge of individual galaxy redshifts (i.e. spectroscopic redshifts) one has:

\begin{equation}\label{eq:wgp_basic}
w^{ij}_{g+} (r_{\rm p}) = -\int \mathrm{d}z \mathcal{W}^{ij}(z) 
\int \frac{\mathrm{d} k_{\perp} k_{\perp}}{2 \pi}  
J_2(k_{\perp} r_{\rm p}) P_{\delta_g \mathrm{I}}(k_{\perp},z),
\end{equation}

\noindent
with the Roman indices indicating the two galaxy samples, and $J_\nu$ being a Bessel function of the first kind of order $\nu$. The projection kernel is given by (see \citealt{mandelbaum10}'s Appendix A)

\begin{equation}\label{eq:zkernel}
\mathcal{W}^{ij}(z) = 
\frac{n^i(z) n^j(z)}{\chi^2(z) \mathrm{d}\chi / \mathrm{d}z} 
\times 
\left [ \int \mathrm{d}z \frac{n^i(z) n^j(z)}{\chi^2(z) \mathrm{d}\chi / \mathrm{d}z} \right ]^{-1}.
\end{equation}

\noindent
In the above, $n^i(z)$ is the estimated redshift distribution for sample $i$, and $\chi(z)$ is the comoving line-of-sight distance corresponding to a redshift $z$. The other two-point correlations follow by analogy as:

\begin{equation}\label{eq:wgg_basic}
w^{ij}_{gg} (\rp) = b^i_g b^j_g \int \mathrm{d}z \mathcal{W}^{ij}(z) 
\int \frac{\mathrm{d} k_{\perp} k_{\perp}}{2 \pi}  
J_0(k_{\perp} \rp) P_{\delta_g}(k_{\perp},z),
\end{equation}

\noindent
and

\begin{multline}\label{eq:wpp_basic}
w^{ij}_{++} (\rp) = \int \mathrm{d}z \mathcal{W}^{ij}(z) \\
\int \frac{\mathrm{d} k_{\perp} k_{\perp}}{2 \pi}  
\left [ J_0(k_{\perp} \rp) + J_4(k_{\perp} \rp) \right ] P_{\rm II}(k_{\perp},z).
\end{multline}

\noindent
In each case, the theory prediction amounts to a projection of a power spectrum along the redshift axis, and then a Bessel integral.

\subsubsection{Modelling IA correlations in the presence of photo-$z$ error}\label{sec:modelling:2pt_theory:photoz}

When dealing with spectroscopic galaxy samples, one can in general safely assume that the associated redshift error is much smaller than the distance scales of interest. This assumption does not hold for photometric samples such as \redmagic, which means the modelling is slightly more complicated. The impact of redshift error is to scatter galaxies along the line of sight; this in effect shuffles galaxies between $\Pi$ bins and so redistributes power out along the line of sight. In principle the effect should wash out when integrating over a sufficiently large range in $\Pi$. In reality, however, one must choose finite $\Pi$ limits, and widening the integration range to large separations is not necessarily desirable, since it can degrade the signal-to-noise of the measurement. This leads to an overall suppression of the measured correlations due to photo$-z$ scatter. Another impact of photo-$z$ error is that it can boost additional (non-IA) signals. That is, galaxy pairs allocated to small $\Pi$ bins may actually be physically quite distant. Such pairs carry little local II signal, but they do tend to increase the lensing and magnification contributions. The consequence of this is that one must account for the $\Pi$ cut-off in the model. To do so we follow the method set out in \citet{joachimi11}, of which we provide an outline below.

To begin, we compute angular spectra from the IA and galaxy-galaxy power spectra. Incorporating all of the magnification and lensing contributions to number counts and shear, one has:
\begin{equation}
C^{ij}_{\hat{\delta}_{g}\hat{\delta}_{g}} = C^{ij}_{\delta_g \delta_g} + C^{ij}_{\delta_\mu \delta_\mu} + C^{ij}_{\delta_{\mu} \delta_g} + C^{ij}_{\delta_g \delta_{\mu}}
\end{equation}
\noindent
\begin{equation}
C^{ij}_{\hat{\delta}_{g} \hat{\gamma}} = C^{ij}_{\delta_g \gamma^{I}} + C^{ij}_{\delta_\mu \gamma^{I}} + C^{ij}_{\delta_g \gamma^{\rm G}} + C^{ij}_{\delta_{\mu} \gamma^{\rm G}}
\end{equation}
\noindent
\begin{equation}
C^{ij}_{\hat{\gamma} \hat{\gamma}} = C^{ij}_{\gamma^I \gamma^I} + C^{ij}_{\gamma^{I}\gamma^{\rm G}} + C^{ij}_{\gamma^{\rm G} \gamma^{I}} + C^{ij}_{\gamma^{\rm G} \gamma^{\rm G}},
\end{equation}
\noindent
where the subscripts $\delta_\mu$, $\gamma^I$, $\delta_g$ and $\gamma^{\rm G}$ indicate magnification, intrinsic shape, gravitationally induced galaxy overdensity and gravitational shear. Implicitly, the II term here is the E-mode autocorrelation, $C_{\gamma^I \gamma^I}^{\mathrm{EE}}$. In principle, one could also include $C_{\gamma^I \gamma^I}^{\mathrm{BB}}$, which can be calculated assuming a particular IA model. We do not include this in our model because (a) typically any IA induced B-modes are small \citep{hirata04, blazek17} and (b) they contribute only to $w_{++}$, where the signal-to-noise of our measurements is low. The Limber integrals used to compute each of the angular power spectra then have the form:

\begin{multline}\label{eq:theory:limber}
C^{ij}_{ab}(\ell | z_1, z_2) = 
\int_0^{\chi_{\rm hor}} \mathrm{d}\chi'\\
\frac{q_a^i\left ( \chi' | \chi(z_1) \right ) q_b^j\left ( \chi' | \chi(z_2) \right ) }{\chi'^{2}}
P_{ab}\left (k=\frac{\ell+0.5}{\chi'}, z(\chi') \right ). \\
\end{multline}

\begin{table}
\begin{centering}
\begin{tabular}{c|ccc}
\hline
Correlation & Kernel & Power Spectrum & Correlation Function \\
\hhline{====}
$\delta_g \delta_g$ & $p^i p^j$ & $P_{\delta_g}$ & $w_{gg}$ \\
$\delta_\mu\delta_\mu$                & $g^i g^j$ & $4(\alpha^i - 1 )(\alpha^j - 1 )P_{\delta}$ & $w_{gg}$ \\
$\delta_\mu \delta_g$         & $g^i p^j$ & $2(\alpha^i - 1 )P_{\delta_{g} \delta}$ & $w_{gg}$ \\  
$\delta_g \gamma^I$        & $p^i p^j$ & $P_{\delta_g \mathrm{I}}$ &  $w_{g+}$ \\
$\delta_\mu \gamma^I$                & $g^i p^j$ & $2(\alpha^i - 1 )P_{\rm GI}$ & $w_{g+}$ \\
$\delta_g \gamma^{\rm G}$   & $p^i g^j$ & $P_{\delta_g \delta}$ & $w_{g+}$ \\
$\delta_\mu \gamma^{\rm G}$           & $g^i g^j$ & $2(\alpha^i - 1 )P_{\delta}$ & $w_{g+}$ \\
$\gamma^I \gamma^I$               & $p^i p^j$ & $P_{\rm II}$ & $w_{++}$ \\
$\gamma^{\rm G} \gamma^I$         & $g^i p^j$ & $P_{\rm GI}$ & $w_{++}$ \\
$\gamma^{\rm G} \gamma^{\rm G}$     & $g^i g^j$ & $P_{\delta}$ & $w_{++}$ \\
\hline
\end{tabular}
\caption{A summary of the various contributing terms to our observables $w_{gg}$, $w_{g+}$ and $w_{++}$. The kernel column lists all the possible combinations for $q_a^i q_b^j$ in Equation~\eqref{eq:theory:limber} (where each $q$ is either the lensing kernel $g$ or the galaxy PDF $p$). For each one we show the kernel (either lensing efficiency or redshift distribution), and the relevant power spectrum included in the Limber integral. The prefactors $\alpha$ are magnification coefficients, which are defined in Section \ref{sec:measurements:magnification}.
}\label{tab:theory:cls}
\end{centering}
\end{table}

\noindent
The kernel $q$ is either the lensing efficiency $g$, or the error distribution $p$, according to Table \ref{tab:theory:cls}. The power spectrum $P_{ab}$ corresponding to a given $C(\ell)$ are also shown in Table \ref{tab:theory:cls}. Here $p_i( z' | z)$ is the conditional probability distribution for the true redshift of a galaxy from sample $i$, which has a best-estimate redshift at $z$. The estimates for $p$ at any given $z$ are obtained using the method described in Section \ref{sec:measurements:redshifts}. Note that this is different from the more common form of the Limber integral in the context of cosmological lensing, which uses the ensemble redshift distribution $n(z)$, not the per-galaxy PDF. One can then transform from harmonic to angular space as follows:

\begin{equation}\label{eq:theory:hankel}
\xi^{ij}_{ab}(\theta | z_1, z_2)
=
\frac{1}{2 \pi}
\int^{\infty}_{0}
\mathrm{d}\ell
\ell
J_{\nu}(\ell \theta)
C_{ab}(\ell | z_1, z_2),
\end{equation}

\noindent
where the order of the Bessel function $\nu$ depends on the type of correlation ($\nu=0$ for $ab= \hat{\delta}_{g}\hat{\delta}_{g}$, $\nu=2$ for $ab= \hat{\delta}_{g} \hat{\gamma}$ or $\nu=(0,4)$ for $ab= \hat{\gamma} \hat{\gamma}$), as in Section \ref{sec:modelling:2pt_theory:specz}.  As argued in \citet{joachimi11}, from here one can obtain the photometric correlation function $\xi^{ij}_{ab}(\rp, \Pi, z_{\rm m})$ using a simple coordinate transformation (see equation A11 in that paper, which also defines $z_{\rm m}=(z_1+z_2)/2$). Finally, the projected correlation function as a function of perpendicular physical separation is expressed as,
\begin{equation}
w^{ij}_{ab}(\rp) = 
\int^{\Pi_{\rm max}}_{-\Pi_{\rm max}} \mathrm{d}\Pi 
\int \mathrm{d} z_{\rm m} \mathcal{W}^{ij}(z_{\rm m})
\xi^{ij}_{ab}(\rp, \Pi, z_{\rm m})
\end{equation}

\noindent
With these ingredients, the recipe for generating a theory prediction for the cross correlation between photometric samples $i$ and $j$ is as follows.
\begin{itemize}
\item {Choose an initial value of $\Pi$ and $z_{\rm m}$. Use \citet{joachimi11}'s equation~A11 to obtain $z_1$ and $z_2$, and evaluate the per-galaxy error distributions for the two samples at these redshift values.}
\item {Carry out the Limber integral in Eq.~\eqref{eq:theory:limber} with these error distributions to obtain $C(\ell | z_1, z_2)$.}
\item {Carry out the Hankel transform in Eq.~\eqref{eq:theory:hankel} with the appropriate Bessel kernel to obtain $\xi(\theta | z_1, z_2)$.}
\item {Perform the coordinate transform, such that $\xi(\theta | z_1, z_2) \rightarrow \xi(\rp | \Pi, z_{\rm m})$.}
\item {Repeat the above steps with varying $\Pi$ and $z_{\rm m}$, to give a three dimensional grid $\xi(\rp, \Pi, z_{\rm m})$.}
\item {Integrate over the redshift kernel $\mathcal{W}(z)$ and then over line-of-sight separation with the appropriate $\Pi_{\rm max}$ to obtain $w_{ab}(\rp)$.}
\end{itemize}

\noindent
We confirm that our implementation of this method returns the same results as Eqs.~\eqref{eq:wgp_basic}-\eqref{eq:wpp_basic} in the limit of narrow photo-$z$ distributions and wide $\Pi$ bounds. We also verify that, with a matching cosmology and set of input parameters, our modelling code can reproduce Fig. 5 from \citet{joachimi11}. Our fiducial modelling setup is to use the steps above to predict $w_{g+}$ and $w_{++}$. 

For $w_{gg}$, however, it is not sufficient to assume redshift space distortions have negligible impact (see Appendix~\ref{app:rsds} and Figure \ref{fig:app:rsds}). For this reason, we instead choose to use a sum over Legendre polynomials to obtain the anisotropic galaxy-galaxy correlation $\xi_{gg}$, which we then integrate over $\Pi$ (eq. \eqref{eq:app:pkbeta}-\eqref{eq:app:wgg}). We do, however, still need to account for lensing, magnification and photo-$z$ suppression. Unlike with $w_{g+}$, where the combined impact of these effects are seen to have some non-trivial scale dependence, this is much less true for $w_{gg}$; using the recipe set out above, we generate theory data vectors for each sample with and without photo-$z$ scatter, lensing, and magnification, finding that correction factor, $a(\rp)=w^{\rm full}_{gg}/w_{gg}$, is flat with $\rp$ to good approximation over scales $2<\rp<70\mpc$. Given this, we derive a single multiplicative factor for each sample, which we apply to the theory predictions as $w_{gg}\rightarrow a w_{gg}$. We obtain $a_{\rm RMH}=0.83$ and $a_{\rm RML} = 0.87$ for our two \redmagic~samples respectively. 

\subsection{Covariance Matrix}\label{sec:measurements:covmat}

We estimate the covariance of our data using an analytic prescription. This approach has a number of advantages over data-based estimators such as jackknife, which have been widely used in the past \citep{hirata07, mandelbaum10, joachimi11, singh15,fortuna21}. For example, it can be used on large scales where jackknife breaks down, and it is unaffected by noise in the data. Note, though, we are assuming here that the covariance of our data is dominated by the Gaussian component, and any connected 4pt and super sample covariance contributions are negligible for our purposes (see e.g. \citealt{takada09,takada13}).

The covariance is assumed to be dominated by two components: one from cosmic variance, and one from shape and shot noise $\mathrm{Cov}=\mathrm{Cov}^{\rm CV} + \mathrm{Cov}^{\rm SN}$. For any two elements of our data vector in scale bins centred on $r_{\mathrm{p}, m}$ and $r_{\mathrm{p}, n}$, the cosmic variance part is given by:

\begin{multline}\label{eq:cov_cv}
\mathrm{Cov}^{\rm CV}\left [ w_{ij}(r_{\mathrm{p},m}) w_{kl}(r_{\mathrm{p},n}) \right ]
= \\
\frac{1}{\mathcal{A}(z_c)}
\int \frac{k \mathrm{d}k}{2 \pi}
\Theta_{ij}(k r_{\mathrm{p},m}) \Theta_{kl}(k r_{\mathrm{p},n}) \times\\
\left [ P_{ik}(k)P_{jl}(k) 
+ 
P_{il}(k)P_{kj}(k) \right ],
\end{multline}

\noindent
where the lower indices define the tracer type (i.e. $g$ or $+$). The term $\Theta_{ij}(x)$ is a Bessel function of the first kind (or a sum of two); specifically $J_2$, $J_0$ and $J_0+J_4$ for $ij = g+, gg$ and $++$ respectively. The power spectra are $P_{\delta_g \mathrm{I}}$ for $ij = g+$, $P_{\delta_g}$ for $gg$ and $P_{\rm II}$ for $++$. Note that we do not include secondary contributions from magnification and lensing, but this is not expected to significantly change our results. The prefactor $\mathcal{A}$ is the projected comoving area of the footprint (including masking), at a characteristic redshift $z_c$.

The noise contribution is simply given by (\citealt{schneider02,hu04,joachimi10}):

\begin{equation}
\mathrm{Cov}^{\rm SN}\left [ w_{g+}(r_{\mathrm{p},m}) w_{g+}(r_{\mathrm{p},n}) \right ]
= \delta_{mn} \frac{\sigma^2_{\mathcal \epsilon}}{N^{g+}_{\rm p}},
\end{equation}

\begin{equation}
\mathrm{Cov}^{\rm SN}\left [ w_{gg}(r_{\mathrm{p},m}) w_{gg}(r_{\mathrm{p},n}) \right ]
= \delta_{mn} \frac{1}{N^{gg}_{\rm p}},
\end{equation}
\begin{equation}
\mathrm{Cov}^{\rm SN}\left [ w_{++}(r_{\mathrm{p},m}) w_{++}(r_{\mathrm{p},n}) \right ]
= \delta_{mn} \frac{\sigma^4_{\mathcal \epsilon}}{N^{++}_{\rm p}},
\end{equation}
\noindent
for our three observables respectively. Since our measurements with \blockfont{TreeCorr} give us the number of galaxy pairs in each bin $N_{\rm p} (r_{\mathrm{p},m})$ without extra computational cost, we use these exact numbers here. The shape dispersion $\sigma_{\mathcal \epsilon}$ is measured for each sample, using the \citet{heymans12} definition, and incorporating the correct response weighting (see \citealt*{y3-shapecatalog} eq. 13).

We perform initial fits using a preliminary covariance matrix, which we then replace with an updated version with best fit values of $A_1$, $b_1$ and $b_2$ entering Equation \eqref{eq:cov_cv} above. Since $w_{g+}$ is shape noise dominated for all samples and on all but the largest scales (and $w_{++}$ entirely so, on all scales), this update makes little difference to the final IA parameter constraints. 

We compare our analytic predictions with jackknife estimates in Appendix \ref{app:lowz}, and find good agreement on scales $2\mpc < \rp < 70 \mpc$.

\subsection{Scale Cuts}\label{sec:theory:scale_cuts}

We impose scale cuts on all three of our measured correlations when fitting, to mitigate model uncertainty. In brief, our minimum scales are $r_{\rm p, min}=(2,6,6)\mpc$ for NLA and $r_{\rm p, min}=(2,2,2)\mpc$ for TATT (where the ordering is $w_{gg}$, $w_{g+}$, $w_{++}$). For the latter two this is driven by the fact that we know our IA models start to break down on certain scales, and rely on assumptions that are valid only in specific regimes (NLA on scales above $\sim 5-10\mpc$, TATT down to $\sim 1-2\mpc$; \citealt{bridle07,blazek15}). The motivation behind the $w_{gg}$ scale cuts is discussed in more detail in Section \ref{sec:theory:scale_cuts:wgg} below.

We also impose an upper cut at $70\mpc$, a choice motivated by the null tests in Appendix \ref{app:null_tests}. This maximum scale is applied to to all three correlations for all samples. Large scale systematics, most prominently PSF modelling error, are known to modulate galaxy number counts at large \rp, but are difficult to model analytically. We thus choose to remove the affected scales.  

\subsubsection{Galaxy clustering}\label{sec:theory:scale_cuts:wgg}

\begin{figure}
\includegraphics[width=1.\columnwidth]{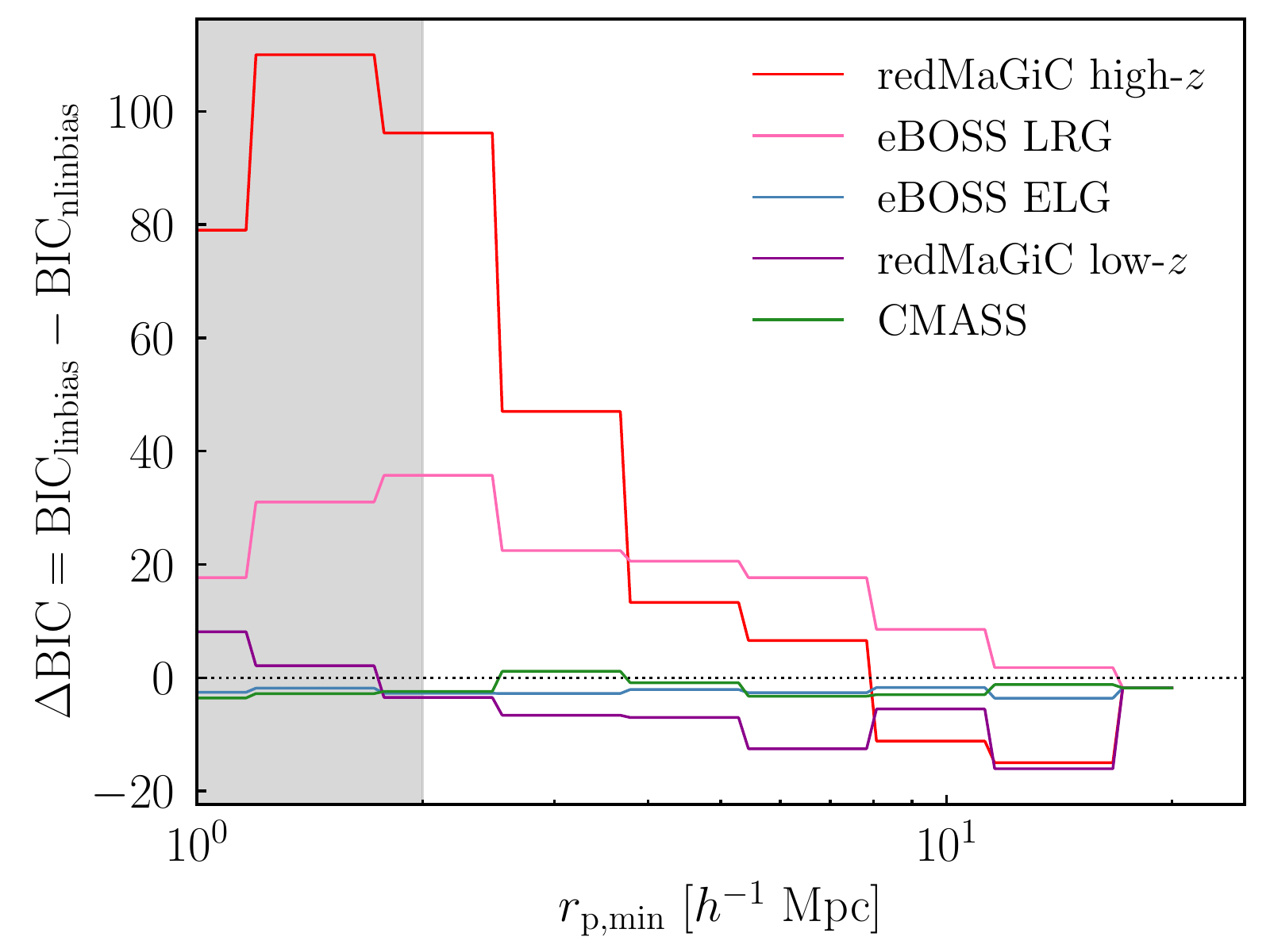}
\caption{The difference in Bayesian Information Criterion between two simple models of effective galaxy bias, as a function of the lower scale cut $r_{p, \mathrm{min}}$ applied to $w_{gg}$. Negative values indicate that the linear bias assumption is justified by the data; positive values, on the contrary, indicate that the dark matter only correlation $w_{\delta \delta}$ with a scale-independent multiplicative factor is not a good representation of $w_{gg}$. The different colours show fits to the galaxy-galaxy correlation from various DES and eBOSS samples. As in Figure \ref{fig:data:2pt}, the shaded region shows scales excluded in our fiducial analysis setup.}
\label{fig:theory:scale_cuts}
\end{figure}

Since scale cuts are designed to mitigate modelling uncertainty, the choice of $r_{\mathrm{p,min}}$ for $w_{gg}$ is unavoidably connected to the choice of galaxy bias model. We first seek to test whether there are a set of cuts that will allow us to use a simple, scale independent linear bias model. For each of our samples, one can estimate an effective bias 
\begin{equation}
b_g^{'}(\rp) = \sqrt{
\frac{\hat{w}_{gg}(\rp)}{w_{\delta \delta}(\rp)}
},
\end{equation}
\noindent
where $\hat{w}_{gg}$ is the measured projected galaxy-galaxy correlation. The matter-matter correlation in the denominator is the theoretical prediction, and so assumes a particular cosmology; we test the impact of switching between reasonably different cosmologies (specifically, the best fitting values from DES Y1 and Planck 2018), and find our results are only very weakly sensitive to this choice. For each sample, we fit $b_g^{'}$ twice, once using a scale independent constant $b_g^{'}(\rp)=b$, and again using a linear-exponential function 
$b_g^{'}(\rp)=a e^{-\rp} + b$.
Although this is not a physically motivated bias model, it has qualitatively the correct behaviour, increasing rapidly on small scales and converging to a constant on large scales. The exact form was motivated by Fig. D1 of \citet{samuroff21}, where the bias in IllustrisTNG is seen to scale roughly as $e^{-\rp}$ plus a constant. In each case, we compute the Bayesian Information Criterion \citep{schwarz78}, 
\begin{equation}
\mathrm{BIC} = k \, \mathrm{log}N_{\rm pts} + \chi^2,
\end{equation}
\noindent
where $k$ is the number of model parameters (either 1 or 2, in the constant/log-linear cases respectively) and $N_{\rm pts}$ is the number of data points included in the fit. The $\chi^2$ for model $M$ is computed using the full data covariance matrix, as $\chi^2_M=[\hat{w}_{gg} - b_{g,M}^{'}w_{\delta \delta}] \mathbf{C}^{-1}[\hat{w}_{gg} - b_{g,M}^{'}w_{\delta \delta}]$. The difference $\Delta \mathrm{BIC}$ then gives us an indicator of which model is preferred by the data -- that is, whether linear bias is sufficient, in a statistical sense, to describe the measured $w_{gg}$. We repeat this process using a range of lower scale cuts $r_{\rm p, min}$, resulting in the curves shown in Figure~\ref{fig:theory:scale_cuts}. Although eBOSS ELGs and \redmagic~low-$z$ appear to be relatively consistent with a linear bias model, even down to small scales, this is not true in all of our samples. The picture is slightly different in the case of \redmagic~high-$z$ and eBOSS LRGs, with the latter in particular preferring the more complicated bias scaling for almost any choice of minimum scale.

The above test indicates that, at least for some of our samples, even at relatively large scales (above $6\mpc$), the linear bias approximation does not provide a good description of the data. Motivated by these findings, our fiducial scale cuts are as follows. We fit $w_{gg}$ for all samples down to $r_{\rm p, min} = 2 \mpc$, with a model that includes nonlinear galaxy bias (as described in Section \ref{sec:modelling:2pt_theory}). At $2\mpc$ we are still well outside the one-halo regime, even for the largest objects in our samples, and so the perturbative expansion in Equation \ref{eq:theory:deltag_terms} may still be sufficiently accurate. To help further validate this choice, we perform additional fits to $w_{gg}$ alone, using very large scales ($>10\mpc$) and linear bias. For each sample, we calculate the shift relative to the $b_1$ value obtained using the fiducial setup, and verify that it is not sufficient to produce an appreciable bias in $w_{g+}$.

\subsubsection{Intrinsic alignment correlations}\label{sec:theory:scale_cuts:wgp}

\begin{figure*}
\includegraphics[width=2.\columnwidth]{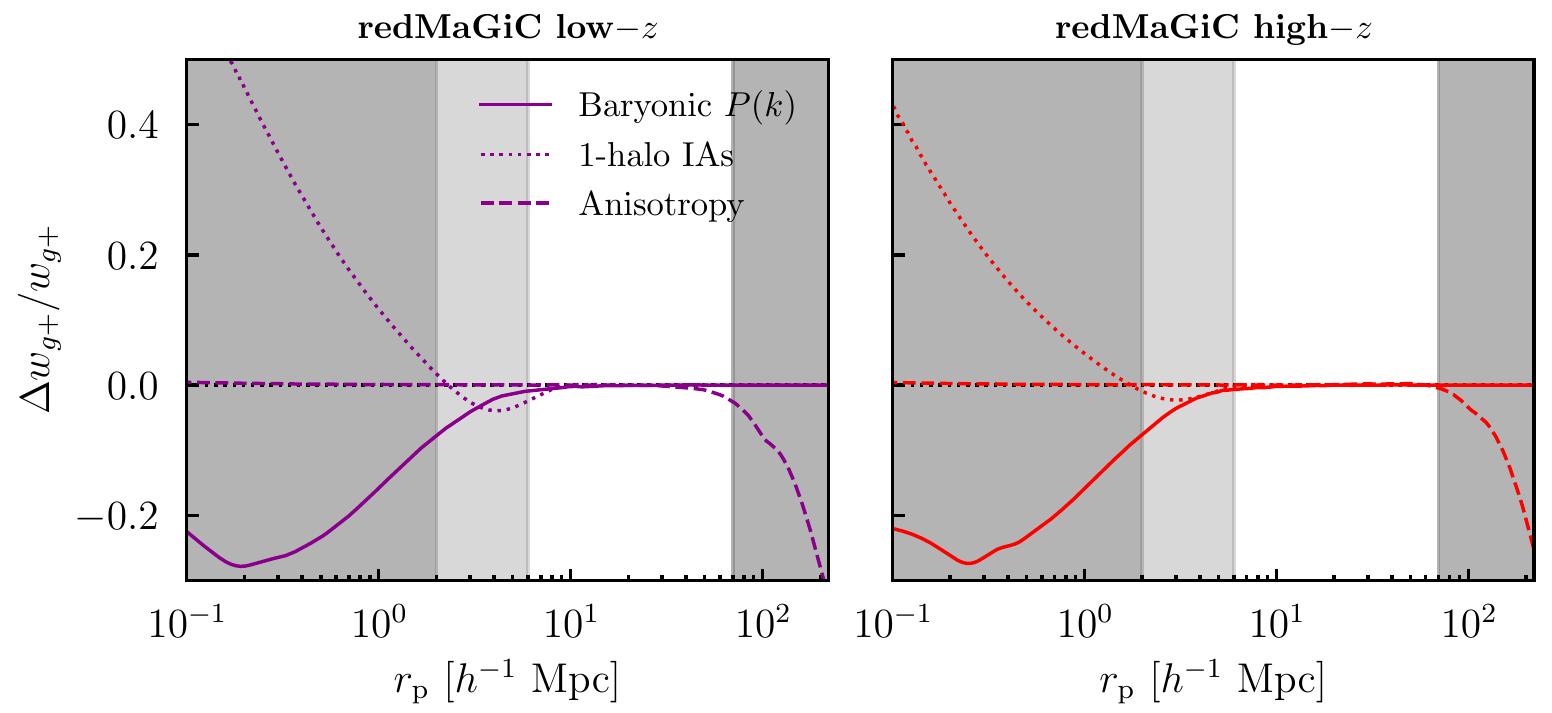}
\caption{The impact of higher order effects on galaxy-shape correlations. We include both \redmagic~high-$z$ and \redmagic~low-$z$ to show that our conclusions hold across the redshift range of our samples. As before, the shaded bands indicate scales removed in our fiducial NLA (light grey) and TATT (dark grey) analyses. We include here: the impact of baryonic feedback, as represented by OWLS-AGN (solid line); the impact of one-halo alignment physics, as represented by the model of \citet{schneider10} (dotted) and signal boosting due to anisotropy in the 3D correlation (\citealt{singh16}; dashed). In all cases, the unmodelled effects appear at the level of a few percent or less on the scales we use, which is below the level of our statistical precision.    }
\label{fig:theory:scale_cuts2}
\end{figure*}

For $w_{g+}$ and $w_{++}$, $r_{\rm p, min}=6\mpc$ for our NLA fits (see Section \ref{sec:results}), primarily driven by IA modelling uncertainty. The TATT model allows us to push to slightly smaller scales, and so here we adopt $r_{\rm p, min} = 2\mpc$. 

We test the robustness of our chosen $w_{g+}$ cuts to a number of unmodelled effects in Figure \ref{fig:theory:scale_cuts2}. Specifically, we generate theory data vectors containing (a) a matter power spectrum contaminated with OWLS-like baryonic feedback (considered as an extreme scenario; see the next paragraph); (b) a one-halo intrinsic alignment signal and (c) projection effects in the 3D correlation function. In each case, we choose a reference IA model $A_1=3.5$ (roughly the NLA best fit for \redmagic~in Section \ref{sec:results}). We do this using both \redmagic~high-$z$ and \redmagic~low-$z$ redshift distributions, since these are representative of the range in $z$ covered by our samples. The fractional differences in Figure \ref{fig:theory:scale_cuts2} are calculated relative to a reference data vector, which does not contain the contamination.

For (a), we follow \citet{y3-generalmethods} in taking the OWLS-AGN scenario \citep{schaye10,vanDaalen11} as an upper limit on the extremity of baryon feedback in the late-time matter power spectrum (see also \citealt*{y3-cosmicshear2} Fig. 5 and the accompanying discussion). The baseline matter power spectrum from CAMB is modified in such a way as to preserve the original cosmology but introduce high$-k$ distortions which mimic the impact of baryons in the OWLS hydrodynamic simulations (\citealt{svcosmology}, eq.~8). As we can see in Figure \ref{fig:theory:scale_cuts2}, baryonic feedback is entirely negligible at $\rp>6\mpc$. Its impact increases rapidly in the intermediate ($2<\rp<6\mpc$) window, but is still only $\sim 2-5\%$ at $\rp=2\mpc$, which is well below the level of statistical error on these scales. If we take the TATT best fit from each \redmagic~sample (fit on scales $\rp>2\mpc$; see Section \ref{sec:results:models}), we obtain $\Delta \chi^2_{\rm RMH} = 0.05$ and $\Delta \chi^2_{\rm RML} = 0.45$.

We carry out a similar exercise with small scale alignments. To get a rough gauge of the impact of one-halo contributions, we use the fitting formulae provided by \citet{schneider10}. We choose to update the overall amplitude of the model to the value found by \citet{singh15} ($a_{h}=0.08$); since this matches LOWZ LRGs, which are somewhat brighter and redder than any of our samples, we expect this to be an upper estimate for the impact of intra-halo physics. Shown by the dotted line in Figure \ref{fig:theory:scale_cuts2}, we again see the effect to be vanishing on scales $\rp>6\mpc$ and slightly larger but still subdominant to errors at $\rp>2\mpc$.

Finally, we also test the impact of a kind of projection effect that induces anisotropy in $\xi_{g+}(\rp,\Pi)$ (dashed lines in Figure \ref{fig:theory:scale_cuts2}). First identified by \citealt{singh16}, the idea is that galaxy alignments along the $\Pi$ direction cannot be measured using shapes measured in 2D projected space; the result is a suppression of the observed alignment signal that scales as $f=\rp/(\rp^2+\Pi^2)^\frac{1}{2}$. As one might intuitively expect from the geometry, $f$ diverges from 1 as $\Pi$ increases, at fixed \rp. Looking at Figure \ref{fig:theory:scale_cuts2} we can see that the impact is primarily at large \rp. To understand this, consider the fact that particular $\Pi$ scales do not contribute equally to the projected correlation at all \rp~(for an illustration of this, see \citealt{singh16} Fig. 10c). That is, at $\rp=1\mpc$, even considerable suppression at large $\Pi$ matters very little; that regime contributes almost nothing to the line of sight integral, since $\xi_{g+}$ scales approximately as $1/r^2$, which at small \rp~and large $\Pi$ is essentially $1/\Pi^2$. At $\rp=30\mpc$, on the other hand, large $\Pi$ scales contribute much more. Although the geometric suppression at a given $\Pi$ is less important for this larger \rp value, the background scaling of $\xi_{g+}$ dominates, and so the overall impact on $w_{g+}$ is larger. This can be modelled in an analogous way to redshift space distortions in galaxy clustering. Although not included in our fiducial model, we can assess the impact in the NLA case using the recipe set out in \citet{singh16} Sec. 2.3. Fortunately we see the impact is largely contained at separations above $\rp>70\mpc$, which are already removed by the upper \rp~cut. Within the range of scales used for our fits, the impact is comfortably smaller than our error bars. 

% ------------- 5. Tests ---------------

\section{Pipeline Testing}\label{sec:validation}

In this section we describe the various tests of the analysis pipeline, and the measurements themselves. These include tests of the theory predictions by comparing different code implementations. We seek to validate the pipeline by reanalysing an existing data set and comparing with published results. Finally we discuss null tests on the data, designed to be sensitive to residual systematics. 

\subsection{Reanalysing LOWZ}\label{sec:tests:lowz}

For the purposes of validating our measurements and demonstrating comparability with previous results, we partially reanalyse the BOSS LOWZ catalogues of \citet{singh15} (see also Section \ref{sec:data}). LOWZ makes a good test data set for several reasons -- not least that it has documented, relatively high signal-to-noise $w_{g+}$ measurements in the literature, and the redshift catalogue is publicly available. We repeat all measurement steps downstream from shape catalogues using our pipeline, and then fit the resulting correlation functions with our modelling setup. At the level of data vectors, our pipeline can reproduce the galaxy clustering and galaxy-shape correlations, $w_{gg}$ and $w_{g+}$, of \citet{singh15} to sub-percent precision on scales $[2,70]$ \mpc.

We also analyse the LOWZ data on large scales, and compare our results to those of \citet{singh15}; when matching their analysis choices exactly (NLA model, linear bias, $\rp>6\mpc$), we recover their reported best fit in the $A_1 - b_1$ plane to $<<1\sigma$. Our fiducial analysis configuration differs from the published LOWZ paper in a number of ways. Most significantly, these include:
\begin{itemize}
	\item Our assumed cosmology is that set out in Section \ref{sec:theory}, instead of WMAP9 \citep{hinshaw13}. This results in a slight increase in the amplitude of the matter power spectrum, which in turn results in a slightly lower alignment amplitude. Note that our fiducial cosmology includes massive neutrinos, which modify the nonlinear $P(k)$ slightly. The difference in $h$ also alters the measurement of the two point functions earlier in the pipeline (via the redshift to distance conversion), although this difference is minimal.
	\item Our fiducial data vector includes galaxy-galaxy, galaxy-shape and shape-shape correlations, whereas \citet{singh15} include only the former two.
	\item We use the \citet{takahashi12} version of halofit to compute the nonlinear matter power spectrum, whereas \citet{singh15} use a slightly older release \citep{smith03}. 
	\item We use an analytic calculation to estimate the data vector covariance matrix, instead of jackknife. While the two agree relatively well, slight differences in the relative weighting of different scales in both $w_{gg}$ and $w_{++}$ are apparent.
	\item We include contributions from lensing and magnification in our model. Although this has little impact on a low redshift spectroscopic data set such as LOWZ, it has a larger bearing on our eBOSS and \redmagic~samples. 
\end{itemize}

We show a more detailed comparison at the parameter level in Appendix \ref{app:lowz}. In short, when matching the analysis choices of \citet{singh15}, we can reproduce their published IA results almost exactly. Switching to our fiducial NLA setup produces a very similar result, with a small reduction in the size of the error bars.

\subsection{Null Tests}\label{sec:tests:null_tests}

A number of systematics (e.g. PSF modelling errors) can lead to a non-zero mean shear. Unlike multiplicative biases, we can look for such effects directly using the data. We find no evidence of such a signal in any of the samples considered here, with $| \left \langle e_i \right \rangle | \sim 10^{-4}$ in all cases. A number of other tests for systematic induced signals are presented in \citet*{y3-shapecatalog}; they find no evidence for correlations between the response-corrected shear and PSF shape and size, or for a statistically significant B-mode signal.

We also measure one additional null signal. Constructing $w_{g\times}$ involves the same basic quantities as $w_{g+}$, but measuring the shape component that is rotated 45 degrees with respect to the radial/tangential direction. Like lensing, astrophysical processes such as intrinsic alignments, to first order, should induce only tangential/radial correlations\footnote{Although some IA models predict a non-zero B-mode contribution (see, e.g., \citealt{catelan01,hirata04,blazek17}), which translates into correlations in the cross ellipticity component, such effects appear only in the II alignment spectra. Given that our constraints are dominated by $g+$ correlations, these terms are thought to be easily subdominant to noise in current surveys.}. Non-zero detection of a cross signal, then, is a red flag for residual measurement systematics. For all samples considered, we find these additional measurements to be consistent with zero within the scales $\rp<70\mpc$. Details of the measurements can be found in Appendix~\ref{app:null_tests}.

% ------------- 6. Results ---------------

\section{Results}\label{sec:results}

This section presents the results of our analyses on the various samples. Although we will focus on intrinsic alignments, it is worth bearing in mind that each analysis also includes two free galaxy bias parameters. The constraints on the bias parameters are strongly dominated by $w_{gg}$, and so they contribute relatively little to the marginal uncertainties on IA parameters. The bias does, however, also enter $w_{g+}$, and so it is important to model it accurately. In every case, the linear bias falls within the bounds of expectation from previous studies ($b_1\sim 1.5-2.0$, depending on the sample), and $b_2$ is small ($b_2\sim 0-0.3$). We note that all samples appear to be fit reasonably well by our model (as quantified by the best $\chi^2$ obtained from fits to the joint $w_{gg}+w_{g+}+w_{++}$ data vector). For more detail on the bias constraints, see Appendix \ref{app:bias_constraints}. It is also worth bearing in mind that all parameters (bias and IA) are constrained within the prior bounds. As we note below, although some samples provide only weak constraints, the priors in Table \ref{tab:theory:models} are sufficiently wide to allow the contours to close in all cases. In Section \ref{sec:results:elgs} below we discuss our results on ELGs, which amount to a null detection. We then move onto the various red samples in Section \ref{sec:results:red_galaxies}, presenting first large-scale results using NLA in Section \ref{sec:results:luminosity_dependence}, and then extending to slightly smaller scales with TATT in Section \ref{sec:results:models}. Section \ref{sec:results:lensmag} then considers more carefully the level of contribution from lensing and magnification.

\subsection{Emission Line Galaxies}\label{sec:results:elgs}

Our first, and perhaps easiest to interpret, results are based on eBOSS Emission Line Galaxies. The data vector is shown in blue in Figure \ref{fig:data:2pt}. We fit the NLA model on large scales (the unshaded region in Figure \ref{fig:data:2pt}), and obtain a null detection,
\begin{equation}
    A_1^{\rm ELG} = -0.42^{+0.50}_{-0.50} \;\;\;\;\;\; (\rp>6\mpc)
\end{equation}

\noindent
with $\chi^2/\mathrm{dof}=1.17$ (with a corresponding $p-$value $p=0.32$). This is expected, given the sample: a non-zero IA signal has never been detected in ELGs (or in any colour-selected sample of blue galaxies more generally; \citealt{mandelbaum10,y1coloursplit,johnston19}). The additional (non $\delta_g$I) terms are also seen to be small, for a number of reasons: first, the magnification coefficient is small $(\alpha-1) \sim 0.1$, for ELGs, which scales down the $\mu$I and $\mu$G contributions. Second, the limits of the line of sight integral tend to suppress the lensing contributions to the signal; integrated out to $\Pi_{\max}=1000\mpc$, $\mu$G tends to dominate on larger scales. In practice, however, with integral limits at $\pm 100 \mpc$, the largest term by some way is $\delta_g$I, with $w_{\delta_g \gamma^{I}}/(w_{\delta_g \gamma^{\rm G}}+w_{\delta_\mu\gamma^{\rm G}}+w_{\delta_\mu\gamma^{I}})\sim14$ at $6\mpc$ (as evaluated at the best fitting parameters). Similarly, for the shape-shape correlation, the ratio of II to other terms is $\sim18$. The end result is a combined best fitting theory prediction that is below the level of shape noise. 

Since the signal-to-noise is relatively low, and there is no visible structure in $w_{g+}$, we also repeat our NLA fits with slightly less stringent cuts, $\rp > 2\mpc$. This tightens the bounds on the alignment amplitude to:

\begin{equation}
    A_1^{\rm ELG} = 0.07^{+0.32}_{-0.42} \;\;\;\;\;\; (\rp>2\mpc)
\end{equation}

\noindent
Indeed, even considering scales down to $0.1\mpc$ in Figure \ref{fig:data:2pt}, we still see no hints of signal in $w_{g+}$ or $w_{++}$. Computing the null $\chi^2$ on all scales $\rp<70\mpc$, we find $\chi^2=17.5$ for 16 data points ($p=0.42)$. This is interesting, since it suggests that there is not a strongly scale dependent one halo (1h) signal of the sort seen in the \redmagic~and CMASS samples (or at least, not one that is detectable above the level of shape noise).

In terms of sample, the closest results in the literature are those of \citet{mandelbaum10} and \citet{tonegawa17}. These both use blue emission line galaxies, from WiggleZ and Subaru respectively, and also make null detections of $A^{\rm M11}_1= 0.15^{+1.03}_{-1.07}$ and $A_1^{\rm T18}=0.49^{+3.56}_{-3.56}$ respectively. Our results tighten the null constraint, imposing an upper limit of $|A_1|<0.78$ at $95\%$ CL. In terms of redshift, our eBOSS ELG measurements sits between the earlier two ($z\sim0.8$, compared with $z\sim0.5$ for WiggleZ and $z\sim1.4$ for Subaru). It is worth exercising some caution here, however, since in both cases it is not clear that the sample matches ours closely. In particular, the \citet{mandelbaum10} sample is a relatively bright selection of starburst galaxies with specific colour cuts (see their Sec 3.1). That said, the best-fitting bias values are relatively similar to our own ($b_1\sim1.4$ for eBOSS ELGs, $b_1\sim1.5$ for WiggleZ). The \citet{tonegawa17} sample on the other hand, has both a complicated spectroscopic selection function, and additional shape catalogue cuts that remove $\sim50\%$ of objects. Although all three results (including our own) present IA results consistent with zero in blue-ish samples across a range of redshifts, it is not clear the results are directly equivalent.

It is also worth stressing here that although very different from (and much bluer than)  an LRG or \redmagic~type sample, eBOSS ELGs are not exactly representative of a typical weak lensing catalogue either. Indeed, the eBOSS ELG selection is designed to facilitate a high S/N BAO measurement, and is based on the presence of particular emission lines, as determined via a complex set of magnitude and colour cuts \citep{raichoor17}. The completeness of the sample given the cut is difficult to quantify \citep{guo19}. In contrast, lensing samples tend to have much simpler (if any) colour selection, and cuts designed to minimise lensing measurement biases and optimise a weak lensing measurement. The two are designed for different scientific purposes, and so we should not expect them to match. For this reason, caution is required when trying to extrapolate these results.

\subsection{Red Galaxies}\label{sec:results:red_galaxies}

\subsubsection{Constraints on Large Scale Intrinsic Alignments }\label{sec:results:luminosity_dependence}

We next consider our other galaxy samples, \redmagic, eBOSS LRGs, and CMASS, which we fit on large scales (again, $>6\mpc$) using NLA. In each case, we find a clear detection, with our three parameter model of the joint data vector $(A_1, b_1, b_2)$ providing a good $\chi^2/\mathrm{dof}$. The constraints and the goodness of fit statistics can be found in the upper four rows of Table \ref{tab:results:nla_luminosity}. Defining the signal-to-noise according to Eq. (2) of \citet{becker16}\footnote{The expression is $S/N=(\hat{\mathbf{w}}\mathbf{C}^{-1}\mathbf{w}^{\rm model})/(\mathbf{w}^{\rm model}\mathbf{C}^{-1}\mathbf{w}^{\rm model})^\frac{1}{2}$, where $\hat{\mathbf{w}}$ is the observed (noisy) data vector, $\mathbf{w}^{\rm model}$ is the best fitting theory prediction, and $\mathbf{C}^{-1}$ is the inverted covariance matrix. This is slightly different from the common definition using $\hat{\mathbf{w}}$ only, which is known to be biased high if noise is present.}, we find $S/N=22$ in \redmagic~low-$z$ and $S/N=18$ in \redmagic~high-$z$. Given the smaller area, the detections in our SDSS samples are slightly weaker, at $S/N=6$ for CMASS and $S/N=5$ for eBOSS LRGs. The best fitting model predictions can be seen in Figure \ref{fig:data:2pt}. As we saw with the ELGs in Section \ref{sec:results:elgs}, CMASS and eBOSS LRGs are dominated by the primary IA signal ($\delta_g$I for $w_{g+}$, II for $w_{++}$). For \redmagic, however, the picture is slightly different; photo-$z$ scatter tends to increase the maximum distance galaxies can be physically separated (by shifting well-separated objects below $\Pi_{\rm max}$), and so boosts the lensing and magnification terms. With \redmagic~low-$z$ this is partly cancelled out by the fact that $\alpha$ is very close to $1$, and that the mean redshift is relatively low. These things are less true for \redmagic~high-$z$, however, and so we see a stronger $\delta_g$G contribution. The $\delta_g$I signal is also slightly stronger, however, and the additional terms still account for only order of a few percent of the total signal.

\begin{figure}
\includegraphics[width=1.\columnwidth]{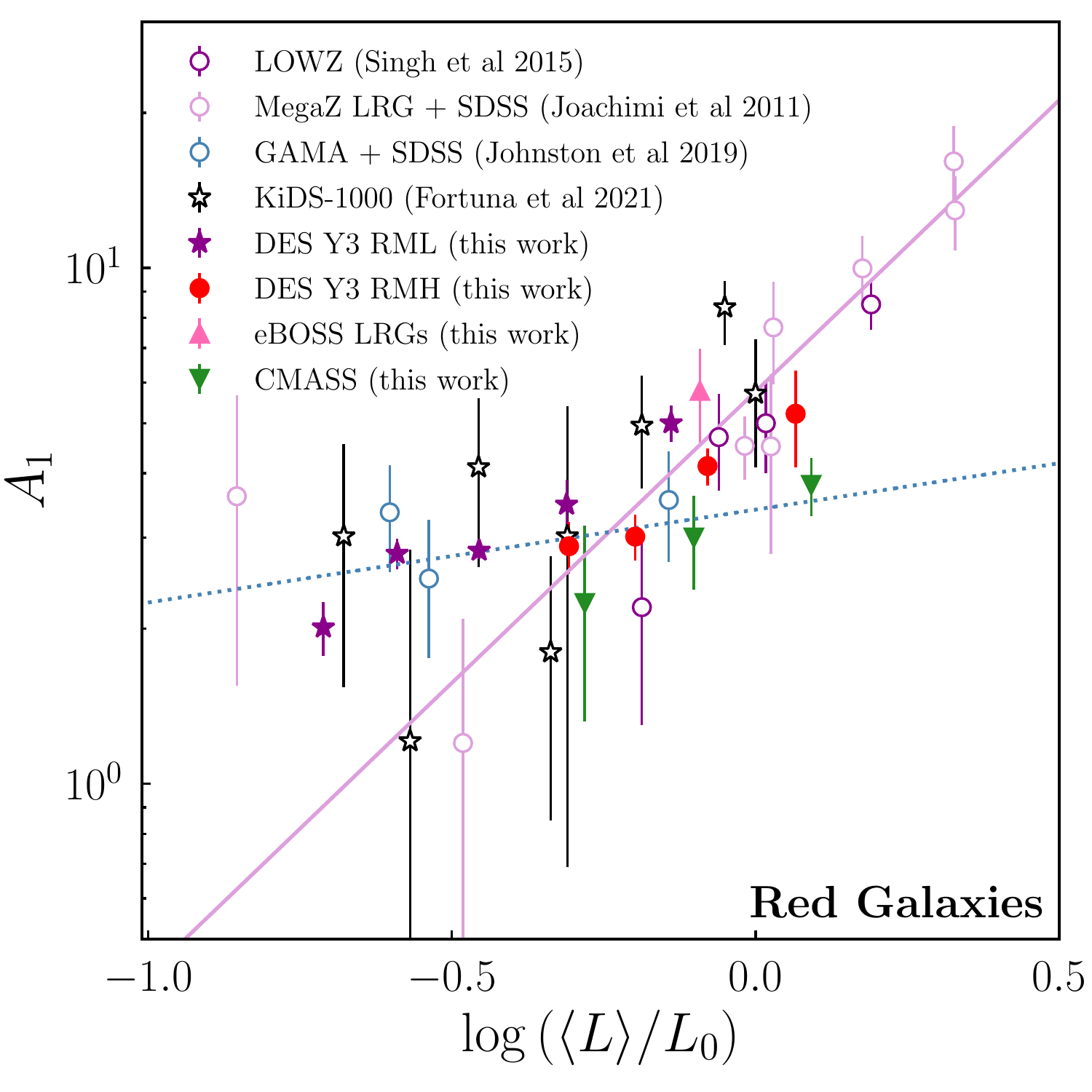}
\caption{Intrinsic alignment strength as a function of $k+e$-corrected $r$-band luminosity in red galaxies. By convention, the luminosities are defined relative to a pivot $L_0$, which corresponds to an absolute magnitude $M_r=-22$. Open points show previous results from the literature (as indicated in the legend). For illustrative purposes we also show two power law fits from the literature, one fit to GAMA+SDSS at low-mid $L$ (dotted blue), and the other to MegaZ at high $L$ (solid purple). The filled points show our measurements on \redmagic~low-$z$ (stars; five points), \redmagic~high-$z$ (red dots, four points), eBOSS LRGs (pink triangle; one point) and CMASS (green upside down triangles; three points). Note that all results shown here assume the one-parameter NLA model. }
\label{fig:results:luminosity_dependence}
\end{figure}

\begin{table}\setlength{\tabcolsep}{3pt}
\begin{center}
\begin{tabular}{l|ccccc}
\hline
Sample                   & $\langle L_r \rangle/L_0 $ & $\langle z \rangle$ & $\langle M_r-M_z \rangle$ & $A_1$  & $p(>\chi^2)$ \\ 
\hhline{======}
RMH (all $L$) & $0.68$    & $0.78$ & 0.43 & $3.54^{+0.18}_{-0.18}$ & $0.20$ \\
RML (all $L$)   & $0.40$   & $0.46$ & 0.43 & $3.34^{+0.13}_{-0.13}$ & $0.12$   \\
CMASS (all $L$) & $0.84$ & $0.52$  & 0.36 & $2.72^{+0.47}_{-0.47}$   & $0.10$ \\
LRGs (all $L$)     & $1.20$   & $0.75$  & 0.39 & $5.78^{+1.19}_{-1.19}$  & $0.89$   \\
\hline
RMH $L0$   & 0.49    & 0.77 & 0.44 & $3.47^{+0.34}_{-0.34}$ & $0.11$ \\
RMH $L1$   & 0.63    & 0.77 & 0.42 & $3.01^{+0.31}_{-0.30}$  & $0.52$  \\
RMH $L2$   & 0.83    & 0.78 & 0.44 & $4.13^{+0.33}_{-0.34}$  & $0.15$  \\
RMH $L3$   & 1.16   & 0.78 & 0.45 & $5.22^{+1.11}_{-1.11}$ & $0.40$  \\
RML $L0$   & 0.19   & 0.32 & 0.50 & $1.95^{+0.23}_{-0.23}$ & $0.28$  \\
RML $L1$   & 0.26   & 0.43 & 0.42 & $2.86^{+0.29}_{-0.29}$ & $0.22$ \\
RML $L2$   & 0.35   & 0.47 & 0.40 & $3.01^{+0.52}_{-0.51}$ & $0.62$  \\
RML $L3$   & 0.49   & 0.50 & 0.40 & $4.39^{+0.20}_{-0.20}$ & $0.79$ \\
RML $L4$   & 0.73  & 0.53 & 0.40 & $5.00^{+0.20}_{-0.21}$ & $0.14$ \\
CMASS $L0$ & 0.52   & 0.48 & 0.42 & $2.23^{+0.92}_{-0.92}$ &  $0.47$ \\
CMASS $L1$ & 0.79   & 0.52 & 0.36 & $3.00^{+0.62}_{-0.61}$ &  $0.47$ \\
CMASS $L2$ & 1.24   & 0.55 & 0.29 & $3.78^{+0.49}_{-0.49}$ &  $0.70$ \\
\hline
\end{tabular}
\caption{NLA model constraints from the various red galaxy samples discussed in this work, as a function of $r$-band luminosity. The upper four rows refer to the full samples, while the lower twelve refer to subsets in luminosity bins, as defined in Section \ref{sec:measurements:absmag} and Figure \ref{fig:data:2pt_lbins}. The first three columns show the mean luminosity, redshift and rest frame colour of each sample. The final two are the posterior mean IA amplitude, and the corresponding $p-$value. In all cases the model is seen to provide a reasonable fit to the data. These fits were performed using large scales only $(\rp>6\mpc)$ in $w_{g+}$ and $w_{++}$. 
}\label{tab:results:nla_luminosity}
\end{center}
\end{table}

Another useful exercise is to divide the samples into luminosity bins, and map out the dependence of the alignment signal. For each of the bins shown in Figures~\ref{fig:data:nofL} and \ref{fig:data:2pt_lbins}, we fit an NLA amplitude (we also fit for galaxy bias, but since we only split the shape sample by luminosity, that doesn't change significantly between $L$ bins). The results are shown in Figure~\ref{fig:results:luminosity_dependence}, with numerical parameter constraints in Table~\ref{tab:results:nla_luminosity}. Note that we also include a selection of previous measurements from the literature, denoted by open points. There are a number of trends worth considering here. First, taken naively, our results are consistent with the qualitative picture of a broken power law dependence on luminosity: the trend in $A_1$ below $\mathrm{log}L/L_0\sim-0.2$ is much shallower than above it. The $(L/L_0)^\beta$ power law parameterisation was first introduced as an empirical scaling by \citet{joachimi11}. Although there is little physical motivation, it has been adopted relatively widely both in direct IA measurements \citep{singh15,johnston19,fortuna21} and in forecasts \citep{krause15,fortuna20}, as it was simple to implement and appeared to fit the available data relatively well. In recent years, the picture has become more complicated, as evidence has begun to emerge of a weaker relationship at low $L$ (see e.g. \citealt{johnston19}). Again, however, this is empirical, and there is no first-principles reason to expect a double power law in particular (or any other form). Our results appear to reinforce that evidence. Our \redmagic~low-$z$ sample in particular provides a significant improvement in the constraints on the fainter end of the $A_1-L$ relation (by a factor of 3 or more in the error bars). At the brighter end, our CMASS, LRG and \redmagic~high-$z$ samples also appear qualitatively consistent with previous results, following a considerably steeper slope. Taken at face value, given the error bars, we could interpret this as ruling out a single power law with relatively high significance. We stress, however, that it is worth being cautious here. Despite all being ``red", there are differences between the composition of the samples, as we will come to below -- it is possible these population differences may be partly responsible for the apparent trends in $L-A_1$ space. For this reason, we do not present best-fitting constraints on $\beta$, but rather a more qualitative discussion of how to interpret our results.

It is noticeable that CMASS (and to a lesser extent \redmagic~high-$z$) tends to lie roughly $\sim 1-3 \sigma$ below the best fit single power law from the literature (the purple line in Figure \ref{fig:results:luminosity_dependence}). Since the error bars here are mostly shape noise dominated and the luminosity bins are disjoint, the points should also be uncorrelated to good approximation, meaning it is unlikely this is random scatter. We can perhaps understand these trends in terms of colour differences. In the upper panel of Figure \ref{fig:data:colour_mag} we can see that CMASS is considerably bluer than LOWZ. Indeed, while most extreme for CMASS, all of our samples tend to peak lower than LOWZ in $M_r-M_z$ space. Not only this, there is also some difference in colour between luminosity bins for a given sample. For example, the galaxies in the upper CMASS bin ($L2$), on average, have slightly bluer rest-frame magnitudes than the lower two bins.  Although for the sake of convenience, it has been useful to split galaxies into binary ``red" and ``blue" categories, our results suggest that this may be an over-simplification for modelling purposes, given the precision of current data sets. They suggest that a more sophisticated modelling may be needed, which accounts for colour and luminosity (and potentially other properties such as satellite fraction) simultaneously. This will be the focus of future work.

We can compare these results with those of \citet{singh15}, who consider colour bins within the LOWZ LRG sample. Although that work reported no clear trend across five bins in $g-i$ rest frame colour, it should be noted that LOWZ covers a fairly narrow range in colour space (see the black contour in Figure \ref{fig:data:colour_mag}). Even the bluest bin in that paper still represents a relatively bright red sample compared with the galaxies considered here. It seems plausible that our wider coverage allows us to see a trend that is not detectable in a relatively homogeneous sample like LOWZ. We thus consider the two results qualitatively consistent.

One other feature worth mentioning, although we do not seek to quantify it, is the behaviour of $w_{g+}$ on very small scales. Both \redmagic~samples appear to exhibit a strongly luminosity-dependent 1h contribution to $w_{g+}$ (Figure \ref{fig:data:2pt}, purple and red). Noticeably, the RML L3 and RMH  L0 bins, though having very similar mean luminosities, have qualitatively different one halo signals. In the case of eBOSS LRGs and CMASS we see no such trends, but this is quite possibly simply the result of low $S/N$, even on smaller scales. In all of these cases, it is worth bearing in mind that the density tracer samples differ. Although the differences between the small scale behaviour of $w_{g+}$ in the various bins/samples \emph{could} be a result of the 1h IA signals, they could also be partly down to differences in, for example, comoving density and galaxy bias.

We also briefly test for redshift evolution in our red galaxy samples. Since we are interested in isolating inherent evolution in the IA signal (as opposed to changes in sample composition), we compare samples at roughly the same luminosity. Specifically, we define two narrow bins in Figure \ref{fig:results:luminosity_dependence} (one in the low luminosity regime, at $\mathrm{log}L/L_0 \sim -0.3$, and the other at higher $L$, $\mathrm{log}L/L_0 \sim -0.05$). Plotting out $A_1(z)$ in these two slices, we find no evidence for redshift evolution over the range $z=[0.25,0.8]$. Although we see the same trend with CMASS being slightly lower than other samples at the same $L$, there is no evidence that this is the result of an underlying redshift trend. Since there is no statistically significant correlation, we do not include the figure in the main body of the paper; for completeness, however, it is shown along with redshift power-law constraints in Appendix \ref{app:redshift_dependence}, Figure \ref{fig:results:redshift_dependence}. 

\subsubsection{Model comparison: NLA \& TATT}\label{sec:results:models}

\begin{figure}
\includegraphics[width=\columnwidth]{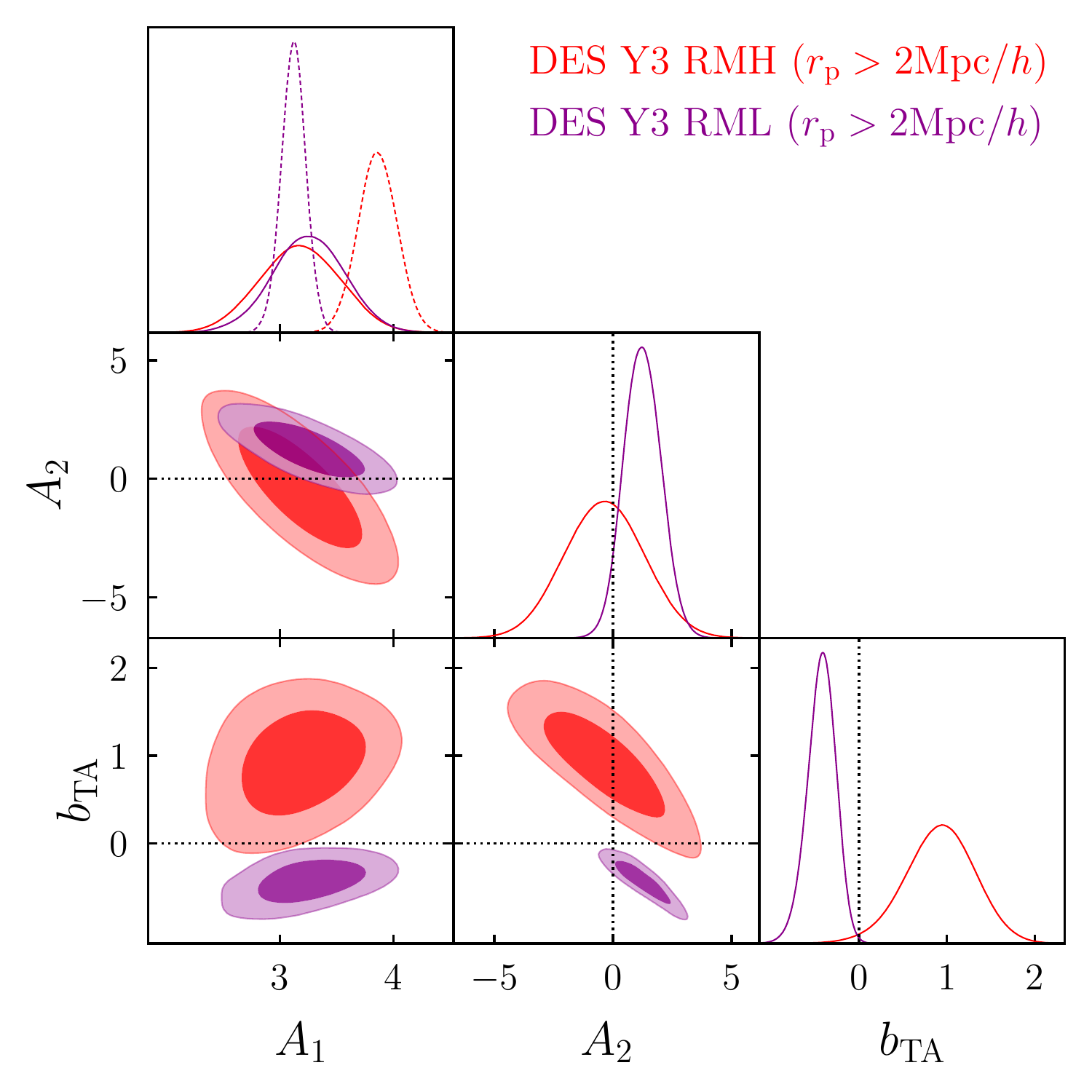}
\caption{$68\%$ and $95\%$ confidence intervals for the TATT model constraints from our two DES Y3 \redmagic~samples, on scales $\rp>2\mpc$. As before, we show \redmagic~high-$z$ in red and \redmagic~low-$z$ in purple. For comparison, the marginalised NLA constraints on $A_1$ are shown as dashed curves in the upper left panel. Note that the scales here differ from the fiducial NLA analyses, and so the best fits are different from those shown in Table \ref{tab:results:nla_luminosity}. The dotted black lines mark the zero points for the non-NLA parameters.
}
\label{fig:results:tatt_contours}
\end{figure}

In Section \ref{sec:results:luminosity_dependence} we explored the behaviour of IAs on very large scales, where the NLA model is thought to be sufficient. We next turn to a slightly different question: on what scales precisely does the simple model break down? Based on theory, there is thought to be an intermediate regime, outside the one halo regime, but where higher order correlations (such as those included in the TATT model) become significant. It is still, however, and open question as to \emph{how} significant and on exactly \emph{which} scales. 

To explore this, we repeat our analysis with the minimum scale reduced slightly to $\rp>2\mpc$ in $w_{g+}$ and $w_{++}$. Note that the cuts on $w_{gg}$ are fixed, and the modelling there does not change (i.e., there are always 2 free galaxy bias parameters). For each sample, we fit both NLA (1 free IA parameter) and TATT (3 IA parameters). Our results are summarised in Figure \ref{fig:results:tatt_contours}. 

Unsurprisingly given Figure~\ref{fig:data:2pt}, CMASS, eBOSS ELGs and eBOSS LRGs provide very broad constraints on the TATT parameters, and so are not shown. The signal-to-noise in these samples is still relatively low. Although one can fit a single amplitude relatively well, there is little constraining power left for the shape of the correlation function. The picture is slightly different, however, in our \redmagic~samples. The cosmological volume here is significantly larger, and the $S/N$ higher. Starting with the slightly larger sample, \redmagic~low-$z$, we find:
\begin{equation}
    A_2^{\rm RML} = 1.24^{+0.74}_{-0.80} \;\;\;\;\;\;  b_{\rm TA}^{\rm RML} = -0.43^{+0.18}_{-0.13}.
\end{equation}
\noindent
In words, our measurements favour (albeit relatively weakly) a combination $A_2>0$ and $b_{\rm TA}<0$. Comparing with the NLA fits on the same scales, we find $p(>\Delta \chi^2)=0.01$ ( $\chi_{\rm NLA}^2=31.2$ and $\chi_{\rm TATT}^2=21.9$). 

In the case of \redmagic~high-$z$, the data appear to prefer a non-zero $b_{\rm TA}$ at the level of roughly $2.5\sigma$, with:
\begin{equation}
    A_2^{\rm RMH} = -0.36^{+1.63}_{-1.64} \;\;\;\;\;\;  b_{\rm TA}^{\rm RMH} = 0.92^{+0.41}_{-0.36}.
\end{equation}
\noindent
Again, comparing the TATT and NLA fits, we obtain $p(>\Delta \chi^2)=0.00004$, suggesting a statistically significant preference for TATT on these scales; the respective goodness-of-fit statistics are $\chi_{\rm NLA}^2=33.9$ and $\chi_{\rm TATT}^2=13.6$, and $\Delta \mathrm{dof}=2$. 

Interestingly, the deviations from NLA manifest in quite different ways at the data vector level (see Figure \ref{fig:data:2pt}). In the case of \redmagic~high-$z$, the positive $b_{\rm TA}$ increases the power on intermediate scales $(2<\rp<10\mpc)$, resulting in a significantly flatter $w_{g+}$. For \redmagic~low-$z$, on the other hand, we see the opposite effect; a reduction in the amplitude of $w_{g+}$ in the $(2<\rp<6\mpc)$ range is accompanied by a slight increase around $20-30\mpc$. This gives a slightly steeper theory prediction, which matches the shape of the measured correlation relatively well. Taken together, this amounts to a $\sim 3.8 \sigma$ difference in $b_{\rm TA}$ between the two \redmagic~samples.

Although interesting, it is worth being cautious here. Since TATT is a relatively flexible model (albeit a physically motivated one), it is possible that non-zero $A_2$ and $b_{\rm TA}$ values could arise due to other untreated systematics. Baryonic physics, for example, tends to appear on small-intermediate scales, and modulates the power spectrum in a scale dependent way. It is also true, however, that baryons tend to \emph{suppress} power on smaller scales; this is the case in (almost) all hydrodynamic simulations at all redshifts. Since \redmagic~high-$z$ prefers a TATT model that does the opposite (relative to NLA), it seems unlikely that baryons are driving the non-zero $b_{\rm TA}$ value here. Even considering the \redmagic~low-$z$ case, it seems very unlikely that we are simply seeing residual baryonic feedback. In Section \ref{sec:theory:scale_cuts}, we saw that the OWLS-AGN scenario (itself an extreme case) produced at most $\Delta \chi^2=0.45$ at the TATT best fits quoted above. The difference between the NLA and TATT goodness-of-fit are more than an order of magnitude larger than this; for baryonic feedback alone to explain the non-zero TATT parameters would require a significantly more extreme scenario than OWLS-AGN. 

Another possible effect here is nonlinear galaxy bias. Our model for it is incomplete in the sense that while we include nonlinear bias in our $w_{gg}$ model, only $b_1$ enters the $w_{g+}$ prediction. Incorporating nonlinear bias, and all the TATT-bias cross terms, into the $w_{g+}$ model is the focus of ongoing work. We can, however, make a rough estimate for the impact based on our $w_{gg}$ fits. For \redmagic~low-$z$, we find the data consistent with linear bias ($b_2 = -0.09 \pm 0.07$). The equivalent value for \redmagic~high-$z$ is slightly larger, but still small $b_2= 0.39 \pm 0.08$. Since the additional terms contributing to $w_{g+}$ will be proportional to $b_2$ multiplied by the various IA coefficients, it seems likely that they should be relatively small compared with the IA-only contributions.

Finally, we also consider the possibility that our results here could be the result of a non-local lensing contribution from small scales. Such contributions add to the galaxy-galaxy lensing ($g$G) term, and tend to boost its power on small to intermediate scales (see e.g. \citealt{baldauf10, maccrann20}). Fortunately, even considerably different halo mass profiles produce approximately the same contribution on scales well outside the virial radius, behaving effectively as an enclosed point mass and scaling as $1/\rp^2$.  
To test this, we generate NLA-only theory data vectors from the NLA fits on large scales; we add a point mass term (Eq. 7-8 of \citealt{maccrann20}), and adjust $\delta M$ until the NLA+PM theory prediction for $w_{g+}$ matches the data on scales $2>\rp>6\mpc$. Although a $1/\rp^2$ scaling \emph{can} match the data in the \redmagic~high$-z$ case, we find the mass required to do this is $\sim 3 \times 10^{15} M_{\odot}/h$, which is much larger than the typical halo mass expected for DES \redmagic~(see, for example, \citealt{y3-2x2ptbiasmodelling,zacharegkas22}). This would also require the point mass contribution to dominate $g$G out to scales of $\sim20\mpc$, which again is not thought to be realistic. Moreover, the point mass explanation should lead to an excess $w_{g+}$ on small scales for both \redmagic\ samples, contrary to the observed behaviour.  For all of these reasons, we conclude that a point mass term cannot explain the deviations from NLA on small scales for \redmagic~high-$z$.

Overall, these tests seem to suggest a real IA signal (or, at least, a significant systematic that we have not considered). This is interesting from a modelling perspective. It implies some dependence in the TATT parameters with galaxy properties ($b_{\rm TA}$ most obviously, going from negative in \redmagic~low$-z$ to positive in \redmagic~high$-z$, but also potentially $A_2$). As we discussed in the previous section, the \redmagic~high- and low$-z$ samples differ in redshift, but also in galaxy properties like colour and luminosity. Although, given this, differences somewhat expected, this is not something on which there are previous results to guide us. Disentangling what exactly is driving the differences is an interesting question, but a potentially difficult one to answer. We leave this for future work.

\subsubsection{Robustness to Cosmology and $X_{\rm lens}$}\label{sec:results:robustness}

In this section we seek to test the robustness of our analysis to various sources of systematic error. One such potential contaminant is the effect known as $X_{\rm lens}$. One can find extensive early discussion in the DES Y3 results papers \citep{y3-3x2ptkp,y3-2x2ptbiasmodelling}, but essentially  $X_{\rm lens}$ is a multiplicative factor of unknown origin between the amplitudes of the galaxy-galaxy lensing and galaxy clustering measurements. This offset was seen to be scale and redshift independent, and to impact only Y3 \redmagic, and not the fiducial magnitude limited lens sample, \blockfont{MagLim}. Subsequent tests have pointed towards a systematic in the photometry, which affects the \redmagic~selection (see \citealt{y3-2x2ptbiasmodelling} Sec. VG). The magnitude of $X_{\rm lens}$ is constrained relatively well by the $3\times2$pt data in \citealt{y3-3x2ptkp}, to $X_{\rm lens} = 0.877^{+0.026}_{-0.019}$, which is roughly the size of the fractional error bar, $(w_{g+} - \sigma_{w_{g+}}) / w_{g+}$, for our \redmagic~samples in the range $2<\rp<70 \mpc$. Since it is scale independent, we expect the impact to be completely degenerate with $A_1$. Given these things, we do not expect $X_{\rm lens}$ to have a qualitative impact on our results. Although it may modulate the best fit $A_1$ in our \redmagic~(not CMASS or eBOSS) at the level of $\sim10\%$, comparison between samples is already uncertain to at least this level due to differences in colour space. Given that the TATT parameters primarily alter the shape of $w_{g+}$, we do not expect $X_{\rm lens}$ to alter the findings of Section \ref{sec:results:models}.

We also briefly consider the impact of our choice of cosmology; to fit for IAs, we need to assume a particular set of cosmological parameters (e.g. for calculating the matter power spectrum). As discussed in Section \ref{sec:theory}, we assume a flat \lcdm~universe with massive neutrinos and a clustering amplitude $\sigma_8$ similar to that reported by Planck. For each of our samples, in addition to the best fitting NLA data vector, we generate a second with a perturbed cosmology; for this we choose the DES Y3 $1\times2$pt best fit. By comparing $w_{gg}$ at the two cosmologies, we can compute an effective shift in large scale bias $\Delta b_1$. This in hand, plus the observed impact on $w_{g+}$, we can estimate the shift in the best fit $A_1$. The end result is a change of at most one or two percent. That is, the difference between plausible cosmologies is not sufficient to significantly affect our results.

\subsection{Assessing the Contribution of Magnification \& Lensing}\label{sec:results:lensmag}

\begin{figure*}
\includegraphics[width=1.8\columnwidth]{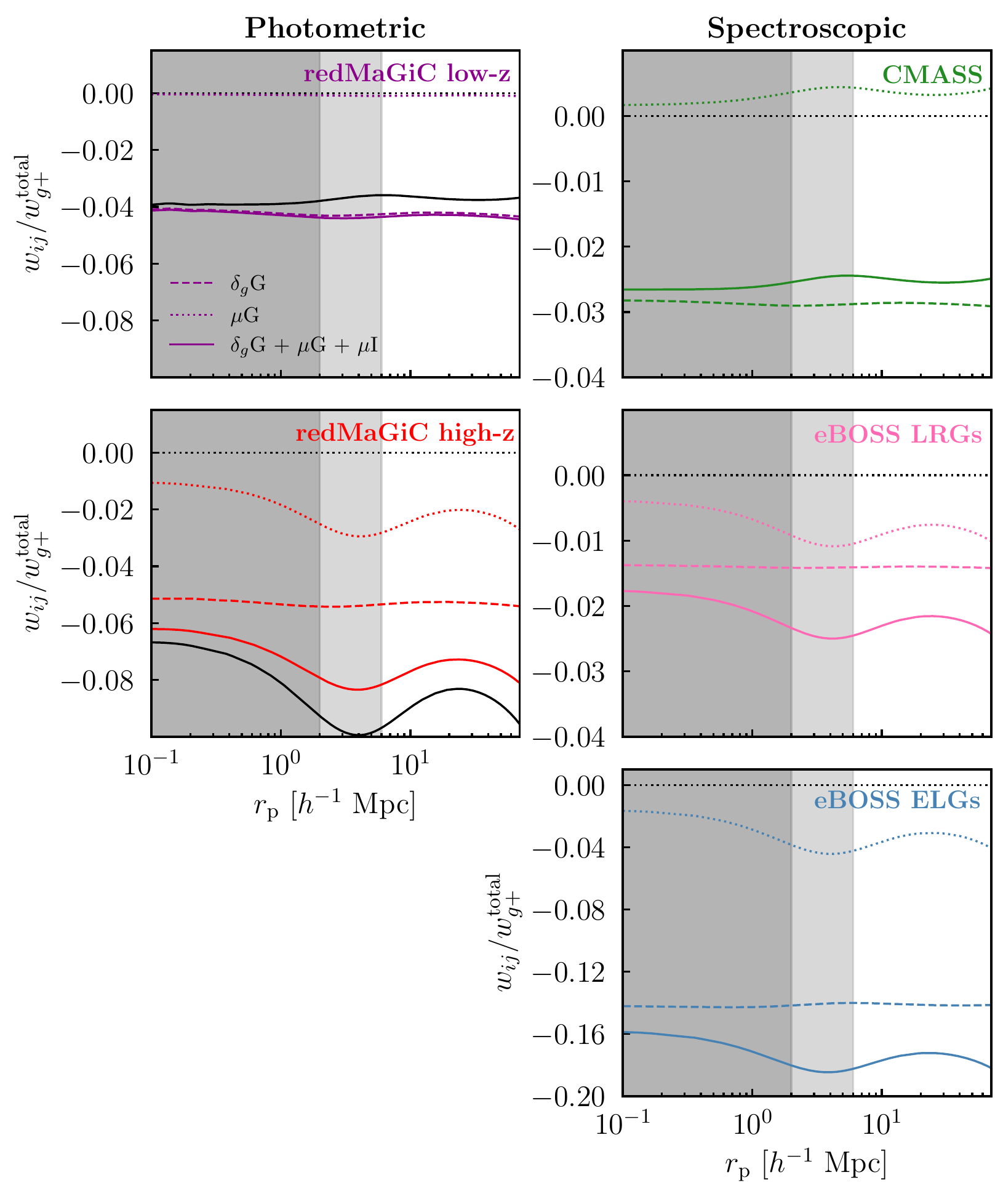}
\caption{The impact of lensing and magnification on our best fitting theory data vectors. Each contribution to $w_{g+}$ is shown as a fraction of the total signal ($\delta_g$I + $\delta_g$G + $\mu$G + $\mu$I). Clockwise from top left, we have: \redmagic~low-$z$ (purple), CMASS (green), eBOSS LRGs (pink), eBOSS ELGs (blue) and \redmagic~high-$z$ (red). In the left-hand column we show our two photometric samples, which are more strongly affected by lensing and magnification. On the right are our three spectroscopic samples. Within each column, the samples are arranged vertically by redshift, starting with low redshift samples at the top. As in previous figures, the shaded bands represent scales excluded from fits to these data vectors. Different line styles indicate different signal components, with the sum of the non $\delta_g$I terms shown as a solid line. Note that the $\mu$I term is negligible on all scales and for all samples, and so we do not show it separately. In the two \redmagic~cases, we include a second solid line (black). This demonstrates the impact of using an alternative estimate for the magnification coefficients $\alpha$, obtained using \blockfont{balrog} source injection. Unlike in the other samples, the CMASS $\mu$G term is positive. This is because the magnification coefficient for this particular sample is small ($\alpha<1$, see Section \ref{sec:measurements:magnification}); in physical terms this means the geometric effect from magnification, which expands the apparent area of a patch of sky, outweighs the boost in number density due to the brightening of the apparent galaxy fluxes. Note that the scale on the vertical axis differs between panels.
}
\label{fig:results:magnification}
\end{figure*}

In addition to the main IA signal, our measurements have contributions from lensing and magnification (see the discussion in Section \ref{sec:theory}). These are always included in our modelling, but it is interesting to briefly discuss their effect. The fractional impact of the various terms is shown in Figure \ref{fig:results:magnification}. Note that the data vectors here are evaluated at the NLA best fit for each sample, and so $b_1$ and $A_1$ differ somewhat between the panels.  

Consider first the two photometric samples, \redmagic~low-$z$ and \redmagic~high-$z$ (upper two panels, purple and red in Figure \ref{fig:results:magnification}).  Here we see a total non-$\delta_g$I contribution of roughly $8\%$ and $4\%$ in the high- and low-$z$ samples respectively. For context, the $1\sigma$ uncertainty on $A_1$ in these samples is $\sim4-5\%$ (see Table \ref{tab:results:nla_luminosity}); including these effects in the modelling of $w_{g+}$ is clearly necessary to avoid bias at the current precision. The closest comparison in the literature is Fig. 5 in \citet{joachimi11}; again, we confirm that our pipeline can reproduce those results. We see a similar ordering of the terms to their figure, with $\delta_g$G dominant, followed by $\mu$G, and with $\mu$I the smallest, at sub-percent level (although note there are some important differences between the comparison in Figure \ref{fig:results:magnification} and that of \citealt{joachimi11}, and so one should not expect the details to match perfectly).

In the two left-hand panels, we also show the results of the same exercise, but using alternative estimates for the magnification coefficients from \blockfont{balrog} (black lines; see Section \ref{sec:measurements:magnification} for the actual values). Note that $\alpha$ is the only quantity that changes here; we do not re-fit the data vectors, and so the dominant $\delta_g$I term in the denominator is fixed. The new $\alpha$ values are thought to include additional effects omitted by the flux-only method described in Section \ref{sec:measurements:magnification}. Unfortunately, however, we do not have \blockfont{balrog} injections covering the whole DES footprint, and so the resulting estimates are noisy. We also cannot easily produce \blockfont{balrog} mocks closely matching our CMASS and eBOSS samples, which somewhat limits its use for our purposes. The overall impact, however, is seen to be relatively small. That is, while ignoring magnification altogether could have a significant impact on our results, the choice of one $\alpha$ estimate over another is unlikely to. 

Although it is common to assume spectroscopic IA measurements are immune to lensing effects, we see in Figure \ref{fig:results:magnification} that this is not entirely the case. The two main terms are the same as with \redmagic: $\mu$G and $\delta_g$G. It is interesting that the latter (dashed) still dominates; even in the case of very narrow per-galaxy redshift distributions, where $p(z) \rightarrow \delta_z$, the ensemble redshift distribution is sufficiently broad to allow a non-negligible galaxy-galaxy lensing contribution. The $\mu$G term (dotted), we also note, does not depend on the quality of the redshifts. Galaxies along the same line of sight are affected by magnification and lensing due to the same foreground structure, which modulates $w_{g+}$ at the level of a few percent at $z\sim0.8$. 

It is worth briefly considering what these results mean for future IA measurements. The Stage IV spectroscopic survey DESI is expected to obtain spectra for a sample of LRGs over a comparable redshift range to ours, but for a considerably wider area and greater number density ($n_c\sim 6 \times 10^{-4} h^3 \mathrm{Mpc}^{-3}$ over 14,000 square degrees; \citealt{zhou20}). Similarly, one can expect at least an order of magnitude increase in the number of ELGs available for the type of measurement we perform here \citep{raichoor20}. Euclid and the Roman Space Telescope will also have spectroscopic instruments, which will further add to the pool of data available for IA measurements. Likewise, Stage IV photometric surveys such as Rubin, Euclid and Roman will probe a similar selection of galaxies to DES, but over a much wider area and to a greater depth (see, for example, \citealt{euclid21}). These will allow measurements using a \redmagic-like sample similar to ours, but with significantly improved S/N and finer binning in colour/redshift/luminosity. Used together -- with photometry providing shape inference, and accurate redshift information from either spectroscopic data or high quality photometric measurements -- the next generation of surveys will provide a powerful tool for studying intrinsic alignments. Given what we see here, it seems very likely that direct IA measurements using these upcoming data will need to account for lensing and magnification.

\section{Conclusions}\label{sec:conclusion}

This paper presents direct constraints on intrinsic alignments from the Dark Energy Survey Year 3 shape catalogues. The Y3 \metacal~catalogue is used to provide shape estimates for 2.4M \redmagic~galaxies from across the DES footprint, as well as $\sim$50,000 CMASS galaxies, $\sim$22,000 eBOSS LRGs and $\sim$100,000 ELGs. We make a high significance detection of intrinsic alignments in all of these samples, with the exception of eBOSS ELGs, where we place upper bounds on the magnitude of the possible alignment amplitude. 

\noindent
The key conclusions of this paper are:
\begin{itemize}
    \item Fitting for $A_1$ in red galaxies, our data support the qualitative  picture of a broken power law in $r$-band luminosity of the form $A_1 \propto L^{\beta}$, with $\beta$ differing between high and low $L$. Our \redmagic~low$-z$ sample provides a significant improvement constraints at the faint end of the $A_1-L$ relation, where the slope is shallower than at the bright end (by a factor of several; see Figure \ref{fig:results:luminosity_dependence}).
    \item Amongst red galaxy samples, however, we find noticeable colour dependence in the IA-luminosity trend. This is most obvious in CMASS and \redmagic~high-$z$, which both lying below the bulk of previous measurements at a similar $L$. These differences can be qualitatively explained by differences in the colour space distributions. This raises potential questions about the sufficiency of a simple red/blue binary split for modelling IAs, and whether joint modelling of luminosity and colour dependence may be needed.
    \item We find no statistically significant signal in our ELG sample on any scale. Using the combination of $w_{g+}$ and $w_{++}$, we impose an upper limit on the large scale NLA amplitude in ELGs at $|A_1|<0.3$ ($68\%$ CL). This is an improvement on the null constraint from WiggleZ at $|A_1|<1.03$ \citep{mandelbaum10}.
    \item The one-parameter NLA model is seen to fit all of our red galaxy samples reasonably on scales $\rp>6\mpc$. In our \redmagic~samples, which give the highest signal-to-noise measurements, we do see deviations from the NLA prediction in the range $2-6\mpc$. These deviations are more pronounced in the higher \redmagic~redshift sample. 
    \item Allowing additional flexibility via the TATT model, we can obtain a good fit to both \redmagic~samples on intermediate scales, $\rp =2-6\mpc$. We thus place constraints on the additional parameters (see Figure \ref{fig:results:tatt_contours}).
    \item We show that lensing and magnification can have a potentially significant impact on direct IA measurements. The extra terms are dominated by a galaxy-galaxy lensing like contribution $\delta_g\delta$, and the magnification-lensing cross correlation, $\delta_\mu \delta$. Together they make up $\sim2-20\%$ of the total signal, depending on the sample. This is relevant for our higher S/N samples, and will certainly be significant for future measurements of a similar kind, even those relying only on spectroscopic samples.
\end{itemize}

A weak lensing cosmology analysis is underway using the Dark Energy Survey Year 6 data, and similar efforts are ongoing on the KiDS legacy and HSC Y3 results. Understanding astrophysical systematics such as intrinsic alignments, on both large and small scales, will clearly be important for the success of these ongoing cosmology projects, as well as future surveys such as the Vera C.\ Rubin Observatory Legacy Survey of Space and Time (LSST), Euclid and the Nancy Grace Roman Space Telescope. Our ability to accurately model IAs and mitigate their impact is, however, still somewhat limited; even given detailed information about the redshift and rest frame colour of a sample (which typically is not available in a photometric survey), we do not have sufficient a priori understanding of the physical processes to predict the IA signal for an arbitrary selection of galaxies. We can, however, make measurements of IAs in a range of samples, and map out the dependence on galaxy properties. In this way, we can start to build up a phenomenological understanding of intrinsic alignments, which will feed into the next generation of analyses. The longer term goal is to develop more accurate models of intrinsic alignments on all scales but also, ideally, to derive informative priors on their parameters. This paper aims to contribute to this task using some of the most constraining current data sets.

Our results provide a small step towards a more complete understanding of intrinsic alignments in lensing surveys. In particular, we present results from new data sets that allow a substantially improved constraint on the faint end of the $L-A_1$ relation, and at intermediate redshifts. This is important, as the extrapolation into this regime is still a significant uncertainty in both model building and model sufficiency testing for future surveys. There are also a number of natural extensions to the work presented here. One obvious example is the development of a simple model that can account for both colour and redshift dependence in our red samples simultaneously. This is the focus of ongoing work.

\section*{Data Availability Statement}

All processed data products created for this work (two point functions, covariance matrices and matched shape catalogues) are available on request. The DES Y3 redMaGiC, \blockfont{metacal} and \blockfont{gold} catalogues have been public since February 2022, and can be found at \url{https://des.ncsa.illinois.edu/releases/y3a2}. Catalogues for the eBOSS LRG and ELG samples (including randoms, but without shapes) are available from \url{https://data.sdss.org/sas/dr16/eboss/lss/catalogs/DR16/}. For the equivalent CMASS catalogues see: \url{https://data.sdss.org/sas/dr12/boss/lss/}.

\section*{Acknowledgements}

The authors would like to thank Anand Raichoor and Rupert Croft for their help navigating SDSS.

SS and JB are partially supported by NSF grant AST-2206563. RM is supported in part by the Department of Energy grant DE-SC0010118 and in part by a grant from the Simons Foundation (Simons Investigator in Astrophysics, Award ID 620789). GR acknowledges support from the National Research Foundation of Korea (NRF) through grants No. 2017R1E1A1A01077508 and No. 2020R1A2C1005655 funded by the Korean Ministry of Education, Science and Technology (MoEST). 

Funding for the DES Projects has been provided by the U.S. Department of Energy, the U.S. National Science Foundation, the Ministry of Science and Education of Spain, 
the Science and Technology Facilities Council of the United Kingdom, the Higher Education Funding Council for England, the National Center for Supercomputing 
Applications at the University of Illinois at Urbana-Champaign, the Kavli Institute of Cosmological Physics at the University of Chicago, 
the Center for Cosmology and Astro-Particle Physics at the Ohio State University,
the Mitchell Institute for Fundamental Physics and Astronomy at Texas A\&M University, Financiadora de Estudos e Projetos, 
Funda{\c c}{\~a}o Carlos Chagas Filho de Amparo {\`a} Pesquisa do Estado do Rio de Janeiro, Conselho Nacional de Desenvolvimento Cient{\'i}fico e Tecnol{\'o}gico and 
the Minist{\'e}rio da Ci{\^e}ncia, Tecnologia e Inova{\c c}{\~a}o, the Deutsche Forschungsgemeinschaft and the Collaborating Institutions in the Dark Energy Survey. 

The Collaborating Institutions are Argonne National Laboratory, the University of California at Santa Cruz, the University of Cambridge, Centro de Investigaciones Energ{\'e}ticas, 
Medioambientales y Tecnol{\'o}gicas-Madrid, the University of Chicago, University College London, the DES-Brazil Consortium, the University of Edinburgh, 
the Eidgen{\"o}ssische Technische Hochschule (ETH) Z{\"u}rich, 
Fermi National Accelerator Laboratory, the University of Illinois at Urbana-Champaign, the Institut de Ci{\`e}ncies de l'Espai (IEEC/CSIC), 
the Institut de F{\'i}sica d'Altes Energies, Lawrence Berkeley National Laboratory, the Ludwig-Maximilians Universit{\"a}t M{\"u}nchen and the associated Excellence Cluster Universe, 
the University of Michigan, NSF's NOIRLab, the University of Nottingham, The Ohio State University, the University of Pennsylvania, the University of Portsmouth, 
SLAC National Accelerator Laboratory, Stanford University, the University of Sussex, Texas A\&M University, and the OzDES Membership Consortium.

Based in part on observations at Cerro Tololo Inter-American Observatory at NSF's NOIRLab (NOIRLab Prop. ID 2012B-0001; PI: J. Frieman), which is managed by the Association of Universities for Research in Astronomy (AURA) under a cooperative agreement with the National Science Foundation.

The DES data management system is supported by the National Science Foundation under Grant Numbers AST-1138766 and AST-1536171.
The DES participants from Spanish institutions are partially supported by MICINN under grants ESP2017-89838, PGC2018-094773, PGC2018-102021, SEV-2016-0588, SEV-2016-0597, and MDM-2015-0509, some of which include ERDF funds from the European Union. IFAE is partially funded by the CERCA program of the Generalitat de Catalunya.
Research leading to these results has received funding from the European Research
Council under the European Union's Seventh Framework Program (FP7/2007-2013) including ERC grant agreements 240672, 291329, and 306478.
We  acknowledge support from the Brazilian Instituto Nacional de Ci\^encia
e Tecnologia (INCT) do e-Universo (CNPq grant 465376/2014-2).

This manuscript has been authored by Fermi Research Alliance, LLC under Contract No. DE-AC02-07CH11359 with the U.S. Department of Energy, Office of Science, Office of High Energy Physics.

% SDSS-IV acknowledgements https://www.sdss.org/collaboration/citing-sdss/
Funding for the Sloan Digital Sky Survey IV has been provided by the Alfred P. Sloan Foundation, the U.S. Department of Energy Office of Science, and the Participating Institutions. SDSS-IV acknowledges support and resources from the Center for High Performance Computing  at the University of Utah. The SDSS website is \url{www.sdss.org}.
SDSS-IV is managed by the Astrophysical Research Consortium for the Participating Institutions of the SDSS Collaboration including the Brazilian Participation Group, the Carnegie Institution for Science, Carnegie Mellon University, Center for Astrophysics | Harvard \& Smithsonian, the Chilean Participation Group, the French Participation Group, Instituto de Astrof\'isica de Canarias, The Johns Hopkins University, Kavli Institute for the Physics and Mathematics of the Universe (IPMU) / University of Tokyo, the Korean Participation Group, Lawrence Berkeley National Laboratory, Leibniz Institut f\"ur Astrophysik Potsdam (AIP),  Max-Planck-Institut f\"ur Astronomie (MPIA Heidelberg), Max-Planck-Institut f\"ur Astrophysik (MPA Garching), Max-Planck-Institut f\"ur Extraterrestrische Physik (MPE), National Astronomical Observatories of China, New Mexico State University, New York University, University of Notre Dame, Observat\'ario Nacional / MCTI, The Ohio State University, Pennsylvania State University, Shanghai Astronomical Observatory, United Kingdom Participation Group, Universidad Nacional Aut\'onoma de M\'exico, University of Arizona, University of Colorado Boulder, University of Oxford, University of Portsmouth, University of Utah, University of Virginia, University of Washington, University of Wisconsin, Vanderbilt University, and Yale University.

Figure \ref{fig:data:footprint} was created using \blockfont{skymapper}, which is available at \url{https://github.com/pmelchior/skymapper}. The contour plots were made using \blockfont{GetDist} (\citealt{lewis19}).

\bibliographystyle{mnras}
\bibliography{refs}

\begin{thebibliography}{}
\makeatletter
\relax
\def\mn@urlcharsother{\let\do\@makeother \do\$\do\&\do\#\do\^\do\_\do\%\do\~}
\def\mn@doi{\begingroup\mn@urlcharsother \@ifnextchar [ {\mn@doi@}
  {\mn@doi@[]}}
\def\mn@doi@[#1]#2{\def\@tempa{#1}\ifx\@tempa\@empty \href
  {http://dx.doi.org/#2} {doi:#2}\else \href {http://dx.doi.org/#2} {#1}\fi
  \endgroup}
\def\mn@eprint#1#2{\mn@eprint@#1:#2::\@nil}
\def\mn@eprint@arXiv#1{\href {http://arxiv.org/abs/#1} {{\tt arXiv:#1}}}
\def\mn@eprint@dblp#1{\href {http://dblp.uni-trier.de/rec/bibtex/#1.xml}
  {dblp:#1}}
\def\mn@eprint@#1:#2:#3:#4\@nil{\def\@tempa {#1}\def\@tempb {#2}\def\@tempc
  {#3}\ifx \@tempc \@empty \let \@tempc \@tempb \let \@tempb \@tempa \fi \ifx
  \@tempb \@empty \def\@tempb {arXiv}\fi \@ifundefined
  {mn@eprint@\@tempb}{\@tempb:\@tempc}{\expandafter \expandafter \csname
  mn@eprint@\@tempb\endcsname \expandafter{\@tempc}}}

\bibitem[\protect\citeauthoryear{{Amon} et~al.,}{{Amon}
  et~al.}{2022}]{y3-cosmicshear1}
{Amon} A.,  et~al., 2022, \mn@doi [\prd] {10.1103/PhysRevD.105.023514}, \href
  {https://ui.adsabs.harvard.edu/abs/2022PhRvD.105b3514A} {105, 023514}

\bibitem[\protect\citeauthoryear{{Asgari} et~al.,}{{Asgari}
  et~al.}{2021}]{asgari20}
{Asgari} M.,  et~al., 2021, \mn@doi [\aap] {10.1051/0004-6361/202039070}, \href
  {https://ui.adsabs.harvard.edu/abs/2021A&A...645A.104A} {645, A104}

\bibitem[\protect\citeauthoryear{{Baldauf}, {Smith}, {Seljak}  \&
  {Mandelbaum}}{{Baldauf} et~al.}{2010}]{baldauf10}
{Baldauf} T.,  {Smith} R.~E.,  {Seljak} U.,   {Mandelbaum} R.,  2010, \mn@doi
  [\prd] {10.1103/PhysRevD.81.063531}, \href
  {https://ui.adsabs.harvard.edu/abs/2010PhRvD..81f3531B} {81, 063531}

\bibitem[\protect\citeauthoryear{{Bautista} et~al.,}{{Bautista}
  et~al.}{2021}]{bautista21}
{Bautista} J.~E.,  et~al., 2021, \mn@doi [\mnras] {10.1093/mnras/staa2800},
  \href {https://ui.adsabs.harvard.edu/abs/2021MNRAS.500..736B} {500, 736}

\bibitem[\protect\citeauthoryear{{Becker} et~al.,}{{Becker}
  et~al.}{2016}]{becker16}
{Becker} M.~R.,  et~al., 2016, \mn@doi [\prd] {10.1103/PhysRevD.94.022002},
  \href {https://ui.adsabs.harvard.edu/abs/2016PhRvD..94b2002B} {94, 022002}

\bibitem[\protect\citeauthoryear{{Blanton} et~al.,}{{Blanton}
  et~al.}{2017}]{blanton17}
{Blanton} M.~R.,  et~al., 2017, \mn@doi [\aj] {10.3847/1538-3881/aa7567}, \href
  {https://ui.adsabs.harvard.edu/abs/2017AJ....154...28B} {154, 28}

\bibitem[\protect\citeauthoryear{{Blazek}, {McQuinn}  \& {Seljak}}{{Blazek}
  et~al.}{2011}]{blazek11}
{Blazek} J.,  {McQuinn} M.,   {Seljak} U.,  2011, \mn@doi [Journal of Cosmology
  and Astro-Particle Physics] {10.1088/1475-7516/2011/05/010}, \href
  {https://ui.adsabs.harvard.edu/abs/2011JCAP...05..010B} {2011, 010}

\bibitem[\protect\citeauthoryear{{Blazek}, {Mandelbaum}, {Seljak}  \&
  {Nakajima}}{{Blazek} et~al.}{2012}]{blazek12}
{Blazek} J.,  {Mandelbaum} R.,  {Seljak} U.,   {Nakajima} R.,  2012, \mn@doi
  [\jcap] {10.1088/1475-7516/2012/05/041}, \href
  {https://ui.adsabs.harvard.edu/abs/2012JCAP...05..041B} {2012, 041}

\bibitem[\protect\citeauthoryear{{Blazek}, {Vlah}  \& {Seljak}}{{Blazek}
  et~al.}{2015}]{blazek15}
{Blazek} J.,  {Vlah} Z.,   {Seljak} U.,  2015, \mn@doi [\jcap]
  {10.1088/1475-7516/2015/08/015}, \href
  {http://adsabs.harvard.edu/abs/2015JCAP...08..015B} {8, 015}

\bibitem[\protect\citeauthoryear{{Blazek}, {MacCrann}, {Troxel}  \&
  {Fang}}{{Blazek} et~al.}{2019}]{blazek17}
{Blazek} J.~A.,  {MacCrann} N.,  {Troxel} M.~A.,   {Fang} X.,  2019, \mn@doi
  [\prd] {10.1103/PhysRevD.100.103506}, \href
  {https://ui.adsabs.harvard.edu/abs/2019PhRvD.100j3506B} {100, 103506}

\bibitem[\protect\citeauthoryear{{Bolton} et~al.,}{{Bolton}
  et~al.}{2012}]{bolton12}
{Bolton} A.~S.,  et~al., 2012, \mn@doi [\aj] {10.1088/0004-6256/144/5/144},
  \href {https://ui.adsabs.harvard.edu/\#abs/2012AJ....144..144B} {144, 144}

\bibitem[\protect\citeauthoryear{{Bridle} \& {King}}{{Bridle} \&
  {King}}{2007}]{bridle07}
{Bridle} S.,  {King} L.,  2007, \mn@doi [New Journal of Physics]
  {10.1088/1367-2630/9/12/444}, \href
  {http://adsabs.harvard.edu/abs/2007NJPh....9..444B} {9, 444}

\bibitem[\protect\citeauthoryear{{Bruzual} \& {Charlot}}{{Bruzual} \&
  {Charlot}}{2003}]{bruzual03}
{Bruzual} G.,  {Charlot} S.,  2003, \mn@doi [\mnras]
  {10.1046/j.1365-8711.2003.06897.x}, \href
  {https://ui.adsabs.harvard.edu/abs/2003MNRAS.344.1000B} {344, 1000}

\bibitem[\protect\citeauthoryear{{Catelan}, {Kamionkowski}  \&
  {Blandford}}{{Catelan} et~al.}{2001}]{catelan01}
{Catelan} P.,  {Kamionkowski} M.,   {Blandford} R.~D.,  2001, \mn@doi [\mnras]
  {10.1046/j.1365-8711.2001.04105.x}, \href
  {http://adsabs.harvard.edu/abs/2001MNRAS.320L...7C} {320, L7}

\bibitem[\protect\citeauthoryear{{Cawthon} et~al.,}{{Cawthon}
  et~al.}{2022}]{y3-wpzlens}
{Cawthon} R.,  et~al., 2022, \mn@doi [\mnras] {10.1093/mnras/stac1160}, \href
  {https://ui.adsabs.harvard.edu/abs/2022MNRAS.513.5517C} {513, 5517}

\bibitem[\protect\citeauthoryear{{Codis} et~al.,}{{Codis}
  et~al.}{2015}]{codis15a}
{Codis} S.,  et~al., 2015, \mn@doi [\mnras] {10.1093/mnras/stv231}, \href
  {http://adsabs.harvard.edu/abs/2015MNRAS.448.3391C} {448, 3391}

\bibitem[\protect\citeauthoryear{{Comparat} et~al.,}{{Comparat}
  et~al.}{2016}]{comparat16}
{Comparat} J.,  et~al., 2016, \mn@doi [\aap] {10.1051/0004-6361/201527377},
  \href {https://ui.adsabs.harvard.edu/\#abs/2016A&A...592A.121C} {592, A121}

\bibitem[\protect\citeauthoryear{{\PRF{Daalen}{van}{van} Daalen}, {Schaye},
  {Booth}  \& {Dalla Vecchia}}{{\PRF{Daalen}{van}{van} Daalen}
  et~al.}{2011}]{vanDaalen11}
{\PRF{Daalen}{van}{van} Daalen} M.~P.,  {Schaye} J.,  {Booth} C.~M.,   {Dalla
  Vecchia} C.,  2011, \mn@doi [\mnras] {10.1111/j.1365-2966.2011.18981.x},
  \href {http://adsabs.harvard.edu/abs/2011MNRAS.415.3649V} {415, 3649}

\bibitem[\protect\citeauthoryear{{Dark Energy Survey Collaboration}}{{Dark
  Energy Survey Collaboration}}{2016}]{svcosmology}
{Dark Energy Survey Collaboration} 2016, \mn@doi [\prd]
  {10.1103/PhysRevD.94.022001}, \href
  {http://adsabs.harvard.edu/abs/2016PhRvD..94b2001A} {94, 022001}

\bibitem[\protect\citeauthoryear{{Dark Energy Survey Collaboration}}{{Dark
  Energy Survey Collaboration}}{2022}]{y3-3x2ptkp}
{Dark Energy Survey Collaboration} 2022, \mn@doi [\prd]
  {10.1103/PhysRevD.105.023520}, \href
  {https://ui.adsabs.harvard.edu/abs/2022PhRvD.105b3520A} {105, 023520}

\bibitem[\protect\citeauthoryear{{Dawson} et~al.,}{{Dawson}
  et~al.}{2013}]{dawson13}
{Dawson} K.~S.,  et~al., 2013, \mn@doi [\aj] {10.1088/0004-6256/145/1/10},
  \href {https://ui.adsabs.harvard.edu/abs/2013AJ....145...10D} {145, 10}

\bibitem[\protect\citeauthoryear{{Dawson} et~al.,}{{Dawson}
  et~al.}{2016}]{dawson16}
{Dawson} K.~S.,  et~al., 2016, \mn@doi [\aj] {10.3847/0004-6256/151/2/44},
  \href {https://ui.adsabs.harvard.edu/abs/2016AJ....151...44D} {151, 44}

\bibitem[\protect\citeauthoryear{{Delubac} et~al.,}{{Delubac}
  et~al.}{2017}]{delubac17}
{Delubac} T.,  et~al., 2017, \mn@doi [\mnras] {10.1093/mnras/stw2741}, \href
  {https://ui.adsabs.harvard.edu/abs/2017MNRAS.465.1831D} {465, 1831}

\bibitem[\protect\citeauthoryear{{Drlica-Wagner} et~al.,}{{Drlica-Wagner}
  et~al.}{2018}]{y1gold}
{Drlica-Wagner} A.,  et~al., 2018, \mn@doi [\apjs] {10.3847/1538-4365/aab4f5},
  \href {http://adsabs.harvard.edu/abs/2018ApJS..235...33D} {235, 33}

\bibitem[\protect\citeauthoryear{{Eisenstein} et~al.,}{{Eisenstein}
  et~al.}{2011}]{eisenstein11}
{Eisenstein} D.~J.,  et~al., 2011, \mn@doi [\aj] {10.1088/0004-6256/142/3/72},
  \href {https://ui.adsabs.harvard.edu/abs/2011AJ....142...72E} {142, 72}

\bibitem[\protect\citeauthoryear{{Elvin-Poole}, {MacCrann}
  et~al.}{{Elvin-Poole} et~al.}{2022}]{y3-2x2ptmagnification}
{Elvin-Poole} J.,  {MacCrann} N.,   et~al., 2022, arXiv e-prints, \href
  {https://ui.adsabs.harvard.edu/abs/2022arXiv220909782E} {p. arXiv:2209.09782}

\bibitem[\protect\citeauthoryear{{Euclid Collaboration}}{{Euclid
  Collaboration}}{2021}]{euclid21}
{Euclid Collaboration} 2021, \mn@doi [\aap] {10.1051/0004-6361/202141061},
  \href {https://ui.adsabs.harvard.edu/abs/2021A&A...655A..44E} {655, A44}

\bibitem[\protect\citeauthoryear{{Everett} et~al.,}{{Everett}
  et~al.}{2022}]{y3-balrog}
{Everett} S.,  et~al., 2022, \mn@doi [\apjs] {10.3847/1538-4365/ac26c1}, \href
  {https://ui.adsabs.harvard.edu/abs/2022ApJS..258...15E} {258, 15}

\bibitem[\protect\citeauthoryear{{Feroz}, {Hobson}, {Cameron}  \&
  {Pettitt}}{{Feroz} et~al.}{2013}]{feroz13}
{Feroz} F.,  {Hobson} M.~P.,  {Cameron} E.,   {Pettitt} A.~N.,  2013, arXiv
  e-prints, \href {https://ui.adsabs.harvard.edu/abs/2013arXiv1306.2144F} {p.
  arXiv:1306.2144}

\bibitem[\protect\citeauthoryear{{Fortuna}, {Hoekstra}, {Joachimi}, {Johnston},
  {Chisari}, {Georgiou}  \& {Mahony}}{{Fortuna} et~al.}{2021a}]{fortuna20}
{Fortuna} M.~C.,  {Hoekstra} H.,  {Joachimi} B.,  {Johnston} H.,  {Chisari}
  N.~E.,  {Georgiou} C.,   {Mahony} C.,  2021a, \mn@doi [\mnras]
  {10.1093/mnras/staa3802}, \href
  {https://ui.adsabs.harvard.edu/abs/2021MNRAS.501.2983F} {501, 2983}

\bibitem[\protect\citeauthoryear{{Fortuna} et~al.,}{{Fortuna}
  et~al.}{2021b}]{fortuna21}
{Fortuna} M.~C.,  et~al., 2021b, \mn@doi [\aap] {10.1051/0004-6361/202140706},
  \href {https://ui.adsabs.harvard.edu/abs/2021A&A...654A..76F} {654, A76}

\bibitem[\protect\citeauthoryear{{Gatti}, {Sheldon}  et~al.}{{Gatti}
  et~al.}{2021}]{y3-shapecatalog}
{Gatti} M.,  {Sheldon} E.,   et~al., 2021, \mn@doi [\mnras]
  {10.1093/mnras/stab918}, \href
  {https://ui.adsabs.harvard.edu/abs/2021MNRAS.504.4312G} {504, 4312}

\bibitem[\protect\citeauthoryear{{Gunn} et~al.,}{{Gunn} et~al.}{2006}]{gunn06}
{Gunn} J.~E.,  et~al., 2006, \mn@doi [\aj] {10.1086/500975}, \href
  {https://ui.adsabs.harvard.edu/\#abs/2006AJ....131.2332G} {131, 2332}

\bibitem[\protect\citeauthoryear{{Guo} et~al.,}{{Guo} et~al.}{2019}]{guo19}
{Guo} H.,  et~al., 2019, \mn@doi [\apj] {10.3847/1538-4357/aaf9ad}, \href
  {https://ui.adsabs.harvard.edu/abs/2019ApJ...871..147G} {871, 147}

\bibitem[\protect\citeauthoryear{{Hamana} et~al.,}{{Hamana}
  et~al.}{2020}]{hamana20}
{Hamana} T.,  et~al., 2020, \mn@doi [\pasj] {10.1093/pasj/psz138}, \href
  {https://ui.adsabs.harvard.edu/abs/2020PASJ...72...16H} {72, 16}

\bibitem[\protect\citeauthoryear{{Heymans} et~al.,}{{Heymans}
  et~al.}{2012}]{heymans12}
{Heymans} C.,  et~al., 2012, \mn@doi [\mnras]
  {10.1111/j.1365-2966.2012.21952.x}, \href
  {https://ui.adsabs.harvard.edu/abs/2012MNRAS.427..146H} {427, 146}

\bibitem[\protect\citeauthoryear{{Heymans} et~al.,}{{Heymans}
  et~al.}{2013}]{heymans13}
{Heymans} C.,  et~al., 2013, \mn@doi [\mnras] {10.1093/mnras/stt601}, \href
  {https://ui.adsabs.harvard.edu/abs/2013MNRAS.432.2433H} {432, 2433}

\bibitem[\protect\citeauthoryear{{Hikage} et~al.,}{{Hikage}
  et~al.}{2019}]{hikage19}
{Hikage} C.,  et~al., 2019, \mn@doi [\pasj] {10.1093/pasj/psz010}, \href
  {https://ui.adsabs.harvard.edu/abs/2019PASJ...71...43H} {71, 43}

\bibitem[\protect\citeauthoryear{{Hilbert}, {Xu}, {Schneider}, {Springel},
  {Vogelsberger}  \& {Hernquist}}{{Hilbert} et~al.}{2017}]{hilbert16}
{Hilbert} S.,  {Xu} D.,  {Schneider} P.,  {Springel} V.,  {Vogelsberger} M.,
  {Hernquist} L.,  2017, \mn@doi [\mnras] {10.1093/mnras/stx482}, \href
  {https://ui.adsabs.harvard.edu/abs/2017MNRAS.468..790H} {468, 790}

\bibitem[\protect\citeauthoryear{{Hildebrandt} et~al.,}{{Hildebrandt}
  et~al.}{2017}]{hildebrandt17}
{Hildebrandt} H.,  et~al., 2017, \mn@doi [\mnras] {10.1093/mnras/stw2805},
  \href {https://ui.adsabs.harvard.edu/abs/2017MNRAS.465.1454H} {465, 1454}

\bibitem[\protect\citeauthoryear{{Hildebrandt} et~al.,}{{Hildebrandt}
  et~al.}{2020}]{hildebrandt20}
{Hildebrandt} H.,  et~al., 2020, \mn@doi [\aap] {10.1051/0004-6361/201834878},
  \href {https://ui.adsabs.harvard.edu/abs/2020A&A...633A..69H} {633, A69}

\bibitem[\protect\citeauthoryear{{Hinshaw} et~al.,}{{Hinshaw}
  et~al.}{2013}]{hinshaw13}
{Hinshaw} G.,  et~al., 2013, \mn@doi [\apjs] {10.1088/0067-0049/208/2/19},
  \href {https://ui.adsabs.harvard.edu/abs/2013ApJS..208...19H} {208, 19}

\bibitem[\protect\citeauthoryear{{Hirata} \& {Seljak}}{{Hirata} \&
  {Seljak}}{2004}]{hirata04}
{Hirata} C.~M.,  {Seljak} U.,  2004, \mn@doi [\prd]
  {10.1103/PhysRevD.70.063526}, \href
  {http://adsabs.harvard.edu/abs/2004PhRvD..70f3526H} {70, 063526}

\bibitem[\protect\citeauthoryear{{Hirata}, {Mandelbaum}, {Ishak}, {Seljak},
  {Nichol}, {Pimbblet}, {Ross}  \& {Wake}}{{Hirata} et~al.}{2007}]{hirata07}
{Hirata} C.~M.,  {Mandelbaum} R.,  {Ishak} M.,  {Seljak} U.,  {Nichol} R.,
  {Pimbblet} K.~A.,  {Ross} N.~P.,   {Wake} D.,  2007, \mn@doi [\mnras]
  {10.1111/j.1365-2966.2007.12312.x}, \href
  {https://ui.adsabs.harvard.edu/abs/2007MNRAS.381.1197H} {381, 1197}

\bibitem[\protect\citeauthoryear{{Hu}}{{Hu}}{1999}]{hu99}
{Hu} W.,  1999, \mn@doi [\apjl] {10.1086/312210}, \href
  {https://ui.adsabs.harvard.edu/abs/1999ApJ...522L..21H} {522, L21}

\bibitem[\protect\citeauthoryear{{Hu} \& {Jain}}{{Hu} \& {Jain}}{2004}]{hu04}
{Hu} W.,  {Jain} B.,  2004, \mn@doi [\prd] {10.1103/PhysRevD.70.043009}, \href
  {https://ui.adsabs.harvard.edu/abs/2004PhRvD..70d3009H} {70, 043009}

\bibitem[\protect\citeauthoryear{{Huff} \& {Mandelbaum}}{{Huff} \&
  {Mandelbaum}}{2017}]{huff17}
{Huff} E.,  {Mandelbaum} R.,  2017, preprint, \href
  {http://adsabs.harvard.edu/abs/2017arXiv170202600H} {} (\mn@eprint {arXiv}
  {1702.02600})

\bibitem[\protect\citeauthoryear{{Hutchinson} et~al.,}{{Hutchinson}
  et~al.}{2016}]{hutchinson16}
{Hutchinson} T.~A.,  et~al., 2016, \mn@doi [\aj] {10.3847/0004-6256/152/6/205},
  \href {https://ui.adsabs.harvard.edu/abs/2016AJ....152..205H} {152, 205}

\bibitem[\protect\citeauthoryear{{Jain} \& {Seljak}}{{Jain} \&
  {Seljak}}{1997}]{jain97}
{Jain} B.,  {Seljak} U.,  1997, \mn@doi [\apj] {10.1086/304372}, \href
  {https://ui.adsabs.harvard.edu/abs/1997ApJ...484..560J} {484, 560}

\bibitem[\protect\citeauthoryear{{Jarvis}, {Bernstein}  \& {Jain}}{{Jarvis}
  et~al.}{2004}]{jarvis04}
{Jarvis} M.,  {Bernstein} G.,   {Jain} B.,  2004, \mn@doi [\mnras]
  {10.1111/j.1365-2966.2004.07926.x}, \href
  {http://adsabs.harvard.edu/abs/2004MNRAS.352..338J} {352, 338}

\bibitem[\protect\citeauthoryear{{Jarvis} et~al.,}{{Jarvis}
  et~al.}{2021}]{jarvis21}
{Jarvis} M.,  et~al., 2021, \mn@doi [\mnras] {10.1093/mnras/staa3679}, \href
  {https://ui.adsabs.harvard.edu/abs/2021MNRAS.501.1282J} {501, 1282}

\bibitem[\protect\citeauthoryear{{Jee}, {Tyson}, {Hilbert}, {Schneider},
  {Schmidt}  \& {Wittman}}{{Jee} et~al.}{2016}]{jee16}
{Jee} M.~J.,  {Tyson} J.~A.,  {Hilbert} S.,  {Schneider} M.~D.,  {Schmidt} S.,
   {Wittman} D.,  2016, \mn@doi [\apj] {10.3847/0004-637X/824/2/77}, \href
  {https://ui.adsabs.harvard.edu/abs/2016ApJ...824...77J} {824, 77}

\bibitem[\protect\citeauthoryear{{Joachimi} \& {Bridle}}{{Joachimi} \&
  {Bridle}}{2010}]{joachimi10}
{Joachimi} B.,  {Bridle} S.~L.,  2010, \mn@doi [\aap]
  {10.1051/0004-6361/200913657}, \href
  {https://ui.adsabs.harvard.edu/abs/2010A&A...523A...1J} {523, A1}

\bibitem[\protect\citeauthoryear{{Joachimi}, {Mandelbaum}, {*******}  \&
  {Bridle}}{{Joachimi} et~al.}{2011}]{joachimi11}
{Joachimi} B.,  {Mandelbaum} R.,  {*******} .,   {Bridle} S.~L.,  2011, \mn@doi
  [\aap] {10.1051/0004-6361/201015621}, \href
  {http://adsabs.harvard.edu/abs/2011A%26A...527A..26J} {527, A26}

\bibitem[\protect\citeauthoryear{{Joachimi} et~al.,}{{Joachimi}
  et~al.}{2015}]{joachimi15}
{Joachimi} B.,  et~al., 2015, \mn@doi [\ssr] {10.1007/s11214-015-0177-4}, \href
  {http://adsabs.harvard.edu/abs/2015SSRv..193....1J} {193, 1}

\bibitem[\protect\citeauthoryear{{Joachimi} et~al.,}{{Joachimi}
  et~al.}{2021}]{joachimi21}
{Joachimi} B.,  et~al., 2021, \mn@doi [\aap] {10.1051/0004-6361/202038831},
  \href {https://ui.adsabs.harvard.edu/abs/2021A&A...646A.129J} {646, A129}

\bibitem[\protect\citeauthoryear{{Johnston} et~al.,}{{Johnston}
  et~al.}{2019}]{johnston19}
{Johnston} H.,  et~al., 2019, \mn@doi [\aap] {10.1051/0004-6361/201834714},
  \href {https://ui.adsabs.harvard.edu/abs/2019A&A...624A..30J} {624, A30}

\bibitem[\protect\citeauthoryear{{Kaiser}}{{Kaiser}}{1987}]{kaiser87}
{Kaiser} N.,  1987, \mn@doi [\mnras] {10.1093/mnras/227.1.1}, \href
  {https://ui.adsabs.harvard.edu/abs/1987MNRAS.227....1K} {227, 1}

\bibitem[\protect\citeauthoryear{{Kiessling} et~al.,}{{Kiessling}
  et~al.}{2015}]{kiessling15}
{Kiessling} A.,  et~al., 2015, \mn@doi [\ssr] {10.1007/s11214-015-0203-6},
  \href {http://adsabs.harvard.edu/abs/2015SSRv..193...67K} {193, 67}

\bibitem[\protect\citeauthoryear{{Kilbinger} et~al.,}{{Kilbinger}
  et~al.}{2013}]{kilbinger13}
{Kilbinger} M.,  et~al., 2013, \mn@doi [\mnras] {10.1093/mnras/stt041}, \href
  {https://ui.adsabs.harvard.edu/abs/2013MNRAS.430.2200K} {430, 2200}

\bibitem[\protect\citeauthoryear{{Kirk} et~al.,}{{Kirk} et~al.}{2015}]{kirk15}
{Kirk} D.,  et~al., 2015, \mn@doi [\ssr] {10.1007/s11214-015-0213-4}, \href
  {http://adsabs.harvard.edu/abs/2015SSRv..193..139K} {193, 139}

\bibitem[\protect\citeauthoryear{{Krause}, {Eifler}  \& {Blazek}}{{Krause}
  et~al.}{2016}]{krause15}
{Krause} E.,  {Eifler} T.,   {Blazek} J.,  2016, \mn@doi [\mnras]
  {10.1093/mnras/stv2615}, \href
  {https://ui.adsabs.harvard.edu/abs/2016MNRAS.456..207K} {456, 207}

\bibitem[\protect\citeauthoryear{{Krause} et~al.,}{{Krause}
  et~al.}{2021}]{y3-generalmethods}
{Krause} E.,  et~al., 2021, arXiv e-prints, \href
  {https://ui.adsabs.harvard.edu/abs/2021arXiv210513548K} {p. arXiv:2105.13548}

\bibitem[\protect\citeauthoryear{{Landy} \& {Szalay}}{{Landy} \&
  {Szalay}}{1993}]{landy93}
{Landy} S.~D.,  {Szalay} A.~S.,  1993, \mn@doi [\apj] {10.1086/172900}, \href
  {http://adsabs.harvard.edu/abs/1993ApJ...412...64L} {412, 64}

\bibitem[\protect\citeauthoryear{{Lemos} et~al.,}{{Lemos}
  et~al.}{2021}]{y3-tensions}
{Lemos} P.,  et~al., 2021, \mn@doi [\mnras] {10.1093/mnras/stab1670}, \href
  {https://ui.adsabs.harvard.edu/abs/2021MNRAS.505.6179L} {505, 6179}

\bibitem[\protect\citeauthoryear{{Lewis}}{{Lewis}}{2019}]{lewis19}
{Lewis} A.,  2019, arXiv e-prints, \href
  {https://ui.adsabs.harvard.edu/abs/2019arXiv191013970L} {p. arXiv:1910.13970}

\bibitem[\protect\citeauthoryear{Lewis, Challinor  \& Lasenby}{Lewis
  et~al.}{2000}]{lewis00}
Lewis A.,  Challinor A.,   Lasenby A.,  2000, \mn@doi [\apj] {10.1086/309179},
  538, 473

\bibitem[\protect\citeauthoryear{{MacCrann}, {Blazek}, {Jain}  \&
  {Krause}}{{MacCrann} et~al.}{2020}]{maccrann20}
{MacCrann} N.,  {Blazek} J.,  {Jain} B.,   {Krause} E.,  2020, \mn@doi [\mnras]
  {10.1093/mnras/stz2761}, \href
  {https://ui.adsabs.harvard.edu/abs/2020MNRAS.491.5498M} {491, 5498}

\bibitem[\protect\citeauthoryear{{Mackey}, {White}  \& {Kamionkowski}}{{Mackey}
  et~al.}{2002}]{mackey02}
{Mackey} J.,  {White} M.,   {Kamionkowski} M.,  2002, \mn@doi [\mnras]
  {10.1046/j.1365-8711.2002.05337.x}, \href
  {https://ui.adsabs.harvard.edu/abs/2002MNRAS.332..788M} {332, 788}

\bibitem[\protect\citeauthoryear{{Mandelbaum} et~al.,}{{Mandelbaum}
  et~al.}{2005}]{mandelbaum05}
{Mandelbaum} R.,  et~al., 2005, \mn@doi [\mnras]
  {10.1111/j.1365-2966.2005.09282.x}, \href
  {https://ui.adsabs.harvard.edu/abs/2005MNRAS.361.1287M} {361, 1287}

\bibitem[\protect\citeauthoryear{{Mandelbaum} et~al.,}{{Mandelbaum}
  et~al.}{2011}]{mandelbaum10}
{Mandelbaum} R.,  et~al., 2011, \mn@doi [\mnras]
  {10.1111/j.1365-2966.2010.17485.x}, \href
  {http://adsabs.harvard.edu/abs/2011MNRAS.410..844M} {410, 844}

\bibitem[\protect\citeauthoryear{{\PRF{Mattia}{de}{de} Mattia}
  et~al.,}{{\PRF{Mattia}{de}{de} Mattia} et~al.}{2021}]{deMattia21}
{\PRF{Mattia}{de}{de} Mattia} A.,  et~al., 2021, \mn@doi [\mnras]
  {10.1093/mnras/staa3891}, \href
  {https://ui.adsabs.harvard.edu/abs/2021MNRAS.501.5616D} {501, 5616}

\bibitem[\protect\citeauthoryear{{McDonald}}{{McDonald}}{2006}]{mcdonald06}
{McDonald} P.,  2006, \mn@doi [\prd] {10.1103/PhysRevD.74.103512}, \href
  {https://ui.adsabs.harvard.edu/abs/2006PhRvD..74j3512M} {74, 103512}

\bibitem[\protect\citeauthoryear{{McEwen}, {Fang}, {Hirata}  \&
  {Blazek}}{{McEwen} et~al.}{2016}]{mcewen16}
{McEwen} J.~E.,  {Fang} X.,  {Hirata} C.~M.,   {Blazek} J.~A.,  2016, \mn@doi
  [\jcap] {10.1088/1475-7516/2016/09/015}, \href
  {https://ui.adsabs.harvard.edu/abs/2016JCAP...09..015M} {2016, 015}

\bibitem[\protect\citeauthoryear{{Morganson} et~al.,}{{Morganson}
  et~al.}{2018}]{morganson18}
{Morganson} E.,  et~al., 2018, \mn@doi [\pasp] {10.1088/1538-3873/aab4ef},
  \href {https://ui.adsabs.harvard.edu/abs/2018PASP..130g4501M} {130, 074501}

\bibitem[\protect\citeauthoryear{{Pandey} et~al.,}{{Pandey}
  et~al.}{2020}]{pandey20}
{Pandey} S.,  et~al., 2020, \mn@doi [\prd] {10.1103/PhysRevD.102.123522}, \href
  {https://ui.adsabs.harvard.edu/abs/2020PhRvD.102l3522P} {102, 123522}

\bibitem[\protect\citeauthoryear{{Pandey} et~al.,}{{Pandey}
  et~al.}{2022}]{y3-2x2ptbiasmodelling}
{Pandey} S.,  et~al., 2022, \mn@doi [\prd] {10.1103/PhysRevD.106.043520}, \href
  {https://ui.adsabs.harvard.edu/abs/2022PhRvD.106d3520P} {106, 043520}

\bibitem[\protect\citeauthoryear{{Porredon} et~al.,}{{Porredon}
  et~al.}{2021}]{y3-2x2maglimforecast}
{Porredon} A.,  et~al., 2021, \mn@doi [\prd] {10.1103/PhysRevD.103.043503},
  \href {https://ui.adsabs.harvard.edu/abs/2021PhRvD.103d3503P} {103, 043503}

\bibitem[\protect\citeauthoryear{{Prakash} et~al.,}{{Prakash}
  et~al.}{2016}]{prakash16}
{Prakash} A.,  et~al., 2016, \mn@doi [\apjs] {10.3847/0067-0049/224/2/34},
  \href {https://ui.adsabs.harvard.edu/abs/2016ApJS..224...34P} {224, 34}

\bibitem[\protect\citeauthoryear{{Prat} et~al.,}{{Prat}
  et~al.}{2022}]{y3-gglensing}
{Prat} J.,  et~al., 2022, \mn@doi [\prd] {10.1103/PhysRevD.105.083528}, \href
  {https://ui.adsabs.harvard.edu/abs/2022PhRvD.105h3528P} {105, 083528}

\bibitem[\protect\citeauthoryear{{Raichoor} et~al.,}{{Raichoor}
  et~al.}{2017}]{raichoor17}
{Raichoor} A.,  et~al., 2017, \mn@doi [\mnras] {10.1093/mnras/stx1790}, \href
  {https://ui.adsabs.harvard.edu/\#abs/2017MNRAS.471.3955R} {471, 3955}

\bibitem[\protect\citeauthoryear{{Raichoor} et~al.,}{{Raichoor}
  et~al.}{2020}]{raichoor20}
{Raichoor} A.,  et~al., 2020, \mn@doi [Research Notes of the American
  Astronomical Society] {10.3847/2515-5172/abc078}, \href
  {https://ui.adsabs.harvard.edu/abs/2020RNAAS...4..180R} {4, 180}

\bibitem[\protect\citeauthoryear{{Raichoor} et~al.,}{{Raichoor}
  et~al.}{2021}]{raichoor21}
{Raichoor} A.,  et~al., 2021, \mn@doi [\mnras] {10.1093/mnras/staa3336}, \href
  {https://ui.adsabs.harvard.edu/abs/2021MNRAS.500.3254R} {500, 3254}

\bibitem[\protect\citeauthoryear{{Reid} et~al.,}{{Reid} et~al.}{2016}]{reid16}
{Reid} B.,  et~al., 2016, \mn@doi [\mnras] {10.1093/mnras/stv2382}, \href
  {https://ui.adsabs.harvard.edu/abs/2016MNRAS.455.1553R} {455, 1553}

\bibitem[\protect\citeauthoryear{{Reyes}, {Mandelbaum}, {Gunn}, {Nakajima},
  {Seljak}  \& {Hirata}}{{Reyes} et~al.}{2012}]{reyes12}
{Reyes} R.,  {Mandelbaum} R.,  {Gunn} J.~E.,  {Nakajima} R.,  {Seljak} U.,
  {Hirata} C.~M.,  2012, \mn@doi [\mnras] {10.1111/j.1365-2966.2012.21472.x},
  \href {https://ui.adsabs.harvard.edu/\#abs/2012MNRAS.425.2610R} {425, 2610}

\bibitem[\protect\citeauthoryear{{Rodr{\'\i}guez-Monroy}
  et~al.,}{{Rodr{\'\i}guez-Monroy} et~al.}{2022}]{y3-galaxyclustering}
{Rodr{\'\i}guez-Monroy} M.,  et~al., 2022, \mn@doi [\mnras]
  {10.1093/mnras/stac104}, \href
  {https://ui.adsabs.harvard.edu/abs/2022MNRAS.511.2665R} {511, 2665}

\bibitem[\protect\citeauthoryear{{Ross} et~al.,}{{Ross} et~al.}{2020}]{ross20}
{Ross} A.~J.,  et~al., 2020, \mn@doi [\mnras] {10.1093/mnras/staa2416}, \href
  {https://ui.adsabs.harvard.edu/abs/2020MNRAS.498.2354R} {498, 2354}

\bibitem[\protect\citeauthoryear{{Rossi} et~al.,}{{Rossi}
  et~al.}{2021}]{rossi21}
{Rossi} G.,  et~al., 2021, \mn@doi [\mnras] {10.1093/mnras/staa3955}, \href
  {https://ui.adsabs.harvard.edu/abs/2021MNRAS.505..377R} {505, 377}

\bibitem[\protect\citeauthoryear{{Rozo} et~al.,}{{Rozo} et~al.}{2016}]{rozo16}
{Rozo} E.,  et~al., 2016, \mn@doi [\mnras] {10.1093/mnras/stw1281}, \href
  {https://ui.adsabs.harvard.edu/abs/2016MNRAS.461.1431R} {461, 1431}

\bibitem[\protect\citeauthoryear{Saito, Baldauf, Vlah, Seljak, Okumura  \&
  McDonald}{Saito et~al.}{2014}]{saito14}
Saito S.,  Baldauf T.,  Vlah Z.,  Seljak U. c.~v.,  Okumura T.,   McDonald P.,
  2014, \mn@doi [Phys. Rev. D] {10.1103/PhysRevD.90.123522}, 90, 123522

\bibitem[\protect\citeauthoryear{{Samuroff} et~al.,}{{Samuroff}
  et~al.}{2019}]{y1coloursplit}
{Samuroff} S.,  et~al., 2019, \mn@doi [\mnras] {10.1093/mnras/stz2197}, \href
  {https://ui.adsabs.harvard.edu/abs/2019MNRAS.489.5453S} {489, 5453}

\bibitem[\protect\citeauthoryear{{Samuroff}, {Mandelbaum}  \&
  {Blazek}}{{Samuroff} et~al.}{2021}]{samuroff21}
{Samuroff} S.,  {Mandelbaum} R.,   {Blazek} J.,  2021, \mn@doi [\mnras]
  {10.1093/mnras/stab2520}, \href
  {https://ui.adsabs.harvard.edu/abs/2021MNRAS.508..637S} {508, 637}

\bibitem[\protect\citeauthoryear{{S{\'a}nchez}, {Prat}  et~al.}{{S{\'a}nchez}
  et~al.}{2022}]{y3-shearratio}
{S{\'a}nchez} C.,  {Prat} J.,   et~al., 2022, \mn@doi [\prd]
  {10.1103/PhysRevD.105.083529}, \href
  {https://ui.adsabs.harvard.edu/abs/2022PhRvD.105h3529S} {105, 083529}

\bibitem[\protect\citeauthoryear{{Schaye} et~al.,}{{Schaye}
  et~al.}{2010}]{schaye10}
{Schaye} J.,  et~al., 2010, \mn@doi [\mnras]
  {10.1111/j.1365-2966.2009.16029.x}, \href
  {https://ui.adsabs.harvard.edu/abs/2010MNRAS.402.1536S} {402, 1536}

\bibitem[\protect\citeauthoryear{{Schmidt}, {Rozo}, {Dodelson}, {Hui}  \&
  {Sheldon}}{{Schmidt} et~al.}{2009}]{schmidt09}
{Schmidt} F.,  {Rozo} E.,  {Dodelson} S.,  {Hui} L.,   {Sheldon} E.,  2009,
  \mn@doi [\apj] {10.1088/0004-637X/702/1/593}, \href
  {https://ui.adsabs.harvard.edu/abs/2009ApJ...702..593S} {702, 593}

\bibitem[\protect\citeauthoryear{{Schneider} \& {Bridle}}{{Schneider} \&
  {Bridle}}{2010}]{schneider10}
{Schneider} M.~D.,  {Bridle} S.,  2010, \mn@doi [\mnras]
  {10.1111/j.1365-2966.2009.15956.x}, \href
  {http://adsabs.harvard.edu/abs/2010MNRAS.402.2127S} {402, 2127}

\bibitem[\protect\citeauthoryear{{Schneider}, {van Waerbeke}, {Kilbinger}  \&
  {Mellier}}{{Schneider} et~al.}{2002}]{schneider02}
{Schneider} P.,  {van Waerbeke} L.,  {Kilbinger} M.,   {Mellier} Y.,  2002,
  \mn@doi [\aap] {10.1051/0004-6361:20021341}, \href
  {https://ui.adsabs.harvard.edu/abs/2002A&A...396....1S} {396, 1}

\bibitem[\protect\citeauthoryear{Schwarz}{Schwarz}{1978}]{schwarz78}
Schwarz G.,  1978, \mn@doi [The Annals of Statistics] {10.1214/aos/1176344136},
  6, 461

\bibitem[\protect\citeauthoryear{{Secco}, {Samuroff}  et~al.}{{Secco}
  et~al.}{2022}]{y3-cosmicshear2}
{Secco} L.~F.,  {Samuroff} S.,   et~al., 2022, \mn@doi [\prd]
  {10.1103/PhysRevD.105.023515}, \href
  {https://ui.adsabs.harvard.edu/abs/2022PhRvD.105b3515S} {105, 023515}

\bibitem[\protect\citeauthoryear{{Sevilla-Noarbe}, {Bechtol}
  et~al.}{{Sevilla-Noarbe} et~al.}{2021}]{y3-gold}
{Sevilla-Noarbe} I.,  {Bechtol} K.,   et~al., 2021, \mn@doi [\apjs]
  {10.3847/1538-4365/abeb66}, \href
  {https://ui.adsabs.harvard.edu/abs/2021ApJS..254...24S} {254, 24}

\bibitem[\protect\citeauthoryear{{Sheldon} \& {Huff}}{{Sheldon} \&
  {Huff}}{2017}]{sheldon17}
{Sheldon} E.~S.,  {Huff} E.~M.,  2017, \mn@doi [\apj]
  {10.3847/1538-4357/aa704b}, \href
  {http://adsabs.harvard.edu/abs/2017ApJ...841...24S} {841, 24}

\bibitem[\protect\citeauthoryear{{Singh} \& {Mandelbaum}}{{Singh} \&
  {Mandelbaum}}{2016}]{singh16}
{Singh} S.,  {Mandelbaum} R.,  2016, \mn@doi [\mnras] {10.1093/mnras/stw144},
  \href {https://ui.adsabs.harvard.edu/abs/2016MNRAS.457.2301S} {457, 2301}

\bibitem[\protect\citeauthoryear{{Singh}, {Mandelbaum}  \& {More}}{{Singh}
  et~al.}{2015}]{singh15}
{Singh} S.,  {Mandelbaum} R.,   {More} S.,  2015, \mn@doi [\mnras]
  {10.1093/mnras/stv778}, \href
  {http://adsabs.harvard.edu/abs/2015MNRAS.450.2195S} {450, 2195}

\bibitem[\protect\citeauthoryear{{Smee} et~al.,}{{Smee} et~al.}{2013}]{smee13}
{Smee} S.~A.,  et~al., 2013, \mn@doi [\aj] {10.1088/0004-6256/146/2/32}, \href
  {https://ui.adsabs.harvard.edu/abs/2013AJ....146...32S} {146, 32}

\bibitem[\protect\citeauthoryear{{Smith} et~al.,}{{Smith}
  et~al.}{2003}]{smith03}
{Smith} R.~E.,  et~al., 2003, \mn@doi [\mnras]
  {10.1046/j.1365-8711.2003.06503.x}, \href
  {https://ui.adsabs.harvard.edu/abs/2003MNRAS.341.1311S} {341, 1311}

\bibitem[\protect\citeauthoryear{{Suchyta} et~al.,}{{Suchyta}
  et~al.}{2016}]{suchyta16}
{Suchyta} E.,  et~al., 2016, \mn@doi [\mnras] {10.1093/mnras/stv2953}, \href
  {https://ui.adsabs.harvard.edu/abs/2016MNRAS.457..786S} {457, 786}

\bibitem[\protect\citeauthoryear{{Takada} \& {Hu}}{{Takada} \&
  {Hu}}{2013}]{takada13}
{Takada} M.,  {Hu} W.,  2013, \mn@doi [\prd] {10.1103/PhysRevD.87.123504},
  \href {http://adsabs.harvard.edu/abs/2013PhRvD..87l3504T} {87, 123504}

\bibitem[\protect\citeauthoryear{{Takada} \& {Jain}}{{Takada} \&
  {Jain}}{2009}]{takada09}
{Takada} M.,  {Jain} B.,  2009, \mn@doi [\mnras]
  {10.1111/j.1365-2966.2009.14504.x}, \href
  {https://ui.adsabs.harvard.edu/abs/2009MNRAS.395.2065T} {395, 2065}

\bibitem[\protect\citeauthoryear{{Takahashi}, {Sato}, {Nishimichi}, {Taruya}
  \& {Oguri}}{{Takahashi} et~al.}{2012}]{takahashi12}
{Takahashi} R.,  {Sato} M.,  {Nishimichi} T.,  {Taruya} A.,   {Oguri} M.,
  2012, \mn@doi [\apj] {10.1088/0004-637X/761/2/152}, \href
  {http://adsabs.harvard.edu/abs/2012ApJ...761..152T} {761, 152}

\bibitem[\protect\citeauthoryear{{Tamone} et~al.,}{{Tamone}
  et~al.}{2020}]{tamone20}
{Tamone} A.,  et~al., 2020, \mn@doi [\mnras] {10.1093/mnras/staa3050}, \href
  {https://ui.adsabs.harvard.edu/abs/2020MNRAS.499.5527T} {499, 5527}

\bibitem[\protect\citeauthoryear{{Tonegawa}, {Okumura}, {Totani}, {Dalton},
  {Glazebrook}  \& {Yabe}}{{Tonegawa} et~al.}{2018}]{tonegawa17}
{Tonegawa} M.,  {Okumura} T.,  {Totani} T.,  {Dalton} G.,  {Glazebrook} K.,
  {Yabe} K.,  2018, \mn@doi [\pasj] {10.1093/pasj/psy030}, \href
  {https://ui.adsabs.harvard.edu/abs/2018PASJ...70...41T} {70, 41}

\bibitem[\protect\citeauthoryear{{Troxel} \& {Ishak}}{{Troxel} \&
  {Ishak}}{2015}]{troxel15}
{Troxel} M.~A.,  {Ishak} M.,  2015, \mn@doi [\physrep]
  {10.1016/j.physrep.2014.11.001}, \href
  {http://adsabs.harvard.edu/abs/2015PhR...558....1T} {558, 1}

\bibitem[\protect\citeauthoryear{{Troxel} et~al.,}{{Troxel}
  et~al.}{2018}]{y1cosmicshear}
{Troxel} M.~A.,  et~al., 2018, \mn@doi [\prd] {10.1103/PhysRevD.98.043528},
  \href {https://ui.adsabs.harvard.edu/abs/2018PhRvD..98d3528T} {98, 043528}

\bibitem[\protect\citeauthoryear{{Vlah}, {Chisari}  \& {Schmidt}}{{Vlah}
  et~al.}{2020}]{vlah20}
{Vlah} Z.,  {Chisari} N.~E.,   {Schmidt} F.,  2020, \mn@doi [\jcap]
  {10.1088/1475-7516/2020/01/025}, \href
  {https://ui.adsabs.harvard.edu/abs/2020JCAP...01..025V} {2020, 025}

\bibitem[\protect\citeauthoryear{{\PRF{Wietersheim-Kramsta}{von}{von}
  Wietersheim-Kramsta} et~al.,}{{\PRF{Wietersheim-Kramsta}{von}{von}
  Wietersheim-Kramsta} et~al.}{2021}]{vonWietersheim21}
{\PRF{Wietersheim-Kramsta}{von}{von} Wietersheim-Kramsta} M.,  et~al., 2021,
  \mn@doi [\mnras] {10.1093/mnras/stab1000}, \href
  {https://ui.adsabs.harvard.edu/abs/2021MNRAS.504.1452V} {504, 1452}

\bibitem[\protect\citeauthoryear{{Wright} et~al.,}{{Wright}
  et~al.}{2010}]{wright10}
{Wright} E.~L.,  et~al., 2010, \mn@doi [\aj] {10.1088/0004-6256/140/6/1868},
  \href {https://ui.adsabs.harvard.edu/abs/2010AJ....140.1868W} {140, 1868}

\bibitem[\protect\citeauthoryear{{Zacharegkas} et~al.,}{{Zacharegkas}
  et~al.}{2022}]{zacharegkas22}
{Zacharegkas} G.,  et~al., 2022, \mn@doi [\mnras] {10.1093/mnras/stab3155},
  \href {https://ui.adsabs.harvard.edu/abs/2022MNRAS.509.3119Z} {509, 3119}

\bibitem[\protect\citeauthoryear{{Zhang}}{{Zhang}}{2010}]{zhang10}
{Zhang} P.,  2010, \mn@doi [\apj] {10.1088/0004-637X/720/2/1090}, \href
  {https://ui.adsabs.harvard.edu/abs/2010ApJ...720.1090Z} {720, 1090}

\bibitem[\protect\citeauthoryear{{Zhou} et~al.,}{{Zhou} et~al.}{2020}]{zhou20}
{Zhou} R.,  et~al., 2020, \mn@doi [Research Notes of the American Astronomical
  Society] {10.3847/2515-5172/abc0f4}, \href
  {https://ui.adsabs.harvard.edu/abs/2020RNAAS...4..181Z} {4, 181}

\bibitem[\protect\citeauthoryear{{Zuntz} et~al.,}{{Zuntz}
  et~al.}{2015}]{zuntz14}
{Zuntz} J.,  et~al., 2015, \mn@doi [Astronomy and Computing]
  {10.1016/j.ascom.2015.05.005}, \href
  {http://adsabs.harvard.edu/abs/2015A%26C....12...45Z} {12, 45}

\bibitem[\protect\citeauthoryear{{Zuntz} et~al.,}{{Zuntz}
  et~al.}{2018}]{y1shearcat}
{Zuntz} J.,  et~al., 2018, \mn@doi [\mnras] {10.1093/mnras/sty2219}, \href
  {https://ui.adsabs.harvard.edu/abs/2018MNRAS.481.1149Z} {481, 1149}

\makeatother
\end{thebibliography}

\section*{Author Affiliations}

$^{1}$Department of Physics, Northeastern University, Boston, MA, 02115, USA\\
$^{2}$McWilliams Center for Cosmology, Department of Physics, Carnegie Mellon University, Pittsburgh, PA 15213, USA\\
$^{3}$ NSF AI Planning Institute for Physics of the Future, Carnegie Mellon University, Pittsburgh, PA 15213, USA\\
$^{4}$ Department of Applied Mathematics and Theoretical Physics, University of Cambridge, Cambridge CB3 0WA, UK\\
$^{5}$ Kavli Institute for Cosmological Physics, University of Chicago, Chicago, IL 60637, USA\\
$^{6}$ Institute of Astronomy, University of Cambridge, Madingley Road, Cambridge CB3 0HA, UK\\
$^{7}$ Kavli Institute for Cosmology, University of Cambridge, Madingley Road, Cambridge CB3 0HA, UK\\
$^{8}$ Department of Astronomy and Astrophysics, University of Chicago, Chicago, IL 60637, USA\\
$^{9}$ Department of Physics and Astronomy, University of Waterloo, 200 University Ave W, Waterloo, ON N2L 3G1, Canada\\
$^{10}$ Center for Cosmology and Astro-Particle Physics, The Ohio State University, Columbus, OH 43210, USA\\
$^{11}$ Argonne National Laboratory, 9700 South Cass Avenue, Lemont, IL 60439, USA\\
$^{12}$ Institute for Astronomy, University of Hawai'i, 2680 Woodlawn Drive, Honolulu, HI 96822, USA\\
$^{13}$ Physics Department, 2320 Chamberlin Hall, University of Wisconsin-Madison, 1150 University Avenue Madison, WI  53706-1390\\
$^{14}$ Department of Physics and Astronomy, University of Pennsylvania, Philadelphia, PA 19104, USA\\
$^{15}$ Instituto de Astrofisica de Canarias, E-38205 La Laguna, Tenerife, Spain\\
$^{16}$ Laborat\'orio Interinstitucional de e-Astronomia - LIneA, Rua Gal. Jos\'e Cristino 77, Rio de Janeiro, RJ - 20921-400, Brazil\\
$^{17}$ Universidad de La Laguna, Dpto. Astrofísica, E-38206 La Laguna, Tenerife, Spain\\
$^{18}$ Center for Astrophysical Surveys, National Center for Supercomputing Applications, 1205 West Clark St., Urbana, IL 61801, USA\\
$^{19}$ Department of Astronomy, University of Illinois at Urbana-Champaign, 1002 W. Green Street, Urbana, IL 61801, USA\\
$^{20}$ Physics Department, William Jewell College, Liberty, MO, 64068\\
$^{21}$ Department of Physics, Duke University Durham, NC 27708, USA\\
$^{22}$ NASA Goddard Space Flight Center, 8800 Greenbelt Rd, Greenbelt, MD 20771, USA\\
$^{23}$ Institut d'Estudis Espacials de Catalunya (IEEC), 08034 Barcelona, Spain\\
$^{24}$ Institute of Space Sciences (ICE, CSIC),  Campus UAB, Carrer de Can Magrans, s/n,  08193 Barcelona, Spain\\
$^{25}$ Kavli Institute for Particle Astrophysics \& Cosmology, P. O. Box 2450, Stanford University, Stanford, CA 94305, USA\\
$^{26}$ Lawrence Berkeley National Laboratory, 1 Cyclotron Road, Berkeley, CA 94720, USA\\
$^{27}$ Universit\'e Grenoble Alpes, CNRS, LPSC-IN2P3, 38000 Grenoble, France\\
$^{28}$ Fermi National Accelerator Laboratory, P. O. Box 500, Batavia, IL 60510, USA\\
$^{29}$ Jet Propulsion Laboratory, California Institute of Technology, 4800 Oak Grove Dr., Pasadena, CA 91109, USA\\
$^{30}$ Institut de F\'{\i}sica d'Altes Energies (IFAE), The Barcelona Institute of Science and Technology, Campus UAB, 08193 Bellaterra (Barcelona) Spain\\
$^{31}$ University Observatory, Faculty of Physics, Ludwig-Maximilians-Universit\"at, Scheinerstr. 1, 81679 Munich, Germany\\
$^{32}$ School of Physics and Astronomy, Cardiff University, CF24 3AA, UK\\
$^{33}$ Mila, 6666 St-Urbain Street, 200 Montreal, QC, H2S 3H1, Canada\\
$^{34}$ Department of Physics, Stanford University, 382 Via Pueblo Mall, Stanford, CA 94305, USA\\
$^{35}$ Instituto de F\'isica Gleb Wataghin, Universidade Estadual de Campinas, 13083-859, Campinas, SP, Brazil\\
$^{36}$ Institute for Astronomy, University of Edinburgh, Edinburgh EH9 3HJ, UK\\
$^{37}$ Department of Physics, University of Genova and INFN, Via Dodecaneso 33, 16146, Genova, Italy\\
$^{38}$ Centro de Investigaciones Energ\'eticas, Medioambientales y Tecnol\'ogicas (CIEMAT), Madrid, Spain\\
$^{39}$ Jodrell Bank Center for Astrophysics, School of Physics and Astronomy, University of Manchester, Oxford Road, Manchester, M13 9PL, UK\\
$^{40}$ SLAC National Accelerator Laboratory, Menlo Park, CA 94025, USA\\
$^{41}$ Department of Astronomy and Space Science, Sejong University, 209, Neungdong-ro, Gwangjin-gu, Seoul, South Korea
$^{42}$ Brookhaven National Laboratory, Bldg 510, Upton, NY 11973, USA\\
$^{43}$ Department of Physics and Astronomy, Stony Brook University, Stony Brook, NY 11794, USA\\
$^{44}$ D\'{e}partement de Physique Th\'{e}orique and Center for Astroparticle Physics, Universit\'{e} de Gen\`{e}ve, 24 quai Ernest Ansermet, CH-1211 Geneva, Switzerland\\
$^{45}$ Department of Physics, University of Michigan, Ann Arbor, MI 48109, USA\\
$^{46}$ Cerro Tololo Inter-American Observatory, NSF's National Optical-Infrared Astronomy Research Laboratory, Casilla 603, La Serena, Chile\\
$^{47}$ Institute of Cosmology and Gravitation, University of Portsmouth, Portsmouth, PO1 3FX, UK\\
$^{48}$ CNRS, UMR 7095, Institut d'Astrophysique de Paris, F-75014, Paris, France\\
$^{49}$ Sorbonne Universit\'es, UPMC Univ Paris 06, UMR 7095, Institut d'Astrophysique de Paris, F-75014, Paris, France\\
$^{50}$ Department of Physics \& Astronomy, University College London, Gower Street, London, WC1E 6BT, UK\\
$^{51}$ Astronomy Unit, Department of Physics, University of Trieste, via Tiepolo 11, I-34131 Trieste, Italy\\
$^{52}$ INAF-Osservatorio Astronomico di Trieste, via G. B. Tiepolo 11, I-34143 Trieste, Italy\\
$^{53}$ Institute for Fundamental Physics of the Universe, Via Beirut 2, 34014 Trieste, Italy\\
$^{54}$ Hamburger Sternwarte, Universit\"{a}t Hamburg, Gojenbergsweg 112, 21029 Hamburg, Germany\\
$^{55}$ Department of Physics, IIT Hyderabad, Kandi, Telangana 502285, India\\
$^{56}$ Institute of Theoretical Astrophysics, University of Oslo. P.O. Box 1029 Blindern, NO-0315 Oslo, Norway\\
$^{57}$ Instituto de Fisica Teorica UAM/CSIC, Universidad Autonoma de Madrid, 28049 Madrid, Spain\\
$^{58}$ School of Mathematics and Physics, University of Queensland,  Brisbane, QLD 4072, Australia\\
$^{59}$ Santa Cruz Institute for Particle Physics, Santa Cruz, CA 95064, USA\\
$^{60}$ Center for Astrophysics $\vert$ Harvard \& Smithsonian, 60 Garden Street, Cambridge, MA 02138, USA\\
$^{61}$ Australian Astronomical Optics, Macquarie University, North Ryde, NSW 2113, Australia\\
$^{62}$ Lowell Observatory, 1400 Mars Hill Rd, Flagstaff, AZ 86001, USA\\
$^{63}$ George P. and Cynthia Woods Mitchell Institute for Fundamental Physics and Astronomy, and Department of Physics and Astronomy, Texas A\&M University, College Station, TX 77843,  USA\\
$^{64}$ Department of Astrophysical Sciences, Princeton University, Peyton Hall, Princeton, NJ 08544, USA\\
$^{65}$ Instituci\'o Catalana de Recerca i Estudis Avan\c{c}ats, E-08010 Barcelona, Spain\\
$^{66}$ Department of Physics and Astronomy and PITT PACC, University of Pittsburgh\\
$^{67}$ Department of Astronomy, University of California, Berkeley,  501 Campbell Hall, Berkeley, CA 94720, USA\\
$^{68}$ Observat\'orio Nacional, Rua Gal. Jos\'e Cristino 77, Rio de Janeiro, RJ - 20921-400, Brazil\\
$^{69}$ School of Physics and Astronomy, University of Southampton,  Southampton, SO17 1BJ, UK\\
$^{70}$ Computer Science and Mathematics Division, Oak Ridge National Laboratory, Oak Ridge, TN 37831\\

\appendix
\section{Modelling Redshift Space Distortions \& IA Anisotropy}\label{app:rsds}

In this appendix we set out the formalism used to estimate the impact of RSDs and anisotropic IAs on our results (see also \citealt{singh16}). One can write the redshift-space galaxy-galaxy power spectrum in terms of
the (isotropic) real space equivalent in the form:

\begin{equation}\label{eq:app:pkbeta}
P_{gg,s}(k) = 
\left ( 1 + \beta_a \mu \right )
\left ( 1 + \beta_b \mu \right )
P_{gg}(k),
\end{equation}

\noindent
with galaxy samples $a,b$ and $\beta_a \equiv f(z)/b_{g,a}$, 
the ratio of the logarithmic growth rate to the linear galaxy bias. The factor $\mu$ is the cosine of the angle between mode $\mathbf{k}$ and the axis of the line of sight $\hat{\mathbf{z}}$, $\mu = \hat{\mathbf{k}} \cdot \hat{\mathbf{z}}$. This is an approximation that applies on linear scales, but begins to break down at large $k$ \citep{kaiser87}. In general, one can decompose Equation~\eqref{eq:app:pkbeta} in terms
of Legendre polynomials $\mathcal{P}_\ell$,

\begin{equation}\label{eq:app:legendre}
P_{gg,s}(k) = \left [ \sum^{2}_{\ell=0} \alpha^{ab}_{2\ell} \mathcal{P}_{2\ell} \right ] P_{gg}(k).
\end{equation}

\noindent
That is, the sum of monopole, quadrupole and hexadecupole
contributions. Note that in the case that $\ell=0, \beta=0$,
the above reverts to the isotropic case and $P_{gg,s}=P_{gg}$.
The coefficients have the same form as equations 48-50 of \citet{baldauf10}
for even values of $\ell$,

\begin{equation}
\alpha^{ab}_{2\ell} = 
\begin{cases}
1 + \frac{1}{3}(\beta_a+\beta_b) + \frac{1}{5}\beta_a\beta_b          \;\;\;\;\;\;\;\;\;\;\;\;\;\;\;\;\;\; \ell=0 \\
\frac{2}{3}(\beta_a+\beta_b) + \frac{4}{7}\beta_a\beta_b    \;\;\;\;\;\;\;\;\;\;\;\;\;\;\;\;\;\;\;\;\;\;\;\; \ell=1 \\
\frac{8}{35}\beta_a\beta_b                                            \;\;\;\;\;\;\;\;\;\;\;\;\;\;\;\;\;\;\;\;\;\;\;\;\;\;\;\;\;\;\;\;\;\;\;\;\;\;\;\;\;\;\;\; \ell=2 \\
\end{cases},
\end{equation} 

\noindent
and are zero otherwise. The configuration space equivalent of Eq.~\eqref{eq:app:legendre} has a similar form:

\begin{equation}\label{eq:app:config_legendre}
\xi_{gg,s}(\rp,\Pi) =  \sum^{2}_{\ell=0} \alpha^{ab}_{2\ell} \mathcal{P}_{2\ell} \xi_{gg,2\ell}(\rp,\Pi),
\end{equation}

\noindent
with 

\begin{equation}\label{eq:app:xi_2ell}
\xi_{gg,2\ell}(\rp,\Pi) = \frac{(-1)^{\ell}}{2\pi} \int k^{2} P_{gg}(k) j_{2\ell}(kr) \mathrm{d}k.
\end{equation}

\noindent
The integration kernel $j_{\mu}$ is a spherical Bessel function
of the first kind of order $\mu$, and it is this that determines the
shape of each term.  
Putting these pieces together, and integrating over line of sight
separation, one finally obtains the expression:

\begin{equation}\label{eq:app:wgg}
w_{gg}(\rp) = \sum^{2}_{\ell=0} \frac{(-1)^{\ell}}{2\pi} \alpha_{2\ell}\mathcal{P}_{2 \ell} 
\int_{- \Pi_{\rm max} }^{\Pi_{\rm max} } 
\int k^{2} P_{gg}(k) j_{2 \ell}(kr) 
\mathrm{d}k 
\mathrm{d}\Pi 
\end{equation}

\noindent
Although RSDs themselves do not have a significant impact on $w_{g+}$ (see \citealt{singh15}), there is an analogous effect due to the projection of 3D shapes into 2D space. This suppresses the observed alignment strength at $|\Pi|>0$, and so alters the shape of $\xi_{g+}$ in the $\rp-\Pi$ plane (see the discussion in Section \ref{sec:theory:scale_cuts:wgp}). The impact can be modelled in a very similar way to with RSDs (\citealt{singh16}, Sec.~2.3 and eq.~13).

\begin{figure}
\includegraphics[width=\columnwidth]{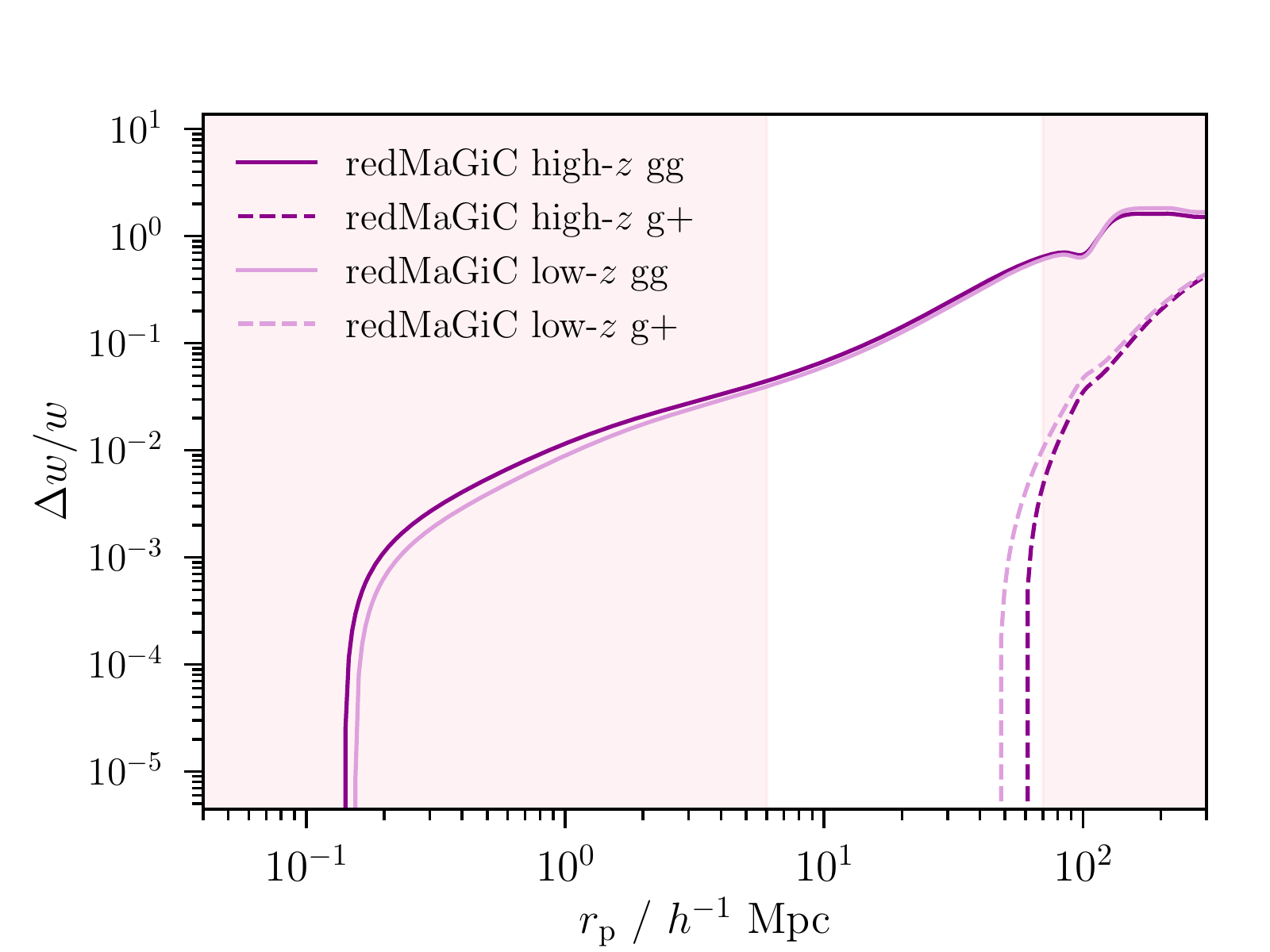}
\caption{The fractional impact of redshift space distortions and projection effects on projected galaxy-galaxy and galaxy-shape correlation functions. Note that the difference is defined as the magnitude of the difference between theory predictions (in either $w_{g+}$ or $w_{gg}$, as labelled) with and without RSDs/projection effects $\Delta w_{ab} = |w^{\rm RSD}_{ab} - w^{\rm no\;RSD}_{ab}  |$. The theory predictions are generated at our fiducial cosmology and $A_1=1$. We show both high- and low-$z$ \redmagic~samples to illustrate the impact of (quite significant) differences in the redshift distributions.  
}
\label{fig:app:rsds}
\end{figure}

Figure \ref{fig:app:rsds} shows the absolute impact of the additional RSD signal and the projection effect described above (note that RSDs and IA anisotropy work in opposite directions, and so the sign of the two eddects in Figure \ref{fig:app:rsds} are different). RSDs have an impact on $w_{gg}$ at the level of tens of percent on scales $\rp>6\mpc$. We thus expect to be sensitive to their impact, and include them in our fiducial model. In the case of $w_{g+}$ we see an impact on very large scales, dropping away below $\sim70\mpc$. Given that we impose an upper scale cut at $\rp=70\mpc$, due to possible large scale systematics, we do not consider it necessary to include IA anisotropy in our fiducial model for $w_{g+}$. 

\section{Null Tests}\label{app:null_tests}

Before carrying out our analysis, we carried our various validation tests. Among those was a null test, constructed by repeating our $w_{g+}$ measurements, but using shapes measured at 45 degrees to the tangential/radial direction. In the absence of systematics, this should return no signal.

\begin{figure*}
\includegraphics[width=2\columnwidth]{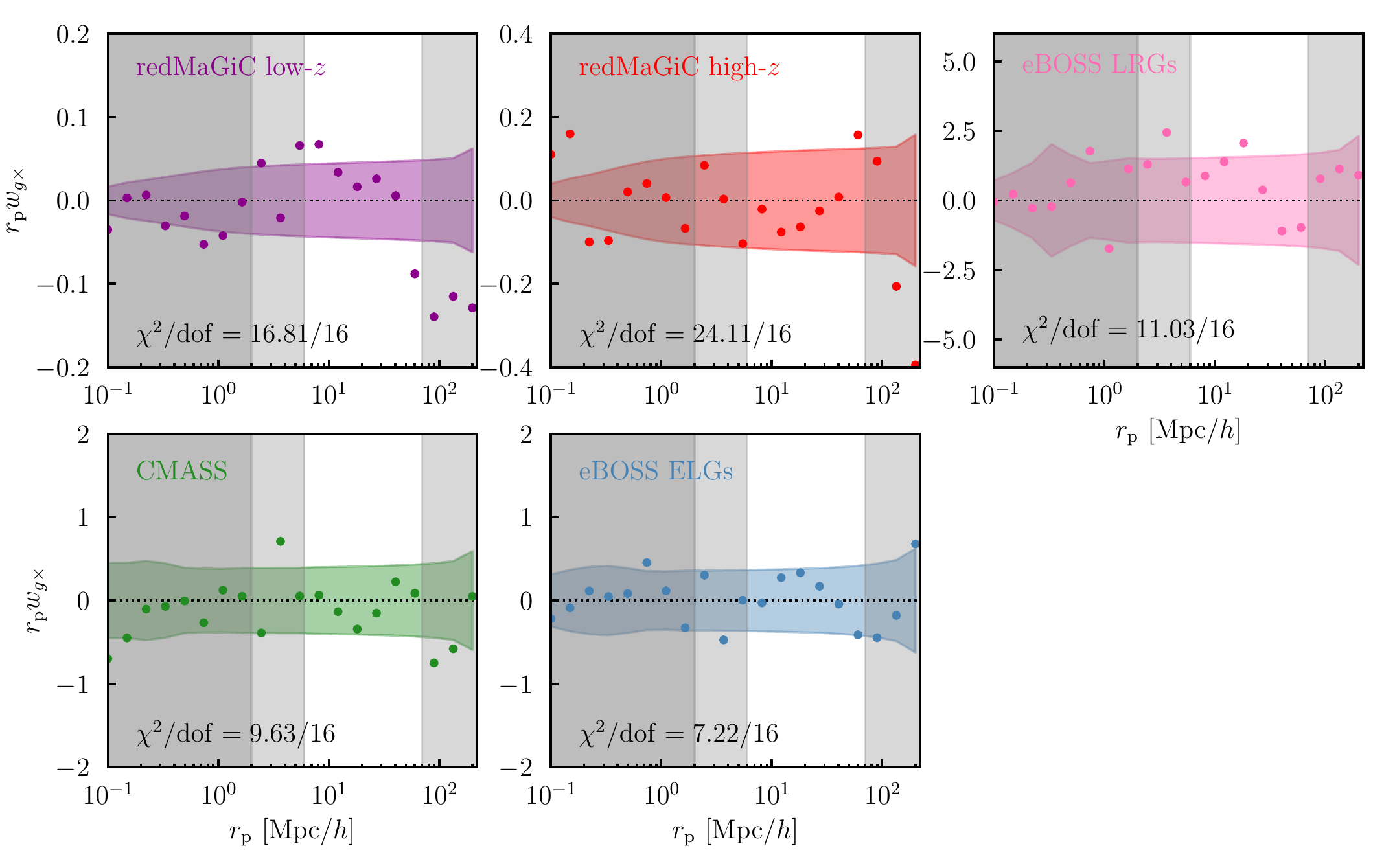}
\caption{Cross shear correlations. The measurements shown are computed in the same way as $w_{g+}$ (see Equation \eqref{eq:wgp_basic}), but with the galaxy shapes rotated by 45 degrees. This is meant as a null test, since to first order neither intrinsic alignments nor gravitational lensing produce such correlations. The coloured shaded regions show the estimated shape+shot noise uncertainty for each sample. The grey shaded bands indicate scales discarded in our two analysis setups. In each case, the quoted null $\chi^2$ is computed on all scales $\rp<70\mpc$.}
\label{fig:null_tests}
\end{figure*}

The results for our five samples are shown in Figure~\ref{fig:null_tests}. The error bands here are calculated assuming shape (and shot) noise only, using the observed number of galaxy pairs in each $\rp$ bin. 
In the case of the two \redmagic~samples, we see a slight increase in (negative) power on very large scales. The reason for this apparent signal is not known for certain. We treat it as an unknown systematic, and simply choose to remove the affected scales. After imposing an upper limit at $\rp<70\mpc$, we find $w_{g\times}$ to be consistent with zero on all surviving scales. The null $\chi^2$ values are shown for each sample in Figure \ref{fig:null_tests}. Even in the case with the worst goodness-of-fit, \redmagic~high-$z$, we find a $\chi^2/\mathrm{dof}=24.1/16$, giving a corresponding $p-$value $p=0.09$.

\section{Comparison with LOWZ}\label{app:lowz}

\begin{figure}
\includegraphics[width=\columnwidth]{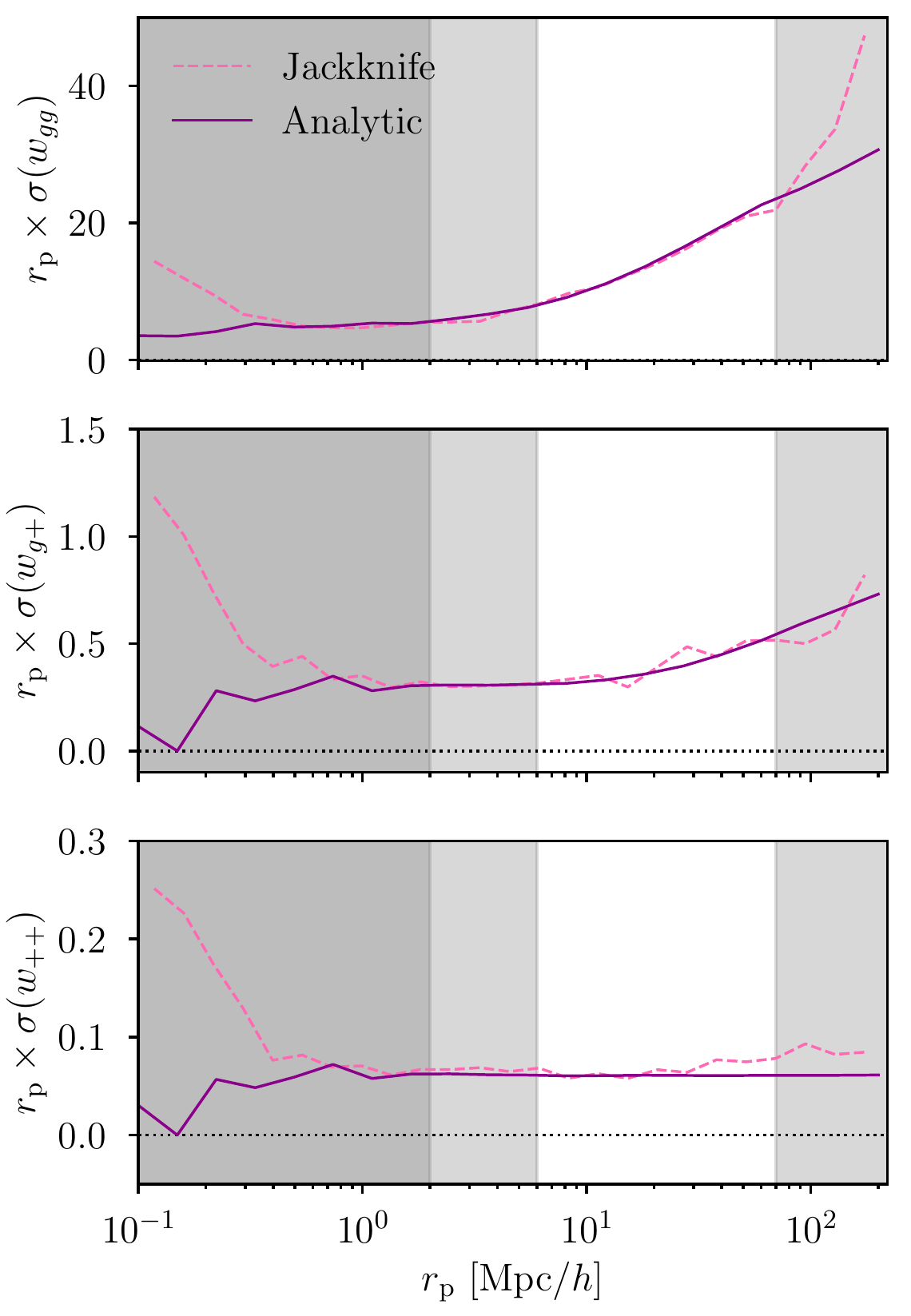}
\caption{A comparison of the square root of the LOWZ covariance diagonals obtained using two methods. From top we show $w_{g+}$, $w_{gg}$ and $w_{++}$.  In each case the shaded purple regions indicate scales excluded from our large scale intrinsic alignment fits. We see very good agreement between the two estimates on the scales of interest. }
\label{fig:validation:lowz_covmats}
\end{figure}

\begin{figure*}
\includegraphics[width=1.8\columnwidth]{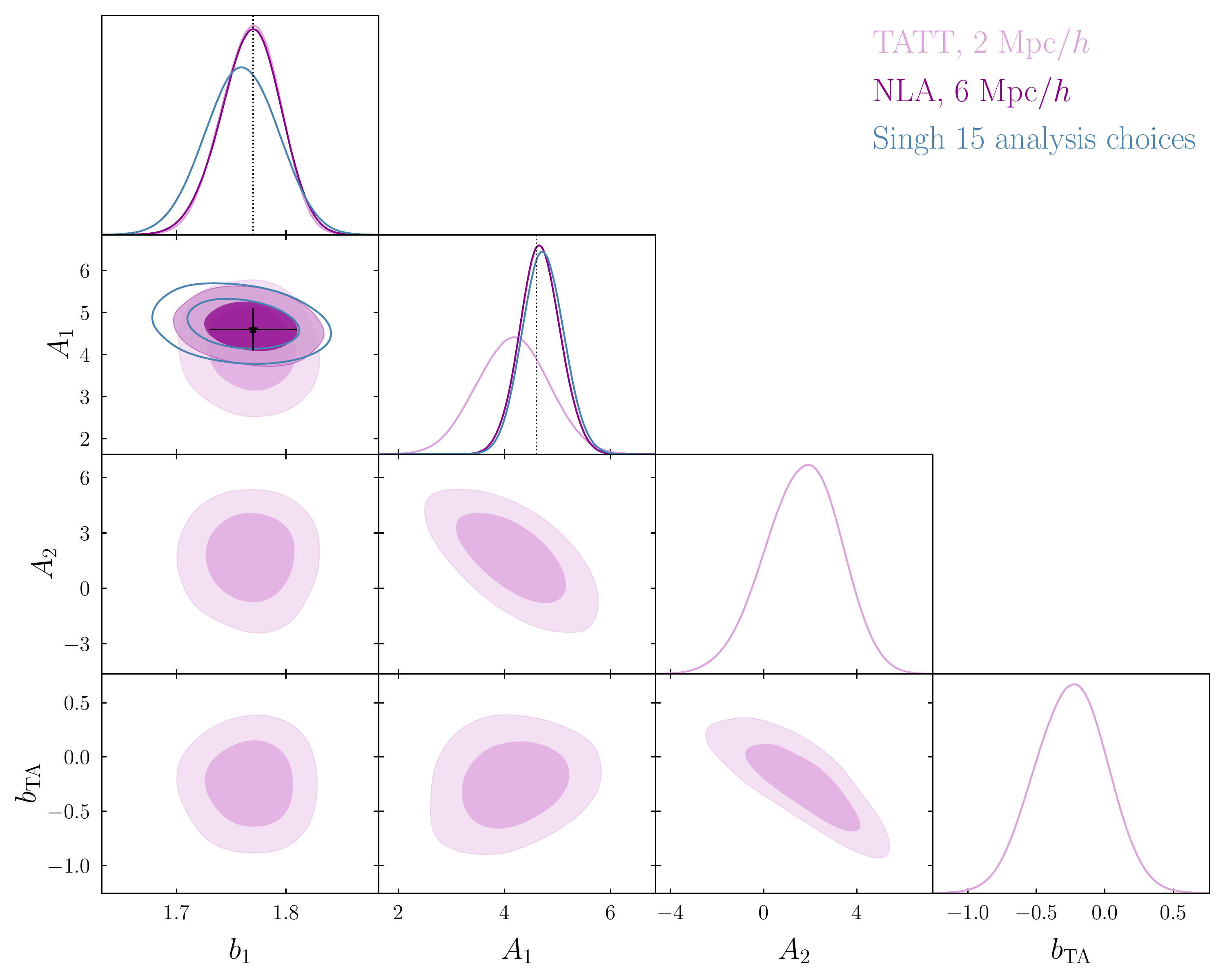}
\caption{$68\%$ and $95\%$ confidence intervals from our reanalysis of the SDSS LOWZ data. The black cross represents the published constraint on $A_1$ and galaxy bias from \citet{singh15}. In blue (open contours) we show the result of fitting the LOWZ data vector using analysis choices matched to those of \citet{singh15}. We recover the best fitting $A_1$ well. The filled contours then show the result of switching to our preferred analysis settings, using the NLA and TATT models.   }
\label{fig:validation:lowz_post}
\end{figure*}

As discussed in Section~\ref{sec:validation}, we carry out several layers of pipeline testing and validation using LOWZ. The LOWZ LRG sample is useful for this, in that it is a relatively well understood data set, which gives a high signal-to-noise $w_{g+}$ signal. Crucially, there are also published IA measurements to which we can compare \citep{singh15}.

In addition to the data vector level comparison described in Section \ref{sec:validation}, we also use the LOWZ data to help validate our analytic covariance estimates. LOWZ covers a broad more or less contiguous footprint, making jackknife estimates viable. We divide that footprint into 100 patches using a $k-$means algorithm, and iteratively re-measure the whole data vector ($w_{gg}+w_{g+}+w_{++}$) in each. The diagonal elements of the resulting jackknife covariance matrix are compared to our analytic estimate in Figure~\ref{fig:validation:lowz_covmats}. As expected the latter is somewhat smoother. The two diverge slightly on very large scales, where the approximations behind the jackknife method break down. On the scales of interest, however, we see very good agreement.

We also carry out an end-to-end reanalysis of LOWZ using our pipeline. Starting with galaxy catalogues and randoms, we remeasure the joint data vector. Using our analytic covariance matrix, and the modelling pipeline set out in Section~\ref{sec:theory}, we obtain parameter constraints. 
The results of this exercise are summarised in Figure~\ref{fig:validation:lowz_post}. In black we show the published IA and bias results from \citet{singh15}; note that the fits for $b_1$ and $A_1$ were performed serially, and so we have a point with error bars instead of a full contour. 
The open blue contour shows the result of analysing the LOWZ data using our pipeline, but with all the analysis choices matched to those of \citet{singh15}. These are detailed in Section~\ref{sec:validation}, but include the choice of cosmology and the version of \blockfont{HaloFit}. We see good agreement in both parameters. 

The filled contours then shown the impact of switching to our analysis choices, assuming the NLA and TATT models. The latter (dark purple) gives a very similar $A_1$ constraint to the original \citet{singh15} analysis. This is reassuring, in the sense that it suggests the new results from our pipeline are readily comparable with those in the literature. 
The lighter purple contours show the impact of opening up the TATT parameter space, and also extending the minimum scale in $w_{g+}$ and $w_{++}$ down to $2\mpc$. The marginalised $A_1$ constraint is broadened and shifted downwards slightly, primarily due to the degeneracy with $A_2$. It is interesting to briefly note here that although the contours on the extra parameters ($A_2$ and $b_{\rm TA}$) are not symmetric about zero, they are totally consistent with zero. That is, the LOWZ data does not appear to require additional terms beyond the nonlinear alignment model to describe scales down to $2\mpc$. 

\section{Constraints on galaxy bias}\label{app:bias_constraints}

\begin{table}
\begin{center}
\begin{tabular}{l|cc}
\hline
Sample & $b_1$ & $b_2$ \\
\hhline{===}
redMaGiC low$-z$ & $1.59^{+0.01}_{-0.01}$ & $-0.09^{+0.07}_{-0.07}$ \\
redMaGiC high$-z$ & $1.81^{+0.01}_{-0.01}$ & $0.39^{+0.08}_{-0.07}$ \\
eBOSS LRGs & $2.20^{+0.03}_{-0.03}$ & $0.34^{+0.17}_{-0.16}$ \\
eBOSS ELGs & $1.37^{+0.06}_{-0.03}$ & $-0.84^{+0.74}_{-0.64}$ \\
CMASS & $1.97^{+0.02}_{-0.02}$ & $0.01^{+0.13}_{-0.15}$ \\
\hline
\end{tabular}
\caption{Constraints on galaxy bias from our various density tracer samples.
}\label{tab:results:bias}
\end{center}
\end{table}

In all samples, and for all fits, we include in our model free parameters for galaxy bias. Although the bias constraints are almost entirely dominated by the $w_{gg}$ part of the data vector, we allow bias to vary alongside our IA parameters. Justified by the exercise in Section~\ref{sec:theory:scale_cuts:wgg}, our model includes two free parameters: $b_1$ and $b_2$ (there are additional terms in the expression for $P_{\delta_g}$, but the model is fully specified by the two values; see Section~\ref{sec:theory:galaxy_power_spectra}).

The main galaxy bias results from each of our samples are presented in Table~\ref{tab:results:bias}. In each case, the model provides a reasonable fit to the joint data vector. 
The redMaGiC numbers here are qualitatively consistent with those presented in the upper panel of \citet{y3-3x2ptkp}'s Fig.~8 (the pale purple points). That there are some small differences in the actual numbers in not surprising, given the different nature of the analysis (e.g., we are assuming a particular fixed cosmology). The values are roughly in line with the expectation for these sorts of galaxy samples.

\section{Redshift dependence of alignments in red galaxies}\label{app:redshift_dependence}

\begin{figure}
\includegraphics[width=1.\columnwidth]{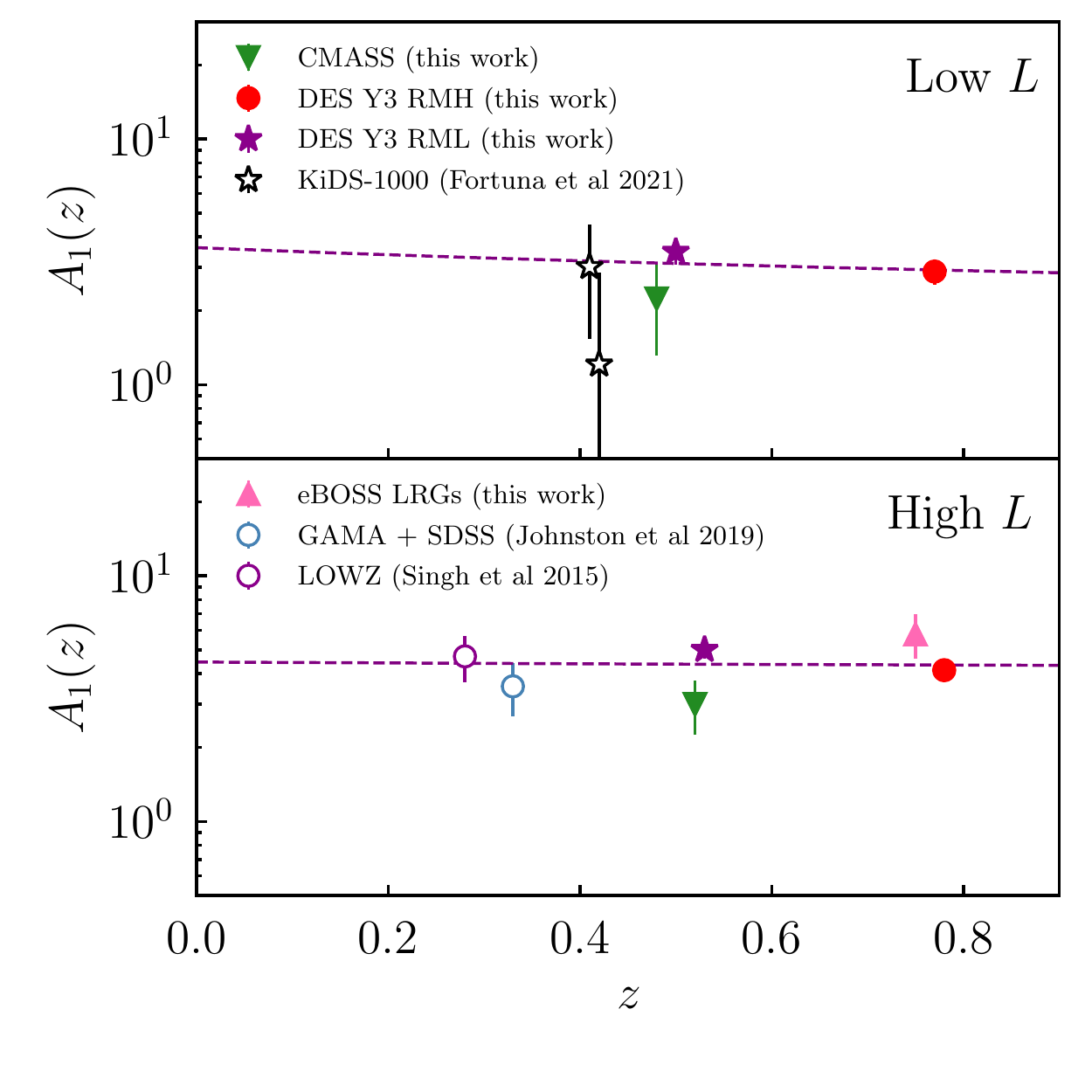}
\caption{Intrinsic alignment strength as a function of estimated mean redshift in red galaxies. To avoid complication due to evolution in galaxy properties, we take only points within two narrow bands in $L/L_0$. See Section~\ref{sec:results:luminosity_dependence} for discussion. We show best fitting power law slopes, fit to each panel; in both cases the power law index $\eta_1$ is consistent with zero to $\ll 1\sigma$.}
\label{fig:results:redshift_dependence}
\end{figure}

In this appendix we illustrate the redshift dependence of our red samples. Figure~\ref{fig:results:redshift_dependence} is the counterpart to Figure~\ref{fig:results:luminosity_dependence}, but showing the trend with redshift rather than luminosity. The colour scheme for the different samples is the same in the two. 

For the sake of comparability, we define two narrow bins in $L_r$, and consider the redshift dependence in each. As explained in Section \ref{sec:results:luminosity_dependence}, these are centred on $\mathrm{log}L/L_0 \sim -0.3$ and $\mathrm{log}L/L_0 \sim -0.05$ respectively. The samples included in these two bins are shown in the upper/lower panels of Figure~\ref{fig:results:redshift_dependence} (labelled ``Low $L$" and ``High $L$"). The idea here is to separate inherent evolution in redshift (i.e. in a fixed sample with unchanging observable properties) from the evolution of galaxy selection with $z$.

As we can see here, there is no clear trend over the baseline of the samples, in either luminosity bin. If one fits a slope in redshift of the form $A_1(z)\propto[(1+z)/(1+z_0)]^{\eta_1}$, where $z_0=0.62$, the results are consistent with $\eta_1=0$. Specifically, we find $\eta_1 = -0.37\pm 0.94$ in the lower $L$ bin, and $\eta_1 = -0.05 \pm 0.73$ in the upper $L$ bin.

\label{lastpage}
\end{document}